\documentclass[12pt]{article}

\usepackage{scicite}
\usepackage{textcomp}
\usepackage{gensymb}
\usepackage{times}
\usepackage{graphicx}
\usepackage{placeins}
\makeatletter
\let\saved@includegraphics\includegraphics
\AtBeginDocument{\let\includegraphics\saved@includegraphics}
\makeatother
\usepackage{xcolor}
\usepackage{amssymb}
\usepackage[normalem]{ulem}
\usepackage{comment}
\usepackage{setspace}
\usepackage{float} 

\newenvironment{sciabstract}{%
\begin{quote} \bf}
{\end{quote}}

\title{Dissipation-enabled hydrodynamic conductivity in a tunable bandgap semiconductor} 

\author
{Cheng Tan$^{1,2,\dagger}$ , Derek Y.~H.~Ho$^{3,4,\dagger}$, Lei Wang$^{5,6}$, Jia I.~A.~Li $^{7}$, Indra Yudhistira$^{4,8}$,\\ 
Daniel A.~Rhodes$^{1}$,Takashi Taniguchi$^{9}$, Kenji Watanabe$^{9}$, Kenneth Shepard$^{2}$,\\ 
Paul L.~McEuen$^{5,6}$, Cory R. Dean$^{10}$, Shaffique Adam$^{3,4,8,11, \ast}$ \& 
James Hone$^{1,\ast}$\\
\\
\normalsize{$^{1}$Department of Mechanical Engineering, Columbia University, New York, NY 10027, USA,}\\
\normalsize{$^{2}$Department of Electrical Engineering, Columbia University, New York, NY 10027, USA,}\\
\normalsize{$^{3}$Yale-NUS College, 16 College Avenue West, 138614, Singapore}\\
\normalsize{$^{4}$Centre for Advanced 2D Materials and Graphene Research Centre,}\\
\normalsize{National University of Singapore, 6 Science Drive 2, 117546, Singapore}\\
\normalsize{$^{5}$Kavli Institute at Cornell for Nanoscale Science, Ithaca, NY 14853, USA}\\
\normalsize{$^{6}$Laboratory of Atomic and Solid State Physics, Cornell University, Ithaca, NY 14853, USA}\\
\normalsize{$^{7}$Department of Physics, Brown University, Providence, RI 02912, USA
 }\\
\normalsize{$^{8}$Department of Physics, National University of Singapore, 2 Science Drive 3, 117551, Singapore}\\
\normalsize{$^{9}$National Institute for Materials Science, 1-1 Namiki, Tsukuba 305-0044, Japan}\\
\normalsize{$^{10}$Department of Physics, Columbia University, New York, NY 10027, USA}\\
\normalsize{$^{11}$ Department of Materials Science and Engineering,} \\
\normalsize{National University of Singapore, 9 Engineering Drive 1, 117575, Singapore} \\
\normalsize{$^{\dagger}$ These authors contributed equally to this work.}\\
\\
\normalsize{$^\ast$To whom correspondence should be addressed; E-mail S.A.: shaffique.adam@yale-nus.edu.sg,}\\
\normalsize{E-mail J.H.: jh2228@columbia.edu.}
}

\date{}

\begin{document} 
\baselineskip24pt

\maketitle

\newpage

\begin{sciabstract}
Electronic transport in the regime where carrier-carrier collisions are the dominant scattering mechanism has taken on new relevance with the advent of ultraclean two-dimensional materials.  Here we present a combined theoretical and experimental study of ambipolar hydrodynamic transport in bilayer graphene demonstrating that the conductivity is given by the sum of two Drude-like terms that describe relative motion between electrons and holes, and the collective motion of the electron-hole plasma. As predicted, the measured conductivity of gapless, charge-neutral bilayer graphene is sample- and temperature-independent over a wide range.  Away from neutrality, the electron-hole conductivity collapses to a single curve, and a set of just four fitting parameters provides quantitative agreement between theory and experiment at all densities, temperatures, and gaps measured. This work validates recent theories for dissipation-enabled hydrodynamic conductivity and creates a link between semiconductor physics and the emerging field of viscous electronics.

\end{sciabstract}

\newpage 
\section*{Introduction}
More than fifty years ago it was predicted that it was possible for electron transport to be described by macroscopic equations of motion similar to those in classical fluid mechanics~\cite{gurzhi_hydrodynamic_1968,narozhny_hydrodynamics_2015}.  This regime is of particular relevance to low-dimensional materials such as graphene, for which interactions are intrinsically strong and disorder can be low. Within this emerging class of hydrodynamic materials, ambipolar conductors with coexisting electrons and holes (such as semimetals or small-gap semiconductors at finite temperature) are of particular interest because electron-hole scattering does not conserve current. Therefore, ambipolar materials can in principle act as \textit{hydrodynamic conductors} in which electron-hole scattering plays a dominant role in determining the conductivity, making them a promising platform for detailed experimental and theoretical exploration of hydrodynamic behavior.  Such systems are predicted~\cite{dyakonov_shallow_1993,predel_effects_2000,govorov_hydrodynamic_2004,muller_quantum-critical_2008,muller_graphene:_2009} to display rich new phenomena beyond the diffusive or ballistic transport of effectively independent carriers seen in most metals, have the potential for technological application -- for example, in the generation of terahertz radiation~\cite{mendl_coherent_2019} --  and are a readily accessible bridge between strongly correlated quantum fluids observed in otherwise unrelated fields such as quark-gluon plasmas in ion colliders and ultra-cold atomic Fermi gasses in optical traps~\cite{zaanen_electrons_2016}. 

 A striking prediction of theory is that at precise charge neutrality, carrier-carrier collisions occur at a quantum critical (Planckian) rate $k_{\rm B} T/\hbar$~\cite{ho_theoretical_2018, zarenia_breakdown_2019-1, GlennWagnerDungX.Nguyen}, that can lead to temperature-independent conductivity when electron-hole scattering is dominant. Planckian dissipation has recently been measured through THz spectroscopy of graphene~\cite{Gallagher2019}, and temperature-independent conductivity has been observed in suspended bilayer graphene~\cite{nam_electronhole_2017}. However, the former study required optical excitation of carriers to observe dominant electron-hole scattering, while the latter yielded inconsistent results across samples and the temperature range was limited to 100 K and below. Therefore there still exists no experimental platform that shows intrinsic hydrodynamic conductivity over a wide temperature range with sufficient repeatability to validate theoretical models. 

The behavior of hydrodynamic conductors away from charge neutrality is less well understood. An observed scaling of conductivity with chemical potential in suspended bilayer graphene was initially interpreted as evidence of electron-hole limited conductivity away from neutrality~\cite{nam_electronhole_2017}. However, from a first-principles viewpoint, electron-hole scattering cannot affect the net current away from perfect neutrality. Instead, more recent theory has pointed to the importance of interaction between electron-hole scattering and momentum non-conserving (dissipative) scattering from defects and phonons. While these mechanisms might naively be expected to add independently to electron-hole scattering as encapsulated in Matthiessen's rule, theory instead predicts that a more complex interplay between these processes determines conductivity, in what has been described as a dissipation-enabled hydrodynamic regime~\cite{zarenia_breakdown_2019-1,GlennWagnerDungX.Nguyen}. This theory has not yet been experimentally tested. Finally, we note that hydrodynamic conductivity is completely unexplored (theoretically or experimentally) for gapped materials.

\begin{figure*}[ht!]
   \includegraphics[width=1\textwidth]{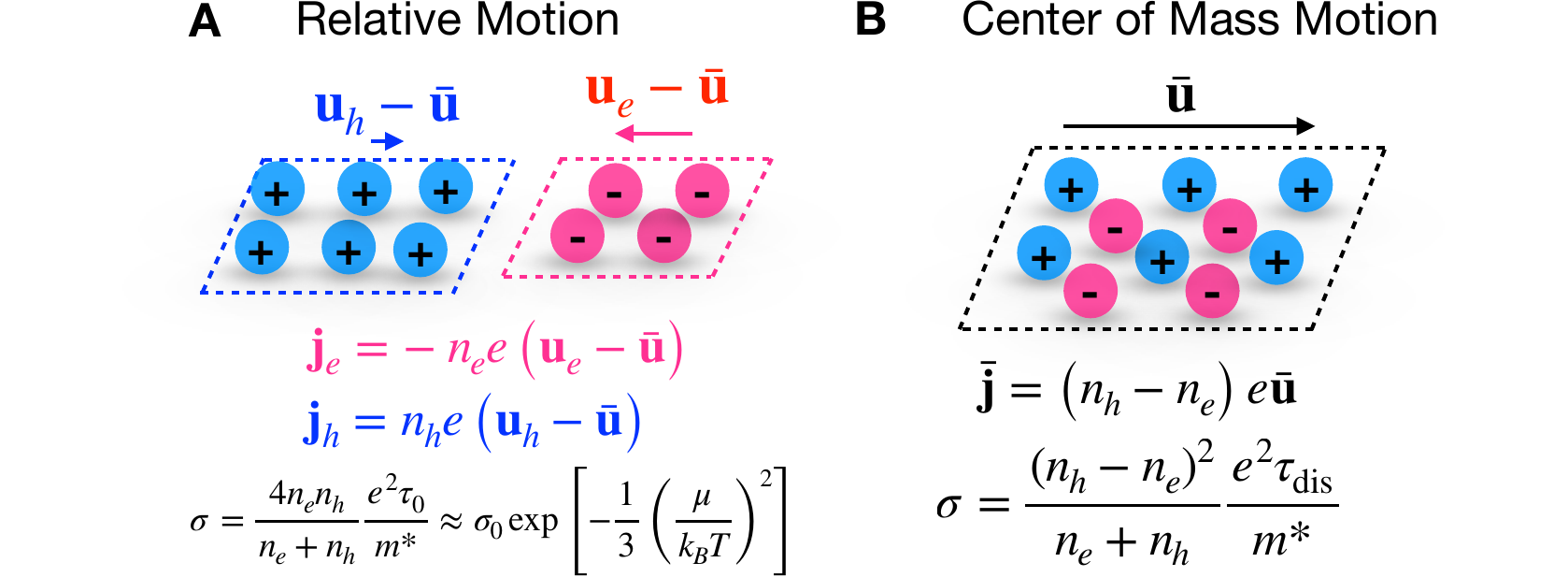}
    \caption{\doublespacing \label{fig_schematic} 
    \textbf{Schematic of Dissipative Hydrodynamics.}  In the limit of strong Coulomb interactions $\tau_0 \ll \tau_{\rm dis}$, the two-fluid model decomposes into two additive components (see Eq.~\protect{\ref{nonsimplesigma}}).  The relative motion between electrons and holes (left panel) is the universal Coulomb drag which dominates at charge neutrality.  Away from neutrality it decays as the number of minority carriers.  The center of mass motion (right panel) is the non-universal linear response of the collective electron-hole plasma under an electric field and reduces to the usual Drude conductivity far from neutrality.  Taken together, these two components fully describe the dissipation-enabled hydrodynamics giving a smooth crossover from universal to non-universal behavior as carrier density is tuned away from neutrality.}
\end{figure*}

\section*{Results and Discussion}
Here we adopt the two-fluid formalism of Ref.~\cite{GlennWagnerDungX.Nguyen}, which assumes that electrons and holes each form a fluid in local equilibrium, a condition that holds in the hydrodynamic regime. The two-fluid model reproduces a numerical solution of the full quantum Boltzmann equation~\cite{nguyen_quantum_2020}, and is largely in agreement with the three-mode ansatz of Ref.~\cite{zarenia_breakdown_2019-1}. For simplicity, we assume that the electron and hole bands are parabolic with the same effective mass  -- a standard approximation for many semi-metals including bilayer graphene.  While the conductivity predicted by this model is in general a complicated function of the carrier densities and relaxation times (see supplemental material), a remarkably simple picture emerges in the limit $\tau_0 \ll \tau_{\rm dis}$, where
$\tau_{\mathrm{0}}$ is the electron-hole relaxation time, and $\tau_{\mathrm{dis}}$ is the relaxation time from dissipative mechanisms. We find
\begin{equation}
    \sigma = \frac{4n_e n_h }{ (n_e + n_h)} \; \frac{e^2}{m^{*}}\tau_{\mathrm{0}}
    + \frac{(n_e - n_h)^2 }{ (n_e + n_h)} \; \frac{e^2 }{m^{*}}\tau_{\mathrm{dis}}, 
    \label{nonsimplesigma}
\end{equation}
where $n_e$ and $n_h$ are the densities of thermally excited electrons and holes, and $m^*$ is the effective mass.

As depicted in Fig.~\ref{fig_schematic}, this equation has a simple physical interpretation, in which the first and second terms describe the relative and center-of-mass motion, respectively, of the electrons and holes (we note that a similar decoupling between relative and center-of-mass motion was conjectured to explain the $\sigma \sim T^2$ dependence of the bulk conductivity of titanium disulfide as a possible signature of electron-hole dominated scattering~\cite{kukkonen_electron-hole_1976}; however, unlike the present case, the conductivity arising from the relative motion is neither temperature independent nor universal, and comparisons between theory and experiment are complicated by imperfect sample stoichiometry disagreeing by over an order of magnitude).  The conductivity due to relative motion is limited by Coulomb drag with relaxation time $\tau_0$, and is maximized at charge neutrality ($n_e=n_h$). Close to charge neutrality, this term can be expressed as $ \sigma_0 \exp\left[ -(1/3)\left(\mu/(k_B T)\right)^2\right]$, where $\mu$ is the chemical potential and $k_B T$ is the product of the Boltzmann constant and temperature (see Eq. S14 of the supplemental material for the more general case).  Here $\sigma_0=(e^2/h)\times 8\log(2)/\alpha_0$, where $h$ is Planck's constant and  $\alpha_0 \sim 0.2$ is a dimensionless constant that characterizes the electron-hole coupling strength~\cite{ho_theoretical_2018, zarenia_breakdown_2019-1,GlennWagnerDungX.Nguyen}. We highlight that this temperature-independent hydrodynamic conductivity $\sigma_0$ is not only independent of the degree of disorder showing no sample-to-sample variation (similar e.g. to mesoscopic universal conductance fluctuations~\cite{lee_universal_1985}), but also because it is insensitive to materials parameters such as $m^*$ in the strongly interacting limit. (see Section 6 of the SI for detailed discussion).

The center-of-mass motion is described by a Drude model for a plasma with charge density $-(n_e-n_h) e$, mass density $(n_e+n_h)m^*$, and momentum non-conserving scattering time $\tau_{\rm dis}$. This term is zero at charge neutrality, and equivalent to the conventional Drude conductivity in the unipolar regime.  Taken together, Eq.~\ref{nonsimplesigma} captures the full crossover from the universal behavior at charge neutrality to non-universal behavior away from charge neutrality.

We next establish that ultra-clean bilayer graphene encapsulated in hexagonal boron nitride (hBN) can act as a model system to compare theory to experiment. Graphene has emerged in recent years as an excellent platform for the study of hydrodynamics~\cite{bandurin_negative_2016,crossno_observation_2016,Berdyugin2019,Gallagher2019} due to its low disorder, weak electron-phonon coupling, and strong carrier-carrier interactions.  Bilayer graphene has electron and hole bands that are well described by hyperbolic bands, with dispersion $ \epsilon_{\pm}(k) = \pm \sqrt{\left( \hbar^2 k^2/ (2 m^{*}) \right)^2 + \left(\Delta/2 \right)^2 }$,
where $\pm$ denote the conduction and valence bands with effective mass $m^* \approx 0.03~m_e $, and a bandgap $\Delta$ that is tunable by an out-of-plane electric field. It is even better suited for the study of hydrodynamic conductivity than monolayer: the hydrodynamic regime is $1000\times$ less sensitive to disorder at low temperature, and should exhibit no high-temperature cutoff, due to weaker coupling between electrons and optical phonons~\cite{ho_theoretical_2018}. hBN-encapsulation~\cite{dean_boron_2010} provides disorder approaching that of suspended graphene while suppressing flexural phonons and providing a wider range of sample geometry. Dual-gated structures offer independent tuning of carrier density and bandgap. 

For this study, five dual-gated devices with Hall bar geometry and channel size from 2$\sim$10 $\mu$m were fabricated, all of which showed substantially identical behavior. Low-temperature conductivity and Hall effect measurements (Fig. S2) were used to calibrate top and bottom capacitances, allowing calculation of $\mu$ and $\Delta$ as a function of the top and bottom gate voltages using the hyperbolic bandstructure (see Methods). Using this calibration, we measured the conductivity $\sigma$ for gapless bilayer graphene as a function of temperature for $\mu=0$, and as a function of $\mu$ at a series of fixed temperatures. This was then repeated for different values of $\Delta$.

\begin{figure*}[ht!]
\begin{center}
    \vspace{-0.75in}
    \includegraphics[width=1\textwidth]{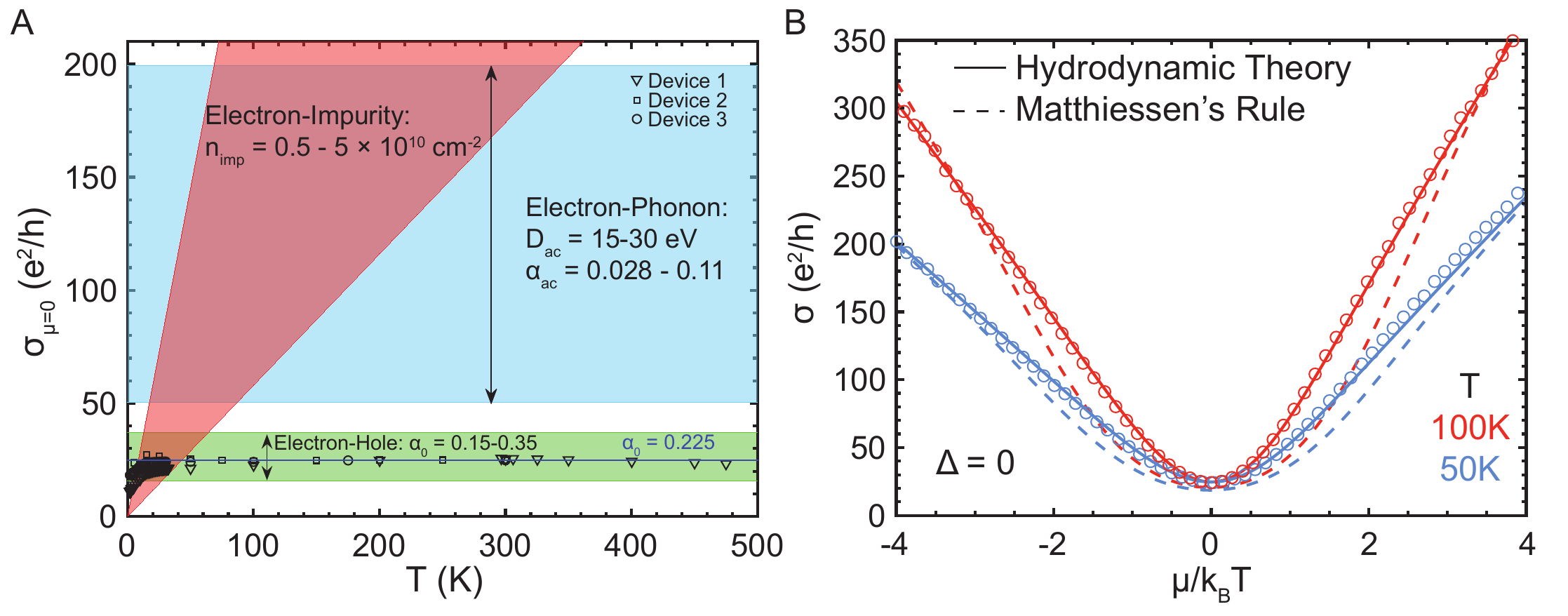}
    \caption{\doublespacing \label{fig_hydro} 
    \textbf{Robust hydrodynamic conductivity in bilayer graphene.} 
     (A) Measured charge neutral conductivity as a function of temperature for the gapless case for three devices (symbols).   The green shaded window shows the expectation from the theoretical literature for the electron-hole limited conductivity $0.15<\alpha_0<0.35$~\protect{\cite{ho_theoretical_2018,zarenia_breakdown_2019-1,GlennWagnerDungX.Nguyen}}.  Similarly, the blue region represents literature estimates of phonon-limited conductivity $0.03<\alpha_{ac}<0.11$
     ~\protect{\cite{borysenko_electron-phonon_2011,huang_hot_2015,efetov_controlling_2010}}, and red for impurity-limited conductivity $0.8 ~\rm{ps} <\tau_{imp}<8~\rm{ps}$, corresponding to a charged impurity density range of $(0.5-5) \times 10^{10}\textrm{ cm}^{-2}$~\cite{das_sarma_electronic_2011,bandurin_negative_2016,crossno_observation_2016}. The solid line is the best fit for the electron-hole limited conductivity $\alpha_0=0.225 \pm 0.002$. (B)  For a given temperature, the experimental data (symbols) can be fit using the dissipative hydrodynamic theory (Eq.~\ref{nonsimplesigma}, solid lines) where to account for a slight electron and hole asymmetry in the data we allow for $\tau_{dis}$ to be different for electrons and holes.  The dashed lines show a fit using a phenomenological ``Matthiessen's rule" where the resistance channels are added together in series. The disagreement with experiment shows that the momentum-conserving and non-conserving scattering do not act independently.  Additionally, the Matthiessen's rule conductivity is below the experimentally observed values, which is unphysical.} 
\end{center}
\end{figure*}
\FloatBarrier

Figure \ref{fig_hydro}A compares experiment to literature estimates of hydrodynamic, phonon-limited, and impurity-limited conductivity for the gapless case ($\Delta=0$) at $\mu=0$. At this point, the system is charge neutral ($n_e = n_h$), and free carriers are generated solely by thermal excitation, with $n_{e,h} \propto T$.  The temperature-independent hydrodynamic conductivity is given by $\sigma_0$ as discussed above (where the range in values for $\alpha_0$ in the theoretical literature does not arise from any expected variation in the experimental value, but rather from the level of approximation in the calculation).  The scattering time due to acoustic and substrate polar optical phonons have been calculated numerically using standard expressions available in the literature (e.g. Refs.~\cite{viljas_electron_2010,li_electron_2011}, see Sec. 3.2 of SI). Unlike the case of monolayer graphene, acoustic phonon scattering is dominant over optical phonons at all temperatures and leads to scattering time of $\tau= (\alpha_{\mathrm{ac}} k_B T )^{-1} \hbar$, where $\alpha_{\mathrm{ac}}$ is the (temperature- and density-independent) bilayer graphene electron-phonon coupling strength~\cite{ho_theoretical_2018} that varies as the square of the deformation potential $D$.  The shaded region shows conductivity for the reported values of $\alpha_{\mathrm{ac}}$ in the literature that correspond to $D$ between 15 and 30 eV. Scattering from charged impurities was calculated using the standard expression~\cite{lv_screening-induced_2010}, yielding a scattering time $\tau_{imp}$ that is nearly temperature- and density-independent (within 20 percent), leading to conductivity that increases linearly with temperature.  $\tau_{imp}$ is inversely proportional to the charged impurity density $n_{imp}$, which can be estimated from Hall effect measurements to fall within the range $5~\times 10^{9} ~\mathrm{cm}^{-2} <n_{imp}<~5\times 10^{10} ~\mathrm{cm}^{-2}$. See Section 3 of the supplemental material for a detailed discussion of all the relevant scattering mechanisms.

The solid points in Fig.~\ref{fig_hydro}A show data for three different devices. All show identical, constant conductivity with a best fit value of $(24.7 \pm 0.2)~e^2/h$ over a remarkably wide temperature range of $50~\mathrm{K}-500~\mathrm{K}$, which falls clearly within the range for electron-hole limited conductivity. This finding confirms the earlier observation in suspended bilayer graphene and extends the temperature range by a factor of five. The magnitude of the conductivity falls well outside the range for acoustic phonon scattering. Likewise, temperature-independent conductivity cannot be explained charged impurity scattering; however, we note that the observed downturn in conductivity below $50$~K is consistent with the calculated impurity-limited conductivity, and that the conductivity at high density (shown below) matches predictions for acoustic phonon scattering. We thus conclude that between $50$ and $500$~K, the charge-neutral conductivity is determined by electron-hole scattering, and we find experimentally that $\alpha_0 = 0.225 \pm 0.002$, indicated by the solid line in the figure. 

We next consider the behavior away from charge neutrality by plotting $\sigma(\mu)$ for two fixed temperatures (Figure~\ref{fig_hydro}B).  The dissipative hydrodynamic theory successfully describes the transition between the hydrodynamic regime near $\mu=0$ and the dissipative regime at large $|\mu|$. In contrast, combining electron-hole scattering with phonon/impurity scattering through Mattheissen's rule \textit{underestimates} the conductivity at intermediate $\mu$, which violates Kohler's theorem~\cite{kohler_1949}; this discrepancy becomes stronger at higher temperature.  This analysis already confirms that: (a) gapless bilayer graphene at $\mu=0$ displays sample-independent hydrodynamic conductivity limited by electron-hole scattering at the Planckian rate $1/ \tau_0= \alpha_{0} k_{B} T /\hbar \sim k_B T / \hbar$~\cite{zaanen_why_2004} over a wide temperature range up to and exceeding room temperature; and (b) its conductivity away from charge neutrality cannot be accounted for by pure electron-hole scattering or by including independent scattering from phonons/impurities. 

\begin{figure*}[t!]
    \includegraphics[width=1\textwidth]{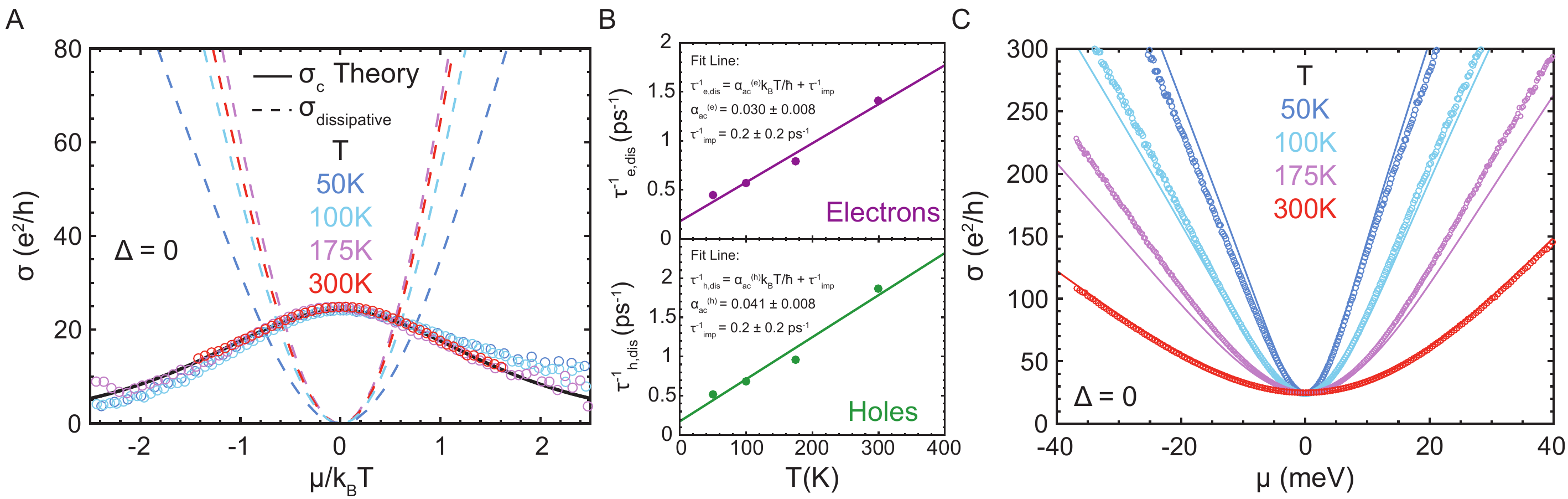}
    \caption{\doublespacing \label{fig_away} 
    \textbf{Ambipolar hydrodynamic conductivity comprises a universal and a dissipative contribution.} (A)  The decay of the universal component of the hydrodynamic conductivity away from neutrality extracted from the experiment (symbols) agrees with the theoretical calculations (solid line). The non-universal dissipative contribution to the hydrodynamic conductivity is also shown (dashed lines).  The sum of the universal and dissipative contributions gives the solid lines in Fig.~\protect{\ref{fig_hydro}B}. 
    (B) The dissipative scattering rates $\tau^{-1}_{e/h,dis}$ extracted at different temperatures are used to obtain a single set of four global fit parameters ($\alpha_0=0.225 \pm 0.002$, $\alpha_{ac}=0.030 \pm 0.008$ for electrons, $\alpha_{ac}=0.041 \pm 0.008$ for holes, and $\tau_{imp}^{-1}= (0.2 \pm 0.2)$ ps$^{-1}$). These four fit parameters are used in the hydrodynamic theory lines in panel C.
    (C) Zero-gap conductivity measurements (symbols) as a function of $\mu$ (meV) for $T = 50, 100, 175, 300$ K. The data is in excellent agreement with hydrodynamic theory developed in this work (solid lines). For the rest of this work, the same set of four global fit parameters mentioned in (B) are used consistently across the full range of carrier densities, temperature, and bandgaps.}
\end{figure*}

We next extract universal Coulomb drag and dissipative contributions to the conductivity (Eq.~\ref{nonsimplesigma}) from the data. At any temperature, we can match the experimental data $\sigma(\mu)$ using the previously determined value $\alpha_0=0.225$ and two fitting parameters, $\tau_{e,dis}$ and  $\tau_{h,dis}$, which represent the dissipative (phonon + impurity) scattering time for electrons and holes, respectively. The observed electron-hole asymmetry in the conductivity data is consistent with previous experiments~\cite{Zou2011} and necessitates fitting separately for electrons and holes.  Following Eq.~\ref{nonsimplesigma}, we can obtain the dissipative component (dashed lines in Fig.~\ref{fig_away}A).  This \textit{dissipative} component collapses onto a single curve when acoustic phonon scattering dominates over impurity scattering, as is seen above 100 K in these devices. This collapse was previously attributed to electron-hole scattering~\cite{nam_electronhole_2017}.  We next subtract the dissipative component from the total measured conductivity.  As seen in the figure, the subtracted experimental data collapse onto the theoretical curve revealing the universal behavior of electron-hole Coulomb drag scattering as a function of carrier density and temperature.  At high temperature where the hydrodynamics is stronger, the agreement is excellent.  To our knowledge, this universal electron-hole scattering contribution to the hydrodynamic conductivity has not been demonstrated previously, in either the theoretical or experimental literature. 

The extracted values of $\tau_{dis}(T)$ can be used to separately determine the phonon and impurity contributions to the dissipative scattering. To do so, we plot $\tau^{-1}_{e/h,dis}$ \textit{vs.} temperature (Fig.~\ref{fig_away}B). Since $\tau_{dis}^{-1}(T)=\alpha_{ac} k_B T / \hbar + \tau_{imp}^{-1}$ (see supplemental material for details), a line fit yields  $\alpha_{ac}$ from the slope and $\tau_{imp}$ from the intercept. Following this procedure, we obtain  $\alpha_{ac}^e=0.030 \pm 0.008$, $\alpha_{ac}^h=0.041 \pm 0.008$, and $\tau_{imp}^{-1}=0.2 \pm 0.2$ ps$^{-1}$. The derived parameters are consistent with theoretical calculations and other experimental estimates in the literature as well as other independent measurements on our samples (see section 4 of the SI for full details). The three parameters above, together with the value of $\alpha_0 = 0.225$ determined earlier, are sufficient to reproduce the entire $\sigma(\mu,T)$ dataset in the hydrodynamic regime. To illustrate this, figure Fig.~\ref{fig_away}C plots $\sigma$ \textit{vs.} $\mu/k_{\mathrm{B}} T$ for four different temperatures. The solid curves, generated by using only these four global parameters, show excellent agreement with the data.

\begin{figure*}[t!]
    \includegraphics[width=1\textwidth]{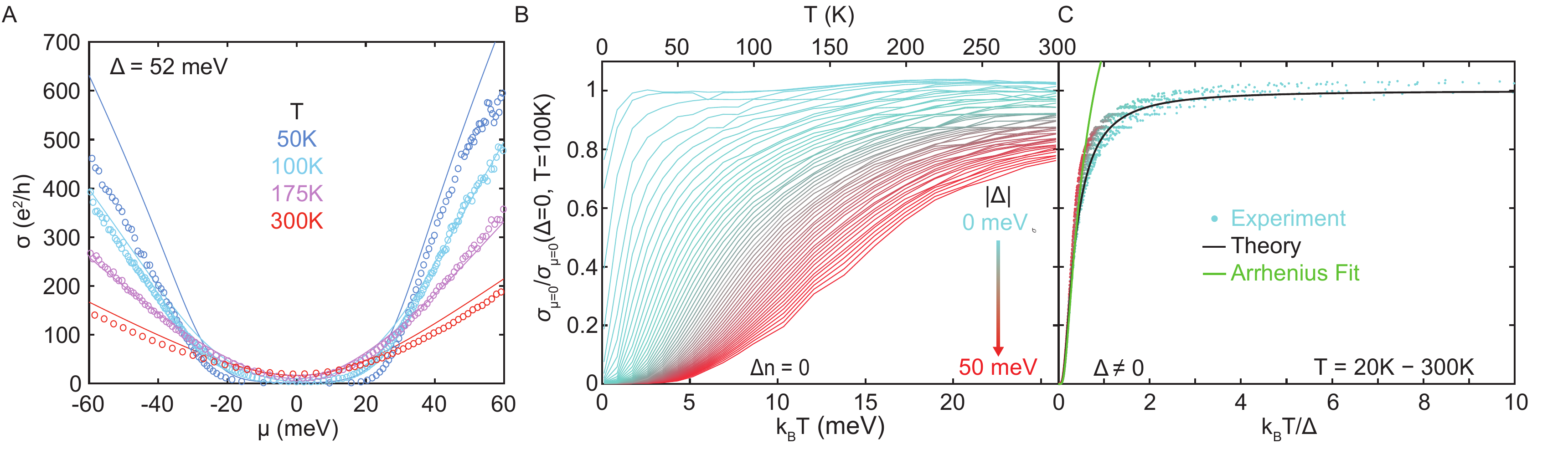}
    \caption{\doublespacing \label{fig_gap}  
    \textbf{Hydrodynamic semiconductor.}  (A) Representative gapped conductivity measurements (symbols) as a function of $\mu$ (meV) for $T = 50, 100, 175, 300$ K. The data is in excellent agreement with hydrodynamic theory developed in this work (solid lines).  The same set of four global fit parameters has been used consistently to fit the full range of carrier densities, temperature, and bandgaps (additional data for $\Delta = 13, 28,$ and $36$~meV are shown in the supplementary material).  (B) Normalized charge-neutral conductivity as a function of $k_BT$ (lower x axis) and $T$ (upper x axis) for varying $\Delta$. The color gradient denotes the magnitude of $|\Delta|$.  (C) Normalized charge-neutral conductivity as a function of $k_BT/\Delta$ for temperatures from 20K to 300K. The data collapse onto a single curve is in agreement with the theoretical prediction (solid line) of Eq.~\protect{\ref{collapse}}. The color scale for the data in (C) matches that in (B).}
\end{figure*}

We now address the effect of a bandgap.  We hypothesize that gap-induced changes in transport scattering times are dictated by changes to the carrier density and group velocity rather than changes to the universal electron-hole coupling strength $\alpha_0$.  In this case,  both terms in Eq.~\ref{nonsimplesigma} are suitably modified.  The thermally activated carrier densities $n_e$ and $n_h$ become functions of both $\mu/k_B T$ and $\Delta/k_B T$, and  we find that $\tau(\Delta)$ is obtained from the gapless $\tau$ by a multiplicative function of $\Delta/k_B T$ (see Sec. 5.3 of SI). Since for $\mu =0$, $k_BT$ and $\Delta$ are the only remaining energy scales (the Coulomb energy drops out since it is present in both $\sigma_{eh}(\Delta)$ and $\sigma_0$), the normalized conductivity for the model hyperbolic bandstructure collapses as a function of $\Delta /  k_B T$:
\begin{eqnarray}
\frac{\sigma_{eh}(\Delta) }{\sigma_0 } &=& 1 + \frac{1}{\log(2)} \left[ \log\left( \cosh\left(\frac{ \Delta}{4 k_B T}\right) \right) - \frac{ \Delta }{4 k_B T} \tanh \left(\frac{ \Delta }{4 k_B T} \right) - \right. \nonumber \\
&& \left. \frac{1}{8}\left(\frac{\Delta }{ k_B T} \right)^2 \exp\left( -\frac{5 \Delta}{8 k_B T}\right) \right]. \label{collapse}
\end{eqnarray}
This temperature-mediated insulating to conducting crossover function is completely different from the usual Arrhenius behavior $\sigma \sim  \exp(- \Delta / 2 k_B T)$ seen in conventional disorder-limited semiconductors within the gap (although it mimics Arrhenius behavior at the lowest temperature).  While this crossover function is specific to our model of two hyperbolic bands, it is only slightly modified for different bandstructures (See section 6.2 of the SI for details).  Making use of the relationship between the top and bottom gates and $\Delta$ (see methods), we plot the resulting function of $k_B T / \Delta$ (solid line) in Fig.~4C alongside the experimental data (dots) of Fig.~4B (omitting the $T<20$K portion that lies in the impurity-limited regime) replotted as a function of $k_B T / \Delta$. As predicted, the experimental data collapse onto a single curve.   The collapse of the experimental data validates our assumptions about $\alpha_0$ and provides strong evidence that transport in bilayer graphene remains electron-hole limited even as we move deep into the insulating regime.

\begin{figure*}[t!]
    \includegraphics[width=1\textwidth]{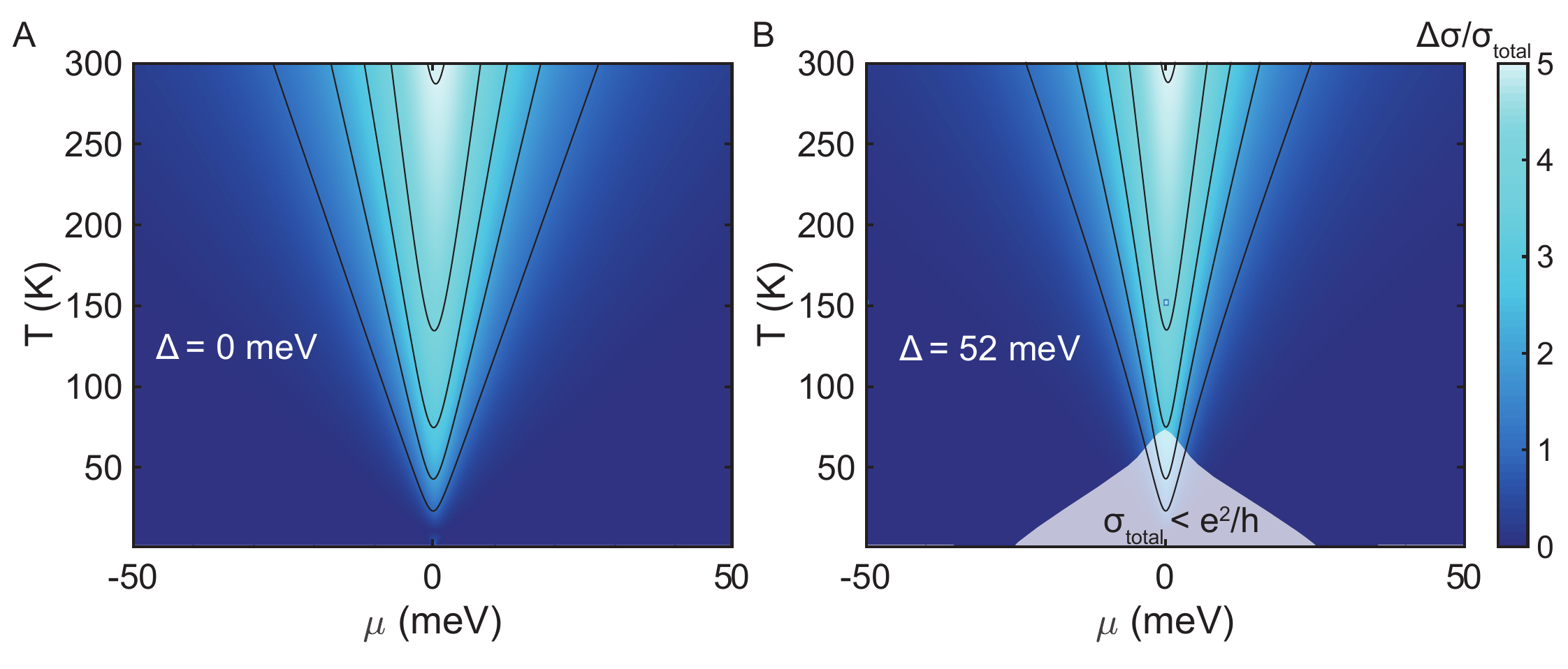}
    \caption{\doublespacing \label{fig_phase}  
    \textbf{Phase space for hydrodynamic conductivity.}  Calculated ratio of $\Delta \sigma = \sigma_\mathrm{ac+i}-\sigma_\mathrm{total}$ to $\sigma_\mathrm{total}$, for $\Delta = 0$ (A) and $52$ meV (B). Contour lines demarcate integer values, incrementing from $1$. The shaded area in (B) shows the insulating regime where $\sigma_\mathrm{total} < e^2/h$.  Remarkably, the degree to which electron-hole scattering dominates transport remains unchanged by the band gap. 
    }
\end{figure*}

Having validated the dissipative hydrodynamics model, it is now possible to quantitatively map out the phase space for hydrodynamic conductivity. To do this, we calculate the net effect of electron-hole scattering by subtracting the conductivity ( Eq.~\ref{nonsimplesigma}) from the conductivity calculated with only phonon and impurity scattering. We plot the ratio of this value to the total conductivity in Fig.~\ref{fig_phase}, for the zero-gap case and for the case with $\Delta = 52$~meV. As expected, electron-hole interactions dominate transport near charge neutrality \textemdash ~even in the presence of a bandgap \textemdash ~with the regions of dominance expanding as temperature increases. 

Our results confirm that an intuitive model (as captured in Eq.~\ref{nonsimplesigma}) provides a complete description of the conductivity of bilayer graphene over a wide range of temperatures, carrier densities and gap sizes.  Our ultraclean samples are dominated by electron-hole scattering achieving both the first room-temperature hydrodynamic conductor including confirmation of Planckian dissipation, and the first realization of a hydrodynamic semiconductor whose properties do not depend on material-specific parameters like the effective mass.   We emphasize that our hydrodynamic theory can be easily adapted to other systems with different bandstructures, electron-phonon coupling, or disorder. For example, we find that the hydrodynamic conductivity seen here is suppressed for $n_{\mathrm{imp}}=10^{11}$~cm\textsuperscript{-2} and disappears completely for $n_{\mathrm{imp}}=10^{12}$~cm\textsuperscript{-2}.  The need for low disorder explains why the hydrodynamic regime went unexplored for so long: the required disorder level of $n_{\mathrm{imp}}\sim 10^{10}$~cm\textsuperscript{-2} (i.e.~$\tau_{\mathrm{dis}} \sim 4$ ps) is only achievable in suspended samples (dielectric constant $\kappa \sim 1$) or hBN-encapsulated samples with graphite gates ($\kappa \sim 4$).  However, once strong hydrodynamics is achieved (i.e.~$\tau_{\mathrm{dis}} \gg \tau_0 $ and $m^*(e^2/\kappa)^2/T \gg 1$), then the  hydrodynamic properties will be universal and material-independent (see supplemental material).  We therefore expect that these insights should be applicable to many ambipolar two-dimensional systems with low disorder and strong electron-hole interactions including gapped monolayer graphene, twisted bilayer graphene, narrow-gap semiconductors and semimetals, and optically excited electron-hole fluids. 

As a room-temperature hydrodynamic conductor, bilayer graphene is an ideal model system for studying more complex hydrodynamic behavior -- including effects of viscosity, flow through constrictions, collective sound modes, high frequency magnetotransport, and shockwaves in supersonic flow -- via a variety of techniques.  Specifically, our experimentally measured values for the electron-hole scattering allows us to conclude that this platform should host more than a factor of two larger violation of the Wiedemann-Franz law compared to monolayer graphene~\cite{crossno_observation_2016}, a large frequency window where one might observe electron-hole sound waves at temperatures extending to room temperature and beyond~\cite{phan_ballistic_2013}, and an ideal system to explore the recently discovered hydrodynamic spin generation effect~\cite{takahashi_spin_2015} for applications in semiconductor spintronics, thereby combining semiconductor physics with viscous electronics.

\section*{Materials and Methods}
Heterostructure devices were fabricated with the van der Waals assembly technique\cite{wang_one-dimensional_2013}. To briefly summarize, a transfer substrate of polypropylene carbonate coated polydimethylsiloxane is used to pick up the top layer of exfoliated hexagonal boron nitride, which is then used to pick up the subsequent layers of the heterostructure. 
Once the heterostructure is assembled it is transferred to the substrate of interest and annealed in vacuum ($\sim 10^{-8}$ Torr) at 350\degree C.  Depending on the gating and contact configurations of interest, different processing steps of electron beam lithography (NanoBeam nB4), etch, and electron beam evaporation are used to etch and define the heterostructure into a dual gated device with multiple terminals for Hall measurements, as outlined in the supplementary material. 
Devices were then wire bonded to a dual-inline package for measurement. An optical image of one device is shown in Figure S1.

Low temperature to room temperature measurements were measured in liquid Helium cryogenic systems capable of temperatures as low as 1.2K, and magnetic fields as high as 14T. 
High temperature measurements were done in a cryostat with a heating stage for elevated temperatures.
Device gates were biased with Keithley 2400 and Yokogawa GS200 DC source meters. 
The device current and voltages were measured with Stanford Research System 830 lock-in amplifiers. The conductivity measurements are performed at currents $\sim 10~\mathrm{nA}-100~\mathrm{nA}$,  well within the range in which electrons may be considered to be in thermal equilibrium with the lattice even in the presence of strong electron-hole scattering~\cite{crossno_observation_2016}. 

For the dual-gated devices used in this study, $\Delta$ and $\mu$ can be independently controlled if the top and bottom gate capacitances are known.  Therefore, we characterize the device by mapping the resistance as a function of top and bottom gate voltages (Fig. S2A).  The peak at $\mu = 0$ in follows a diagonal line whose slope is the ratio of the two capacitances. This is combined with Hall effect measurements to determine each capacitance individually, allowing us to define two experimental parameters: the interlayer potential energy difference $\Delta_{\mathrm{ext}}$, which sets $\Delta$; and an effective voltage $V_{\mathrm{eff}}$, which tunes $\mu$ at constant $\Delta_{\mathrm{ext}}$.  For the range considered in this work, $\Delta_\mathrm{ext} \approx 2.6\Delta$ as determined experimentally from Arrhenius fittings, in good agreement with tight-binding models~\cite{mccann_asymmetry_2006}. The inset in Fig.~S2D inset shows the induced carrier density $\Delta n$ determined from low-temperature Hall effect measurements, taken along contours of fixed $\Delta_{\mathrm{ext}}=0,150$ meV, as depicted in Fig.~S2C. These measurements confirm that: (i) $\Delta n$ increases linearly with $V_{\mathrm{eff}}$; (ii) the samples are in the low-disorder limit with charge disorder below $\sim3\times10^{10}$cm$^{-2}$; and (iii) a gap opens between the electron and hole branches for nonzero $\Delta_{\mathrm{ext}}$. At higher temperatures, the Hall data show thermal excitation of electrons and holes (Fig.~S8). Details of device characterization and determination of carrier density, chemical potential, and bandgap are provided in the SI (sections 5.1 and 5.2).   

\FloatBarrier

\newpage

\section*{Acknowledgments}
We acknowledge helpful discussions with Michael Fuhrer, Peijie Ong, Yinming Shao, Oleg Sushkov, James Teherani, Giovanni Vignale, Glenn Wagner and Mohammad Zarenia. \textbf{Funding:} The experimental work at Columbia was primarily supported by the National Science Foundation program for Emerging Frontiers in Research and Innovation (EFRI-1741660). The theoretical work in Singapore was supported by the Singapore Ministry of Education (MOE2017-T2-1-130) and the Singapore National Research Foundation Investigator Award (NRF-NRFI06-2020-0003).
Samples were fabricated at the Columbia Nano Initiative Shared Facilities. Growth of hexagonal boron nitride crystals was supported by the Elemental Strategy Initiative conducted by the MEXT, Japan and the CREST (JPMJCR15F3), JST. We also acknowledge use of the dedicated research computing resources at CA2DM. Data analysis was partly supported by the National Science Foundation (NSF) MRSEC program through Columbia in the Center for Precision Assembly of Superstratic and Superatomic Solids (DMR-1420634). C.T. acknowledges support from a National Defense Science and Engineering Graduate (NDSEG) Fellowship: Contract FA9550-11-C-0028, awarded by the U.S. Department of Defense. \textbf{Author contributions:} C.T., L.W., and J.H. conceptualized the experiment. D.Y.H.H., I.Y., and S.A. provided theoretical understanding for the experiment. C.T., L.W., and J.I.A.L. fabricated the devices and collected the data. C.T., D.Y.H.H., I.Y., D.A.R., S.A. and J.H. curated and analyzed the data. T.T. and K.W. synthesized hBN crystals. K.S., P.L.M., C.D., S.A., and J.H. supervised and provided resources for the project. C.T., D.Y.H.H., S.A., and J.H. wrote the original draft. All authors reviewed and edited this work.
\textbf{Competing interests:} The authors declare that they have no competing interests. 
\textbf{Data and materials availability:} All data needed to evaluate the conclusions in the paper are present in the paper and Supplementary Materials. \\

\FloatBarrier

\section*{Supplementary Materials}
The Supplementary Materials PDF file includes \\
Supplementary Text \\
Figs. S1 to S15 \\



\newpage 

\Huge{Supplementary Information}
\normalsize

\section{Fabrication of Devices}
\begin{figure}[h!]
\begin{center}$
\begin{array}{c}
\includegraphics[height=!,width=8cm]{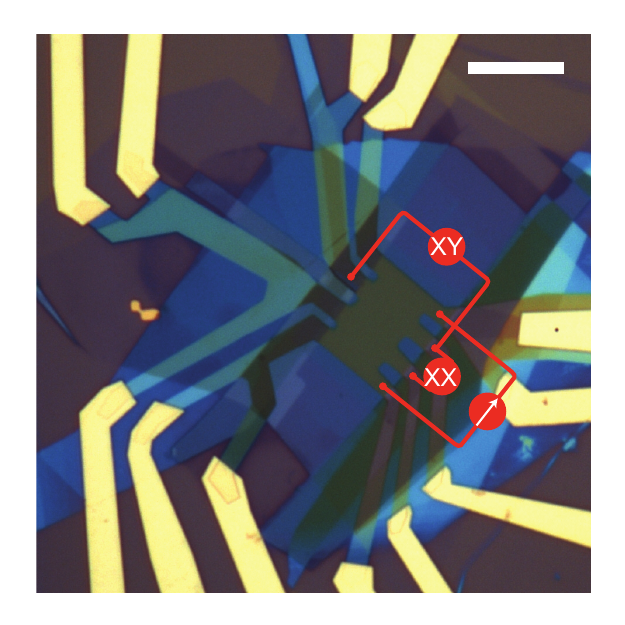} 
\end{array}$
\end{center}
\caption{\textbf{Device schematic.} A fabricated device with dual graphite gates and graphite contacts. The measurement scheme is shown in red. Scale bar is 10 $\mu$m. } \label{figdevice}
\end{figure}

Of the five total devices measured in this work, four were made with graphite gates, and one with metallic gates. 
All showed closely similar behavior.
To fabricate dual graphite gate devices, we first assemble a stack with hBN, top graphite, hBN, graphite contacts (optional), BLG, hBN, and bottom graphite in that order. 
The stack is then etched twice with 40 sccm CHF\textsubscript{3} + 4 sccm O\textsubscript{2} to first shape the top gate and then the channel; Cr/Pd/Au  (2 nm/40 nm/50 nm) is evaporated to make contact to the gates and channel. 
The metallic gate device was made by placing down a hBN, BLG, hBN stack on to a pre-patterned Pd back gate. Cr/Pd/Au (2 nm/20 nm/50 nm) was evaporated as top gate before the stack was then etched to shape the device. Finally,  Cr/Pd/Au (2 nm/20 nm/50 nm) was evaporated to make contacts to the channel.
We list in Table 1 the device dimensions and figures they correspond to. 
A typical device with dual graphite gates and graphite contacts is shown in Fig.~\ref{figdevice}, and the cross section schematic is presented in Fig.~\ref{fig_schematic}B.

\begin{figure*}[ht!]
    \includegraphics[width=1\textwidth]{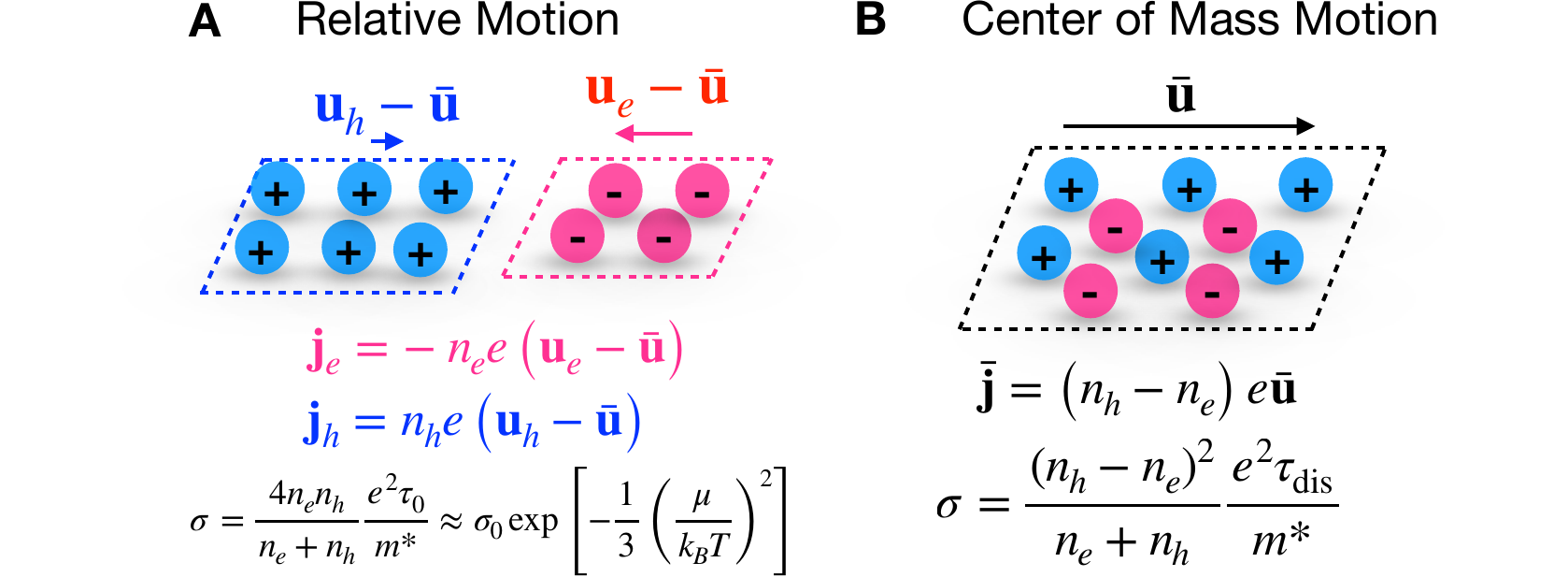}
    \caption{
    \textbf{Experimental characterization of dual-gated bilayer graphene.} (A) Schematic bandstructure consisting of two hyperbolic bands separated by a tunable band gap $\Delta$, with chemical potential $\mu$ and thermal energy $k_BT$.
    (B) Cross sectional schematic of an encapsulated BLG device with dual local gates. (C) Resistance as a function of independent top and bottom gate voltages $V_{\mathrm{TG}}$ and $V_{\mathrm{BG}}$. Dashed lines show cuts at $\Delta_{\mathrm{ext}}$ = 0 and 150 meV.
    (D) Measured conductivity as a function of $V_{\mathrm{eff}}$ at $\Delta_{\mathrm{ext}} = 0$ (top) and  $\Delta_{\mathrm{ext}} = 150$ meV (bottom) for $T$ from 5K to 300K.  Insets show the Hall density.}
\label{fig_schematic} 
\end{figure*}

\begin{table}[h!]
\caption{\label{tab:widgets} \textbf{Device Parameters.} List of measured devices and their corresponding figures in the main text.}
\vspace{0.1in}
\centering
\begin{tabular}{l|l|l|l|l|l}
Device & Width (um) & Length (um) & Gates & Contacts & Figures \\\hline
1  & 2 & 11 & Graphite & Metallic & 2A  \\
 2 & 5.5 & 3.5 & Graphite & Graphite & 2A \\
 3 & 6.26 & 10.5 & Graphite & Metallic & 2A-B, 3, 4 \\
 4 & 8 & 8 & Graphite & Metallic &  NA \\
5 & 9.6 & 3.85 & Metallic & Metallic & NA\\
\end{tabular}
\end{table}


\section{Two-fluid model for bilayer graphene in the hydrodynamic regime}

The transport properties of bilayer graphene in the hydrodynamic can be determined using a two-fluid model (see e.g. Ref.~\cite{GlennWagnerDungX.Nguyen,nguyen_quantum_2020}).  The two-fluid model describes the evolution of the average drift velocity of electrons and holes $\vec{u}_{e/h}$. It consists of two equations of motion, one for each carrier species: 
\begin{eqnarray}
    \frac{d \vec{u}_{e}}{dt} &=& -\frac{\vec{u}_{e} - \vec{u}_{h} }{  \tau_{\mathrm{e,eh}}  }  -\frac{\vec{u}_{e}}{ \tau_{\mathrm{e,\mathrm{dis}}} } - \frac{e \vec{E}}{m^{*}_{e}}, \nonumber \\
    \frac{d \vec{u}_{h}}{dt} &=& -\frac{\vec{u}_{h} - \vec{u}_{e} }{  \tau_{\mathrm{h,eh}} } -\frac{\vec{u}_h}{ \tau_{\mathrm{h,\mathrm{dis}}} } + \frac{e \vec{E}}{m^{*}_{h}}, \label{diffeqs}
\end{eqnarray} 
where $e > 0$ is the magnitude of the electron charge.
Here, $\tau_{e(h),eh}^{-1}$ is the average rate of collisions with holes (electrons) per electron (hole), while $\tau_{e(h),\mathrm{dis}}^{-1}$ is the average rate of collisions with surrounding impurities and phonons per electron (hole). 
$m^{*}_{e/h}$ are the effective masses and $\vec{E}$ is an external electric field.
We detail the method for calculating the various relaxation times $\tau_{e/h}$ in the subsection below.
The electron and hole drift velocities $\vec{u}_{e,h}$ are the average velocities of the electrons and holes and given by 
\begin{equation}
\vec{u}_{e/h} = \left(\int \frac{ d^{2} k}{(2 \pi)^2} \frac{\hbar \vec{k}}{m^{*}_{e/h}} f_{e/h}(\epsilon_{\vec{k}})\right) \Bigg/ \left(\int \frac{ d^{2} k}{(2 \pi)^2} f_{e/h}(\epsilon_{\vec{k}})  \right) ,
\end{equation}
where $f(\epsilon)$ is the standard Fermi distribution function $1/\left( \exp[(\epsilon - \mu) / (k_B T)]+1 \right)$, with $\epsilon_{\vec{k}} = \hbar^2 k^2 / 2 m^{*}$, and the integrals are over the entire Brillouin zone. 
The above equations are obtained from the Boltzmann kinetic equations using standard techniques~\cite{GlennWagnerDungX.Nguyen}. Physically, they describe the electrons (holes) as moving with one effective drift velocity $\vec{u}_{e(h)}$  This description works well in the hydrodynamic regime where the electron-electron collision rate is the largest scattering rate in the system.  It is likely to hold even when electron-hole scattering rate is larger than but still comparable to that of electron-electron~\cite{alekseev_magnetoresistance_2017}.

Note that since electron-hole collisions preserve total momentum density, $m_{e}^{*} n_e d\vec{u}_e / dt + m_{h}^{*} n_h d\vec{u}_h / dt = 0 $ must be true in the absence of external forces (i.e. $\tau_{e/h,\mathrm{dis}}^{-1} = 0$ and $\vec{E} = 0$). Combining equations~(\ref{diffeqs}) with this condition yields this constraint on electron-hole relaxation times: $n_e m_{e}^{*} \tau_{h,eh} = n_h m_{h}^{*} \tau_{e,eh}$.  Taking $ \tau_{e,eh} = \frac{m_{e}^{*} n_e + m_{h}^{*} n_h}{m_{h}^{*} n_h} \tau_0$ and $\tau_{h,eh} = \frac{m_{e}^{*} n_e + m_{h}^{*} n_h}{m_{e}^{*} n_e} \tau_0$ ensures that the constraint is satisfied, where $\tau_0^{-1} \equiv \tau_{e,eh}^{-1} + \tau_{h,eh}^{-1}$ evaluated at charge neutrality $n_e = n_h$.  To obtain the conductivity, one may solve Eqs.~(\ref{diffeqs}) for the steady-state $\vec{u}_{e/h}$, substitute these into the total current density $\vec{j} = n_e (-e) \vec{u}_e + n_h e \vec{u}_h$, and read off the conductivity $\sigma$ in $\vec{j} = \sigma \vec{E}$.  It is more instructive however to work instead in terms of the center-of-mass (COM) velocity 
\begin{equation}
    \vec{\bar{u}} \equiv \frac{n_e m^{*}_e \vec{u}_e +n_h m^{*}_h \vec{u}_h  }{n_e m^{*}_e  +n_h m^{*}_h } \label{vcom}
\end{equation} 
and the relative velocity
\begin{equation}
    \vec{v} = \vec{u}_e - \vec{u}_h. \label{vrel}
\end{equation}
Using these variables, current density becomes 
\begin{equation}
    \vec{j} = - \frac{n_e n_h}{n_e m^{*}_e  +n_h m^{*}_h } (m^{*}_e  + m^{*}_h ) e \vec{v} + (n_h - n_e ) e \vec{\bar{u}},  \label{jcomrel}
\end{equation}
in which the first term represents the contribution from the electrons and holes moving in opposite direction due to the opposite forces exerted on them by the electric field and the second represents the contribution from electrons and holes moving in unison in the same direction due to the Coulomb drag ``friction" between electrons and holes, the strength of which is quantified by $\tau_0^{-1}$.
Rewriting Eqs.~(\ref{diffeqs}) in terms of $\vec{\bar{u}}$ and $\vec{v}$ and performing rearrangements to make the time-derivatives of $\vec{\bar{u}}$ and $\vec{v}$ the subjects, we find
\begin{eqnarray}
\frac{d \vec{\bar{u}} }{dt} = && \left[ - \vec{\bar{u}} \left( \frac{n_e m^{*}_e}{\tau_{e,\mathrm{dis}}} + \frac{n_h m^{*}_h}{\tau_{h,\mathrm{dis}}} \right) -  \right. \nonumber \\
 && \left. \frac{n_e m^{*}_e n_h m^{*}_h}{n_e m^{*}_e + n_h m^{*}_h} \left( \frac{1}{\tau_{e,\mathrm{dis}}} - \frac{1}{\tau_{h,\mathrm{dis}}} \right)\vec{v} -e \vec{E} \left( n_e - n_h \right) \right] \left( n_e m^{*}_e +n_h m^{*}_h \right)^{-1}, \label{vcomeom} \\
\frac{d \vec{v}}{dt} = && -\frac{\vec{v}}{\tau_0} - \vec{\bar{u}} \left( \frac{1}{\tau_{e,\mathrm{dis}}} -\frac{1}{\tau_{h,\mathrm{dis}}}\right) - \frac{\vec{v}}{n_e m^{*}_e  +n_h m^{*}_h } \left( \frac{n_h m^{*}_h }{\tau_{e,\mathrm{dis}}} +  \frac{n_e m^{*}_e }{\tau_{h,\mathrm{dis}}} \right) \nonumber \\ 
&& - e \vec{E} \left(\frac{1}{m^{*}_e} + \frac{1}{m^{*}_h} \right), \label{vreleom}
\end{eqnarray}
where we have made use of relationships above.  Several insights may be drawn here. 
From Eq.~(\ref{vcomeom}), the COM velocity is unaffected by the electric field at charge neutrality $n_e = n_h$ since the net force from the electric field is zero. 
Away from neutrality, it increases without bound in the absence of external dissipative scattering mechanisms and the magnitude of $\vec{\bar{u}}$ tends to infinity as $t \rightarrow \infty$.
Put differently, there exists no steady-state (i.e. time-independent) solution $\vec{\bar{u}}$ satisfying Eq.~(\ref{vcomeom}) with $\tau_{e,\mathrm{dis}}^{-1} = \tau_{h,\mathrm{dis}}^{-1} = 0$ and $d \vec{\bar{u}}/ dt = 0$.
From Eq.~(\ref{vreleom}) on the other hand, the relative velocity is finite even when $\tau_{e,\mathrm{dis}}^{-1} = \tau_{h,\mathrm{dis}}^{-1} = 0$, since it still admits a steady-state solution $\vec{v} = -e \vec{E} \tau_0 ( 1/m_{e}^{*} + 1/m_{e}^{*} )$. 
Noting that 
\begin{equation}
    \vec{u}_{e/h} = \vec{\bar{u}} \pm \frac{n_{h/e} m_{h/e}^{*}}{ n_{e} m_{e}^{*} + n_{h} m_{h}^{*}}\vec{v},
\end{equation}
the above statements together imply that current density $\vec{j}$ and conductivity are formally infinite in the absence of external momentum dissipation if $n_e \neq n_h$.
It is only precisely at $n_e = n_h$ that the current density is finite even in the absence of external dissipation (i.e. electron-hole scattering alone can relax a net current) since the electric field is unable to accelerate the center of mass. 
Finally, we note from both Eqs.~(\ref{vcomeom}) and (\ref{vreleom}) that the COM and relative velocities are completely decoupled from each other when the dissipative relaxation times  $\tau_{e,\mathrm{dis}}$ and $\tau_{h,\mathrm{dis}}$ are equal. 
In this case, changing the relative velocity $\vec{v}(t)$ whilst maintaining constant COM velocity $\vec{\bar{u}}(t)$ leads to changes in the external dissipative frictional forces on electrons and holes respectively that exactly cancel each other.

Solving Eqs.~(\ref{vcomeom}) and (\ref{vreleom}) for the steady-state $\vec{\bar{u}}$ and $\vec{v}$ by setting the time-derivatives to zero, we obtain
\begin{eqnarray}
    \vec{\bar{u}} = -e \vec{E} \frac{(n_e  \tau_{e,\mathrm{dis}} -n_h \tau_{h ,\mathrm{dis}})\tau_0  + (n_e - n_h )\tau_{e,\mathrm{dis}}\tau_{h,\mathrm{dis}} }{(n_e m_{e}^{*} + n_h m_{h}^{*})\tau_0 +  n_e m_{e}^{*} \tau_{h,\mathrm{dis}} + n_h m_{h}^{*} \tau_{e,\mathrm{dis}}}, \\
    \vec{v} = -e \vec{E} \frac{(n_e m_{e}^{*}+n_h m_{h}^{*})(m_{e}^{*}\tau_{h,\mathrm{dis}} + m_{h}^{*}\tau_{e,\mathrm{dis}} )\tau_0 }{m_{e}^{*}m_{h}^{*} \left[ (n_e m_{e}^{*} + n_h m_{h}^{*})\tau_0 +  n_e m_{e}^{*} \tau_{h,\mathrm{dis}} + n_h m_{h}^{*} \tau_{e,\mathrm{dis}} \right]  }.
\end{eqnarray}
One may substitute these equations into Eq.~(\ref{jcomrel}) and directly read off the conductivity from $\vec{j} = \sigma \vec{E}$. 
To simplify the expression, we make the common assumption of equal electron and hole effective masses $m_{e}^{*} = m_h^{*} \equiv m^{*}$, resulting in 
\begin{eqnarray}
    \vec{j} = && \left[ \frac{e^{2}}{m^{*}} \frac{2 n_e n_h  \tau_0 (\tau_{e,\mathrm{dis}} + \tau_{h,\mathrm{dis}} ) }{ n_e (\tau_0 +\tau_{h,\mathrm{dis}})+ n_h (\tau_0 +\tau_{e,\mathrm{dis}})  } \right]\vec{E} \nonumber \\ 
    && + \left[ \frac{e^{2}}{m^{*}}\frac{(n_h - n_e)\left(n_h \tau_{h,\mathrm{dis}} (\tau_0 + \tau_{e,\mathrm{dis}}) - n_e \tau_{e,\mathrm{dis}} (\tau_0 + \tau_{h,\mathrm{dis}}) \right) }{n_h (\tau_0 + \tau_{e,\mathrm{dis}}) + n_e (\tau_0 + \tau_{h,\mathrm{dis}}) } \right] \vec{E}, 
\end{eqnarray}
where the first and second terms represent contributions from the relative and COM motions respectively of the electron-hole plasma. 

Thus far, our calculation has been exact and no approximations have been made. 
We now consider the limit of strong electron-hole scattering $\tau_0 / \tau_{e, \mathrm{dis}}, \tau_0 / \tau_{h, \mathrm{dis}} \rightarrow 0 $, and Taylor expand the current density to zeroth order in $\tau_0 / \tau_{e/h, \mathrm{dis}}$ and obtain $\vec{j} = \sigma_c \vec{E} +  \sigma_{\mathrm{dis}} \vec{E}$, where 
\begin{equation}
   \sigma_{c} = \frac{e^2}{m^{*}} \frac{n_e n_h (n_e + n_h ) (\tau_{e,dis}^{-1}+\tau_{h,dis}^{-1})^{2} }{ \left( n_e \tau_{e,dis}^{-1} + n_h \tau_{h,dis}^{-1} \right)^{2} } \tau_0 \label{sigmacasymm}
\end{equation}
represents the Coulomb drag conductivity arising from the Coulombic friction between electrons and holes, and
\begin{equation}
   \sigma_{\mathrm{dis}} = \frac{e^2}{m^{*}} \frac{(n_e - n_h)^2}{n_e \tau_{e,dis}^{-1} + n_h \tau_{h,dis}^{-1} }. \label{sigmadis}
\end{equation}
represents the conductivity due to external dissipative forces.
Eqs.~(\ref{sigmacasymm}) and (\ref{sigmadis}) are used for all the plots calculated in the main text.
 
Evidently from Eq.~(\ref{sigmacasymm}), the drag conductivity $\sigma_c$ is determined by electron-hole scattering time $\tau_0$ and the \emph{ratio} of electron and hole scattering times from external dissipative mechanisms (i.e. if $\tau_{e,dis} = \tau_{h,dis}$, the dependence on both $\tau_{e,dis}$ and $\tau_{h,dis}$ vanishes). 
A particular exception is at the CNP $n_e = n_h$, at which the dissipative times drop out and $\sigma_c$ depends only on $\tau_0$.  We find  
\begin{equation}
     \sigma_{c} = \sigma_0 \times \left( \frac{2}{\log(2)}\frac{\log(1+\exp(\mu/k_B T))\log(1+\exp(-\mu/k_B T))}{\log(1+\exp(\mu/k_B T)) + \log(1+\exp(-\mu/k_B T))}\right),
\end{equation}
where $\sigma_0=(e^2/h)\times 8\log(2)/\alpha_0$.  Near charge neutrality, the term in large brackets above asymptotes to  $\exp\left[ -(1/3)\left(\mu/(k_B T)\right)^2\right]$, and this is what we show in the main text for simplicity since it is in good agreement with experimental data (See e.g. Fig. 3A).

\section{Scattering times in gapless bilayer graphene}

In this section we discuss the scattering times used in the main text.  Electronic scattering times in bilayer graphene have been thoroughly studied for well over a decade and comprehensive reviews may be found in Refs.~\cite{das_sarma_electronic_2011, mccann_electronic_2013}. 
In this section we summarize the relevant aspects of the subject for our experiment. 

Electrons in gapless bilayer graphene are well-described by a parabolic band dispersion at energies below $0.4$~eV \cite{mccann_landau-level_2006}, corresponding to density $\sim 10^{13}$~cm$^{-2}$ and temperature $\sim 4600$~K. 
The system has been studied experimentally in two configurations- mounted on a hexagonal boron nitride substrate or suspended between supports, the former of which is the focus of this work.

\begin{figure}[h!]
\begin{center}$
\begin{array}{cc}
\includegraphics[height=!,width=8cm]{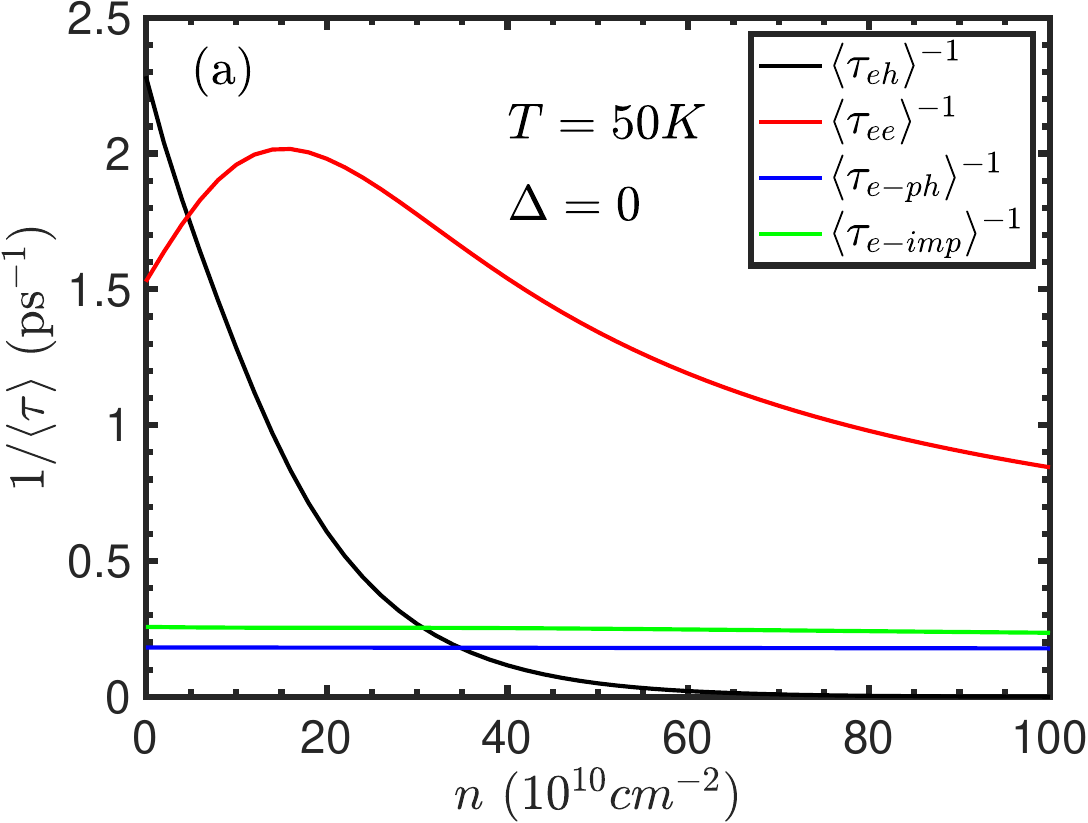} &
\includegraphics[height=!,width=8cm]{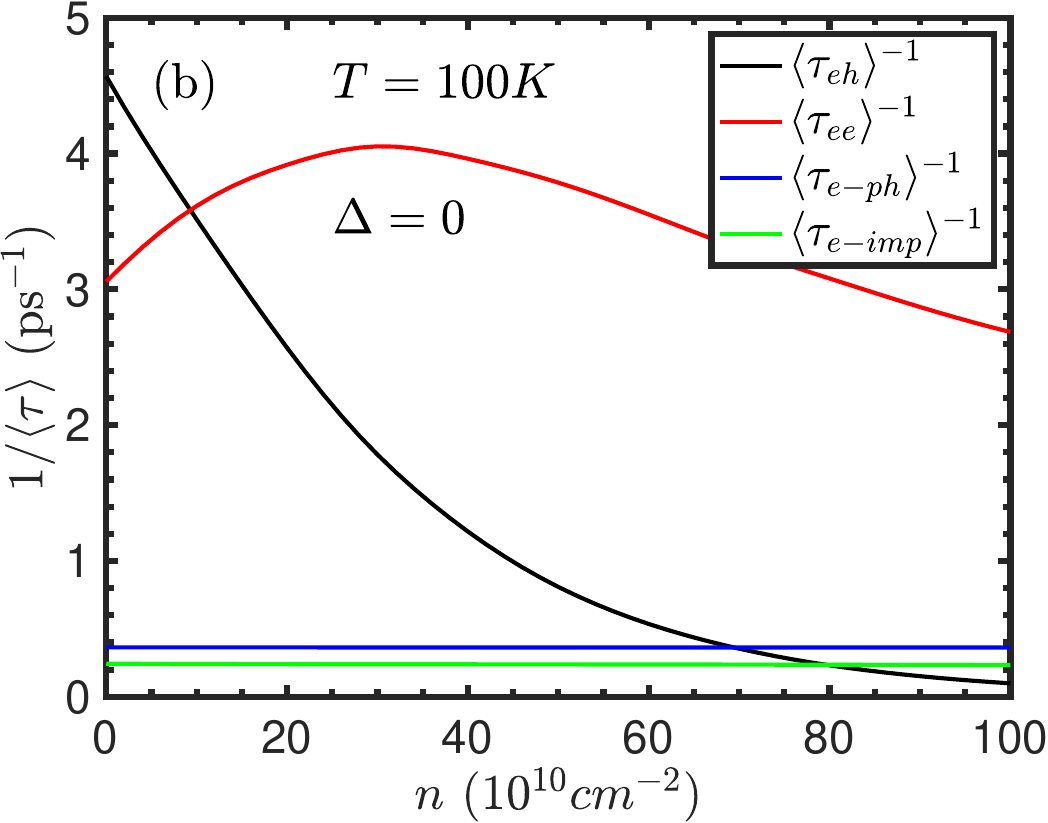} \\
\includegraphics[height=!,width=8cm]{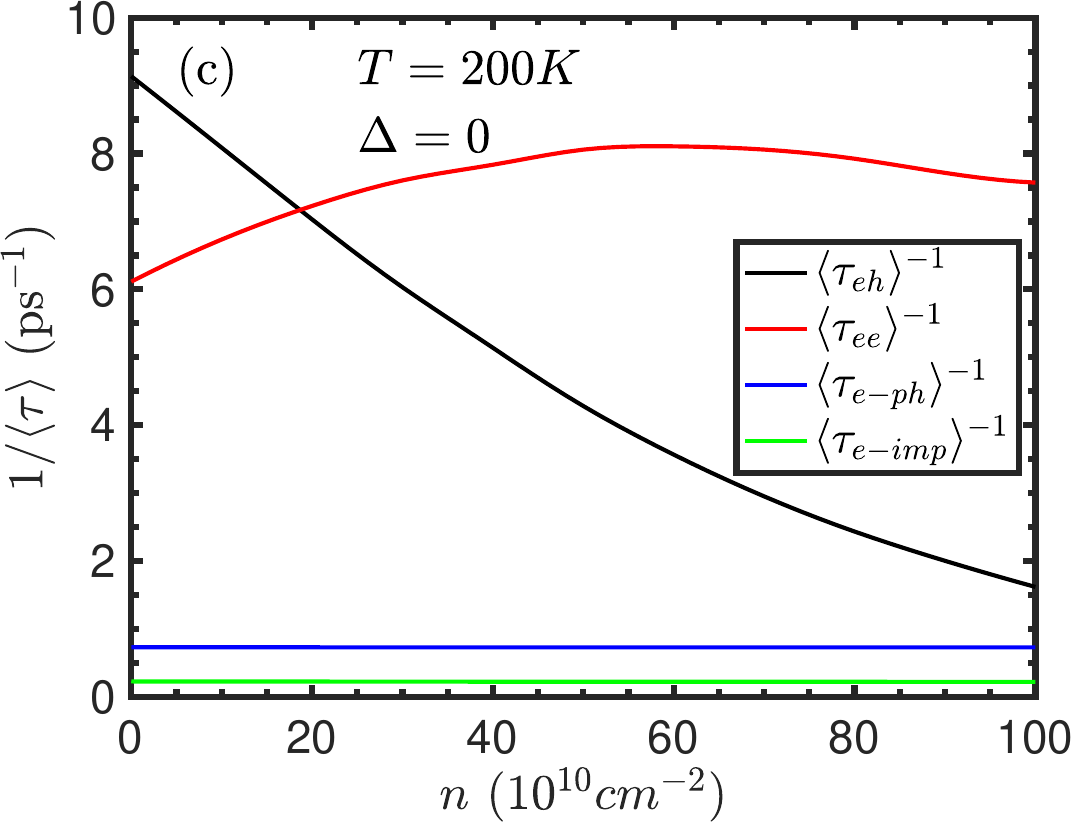}&
\includegraphics[height=!,width=8cm]{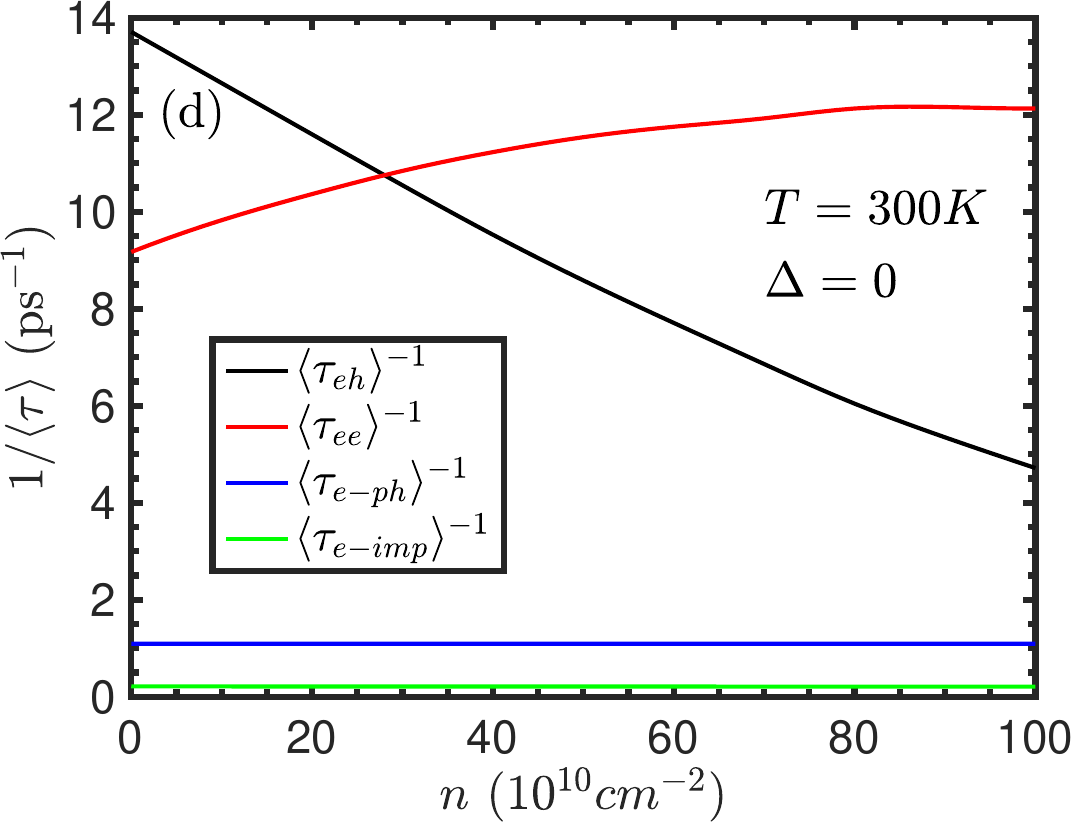}
\end{array}$
\end{center}
\caption{\textbf{Comparison of the relevant scattering rates}. Here we use $n_{\mathrm{imp}} =10^{10}\mathrm{cm}^{-2}$, $D_{ac} = 15 eV$ throughout (see text for details). A sizable hydrodynamic density window in which $\tau_{ee}^{-1}$ dominates occurs at all temperatures above $50$K. At low densities,  $\tau_{eh}^{-1}$ dominates over a window of densities that grows in size with temperature.
}\label{invtausvsn}
\end{figure}

Experimental measurements of charge conductivity in hBN-supported samples have been explained in terms of charged impurity scattering~\cite{dean_boron_2010} as well as in-plane acoustic  phonons~\cite{ochoa_temperature_2011}.   More recently, samples of hBN-supported bilayer graphene were reported~\cite{bandurin_negative_2016} to have such high levels of purity as to be in the hydrodynamic regime~\cite{narozhny_hydrodynamics_2015,ho_theoretical_2018,zarenia_breakdown_2019-1}, in which the scattering rate of electrons with one another exceeds that with impurities and phonons. Here, we give a detailed comparison of the above mentioned scattering rates.  We will in-turn discuss (i) Acoustic phonon scattering, \cite{viljas_electron_2010,ochoa_temperature_2011,borysenko_electron-phonon_2011} (ii) Optical phonons \cite{li_electron_2011,viljas_electron_2010} , (iii) Charged-impurity scattering\cite{das_sarma_electronic_2011}, and (iv)  Electron-hole scattering ~\cite{GlennWagnerDungX.Nguyen,nam_electronhole_2017,svintsov_hydrodynamic_2012} (note that the electron-electron scattering rate is not relevant~\cite{GlennWagnerDungX.Nguyen} for current relaxation and is included only for completeness).  
Our comparison is summarized in Fig.~\ref{invtausvsn}, where we find that at low carrier density (n $\lesssim 3\times 10^{11} \rm{cm}^{-2}$), for all temperatures considered (50-300K), electron-hole scattering dominates as the largest relevant (i.e.~current-relaxing) scattering rate.  
The forms of the equations used to fit to the scattering rates from experimental data in the main text are consistent with those found in the large body of calculations in the literature on electron scattering rates in bilayer graphene (which we simply summarize and reproduce briefly below).

\subsection{Acoustic phonon scattering}

We calculate the scattering time for electron-acoustic-phonon scattering using the inverse quasiparticle lifetime (see e.g. Ref.~\cite{viljas_electron_2010, kaasbjerg_unraveling_2012}).  
The quasiparticle lifetime $\tau \left( \epsilon_{k,\gamma} \right)$ of a quasiparticle at energy $\epsilon_{k,\gamma}$ with acoustic phonons is given by
\begin{eqnarray}
\frac{1}{\tau \left( \epsilon_{k,\gamma} \right)} = && \frac{2\pi}{\hbar} \int \frac{d^{2} q}{(2\pi)^2} |g_{\mathbf{k}+\mathbf{q},\mathbf{k}}|^{2} \times [ \left( N_{q} + 1 - f(\epsilon_{k,\gamma} - \hbar v_{s} q ) \right) \delta(\epsilon_{\mathbf{k}+\mathbf{q},\gamma} - \epsilon_{\mathbf{k},\gamma} + \hbar v_{s} q) + \nonumber \\
&& \left( f(\epsilon_{k,\gamma} + \hbar v_{s} q ) + N_{q} \right) \delta(\epsilon_{\mathbf{k}+\mathbf{q},\gamma} - \epsilon_{\mathbf{k},\gamma} - \hbar v_{s} q) ] , \label{ephtau}
\end{eqnarray} 
where 
\begin{equation}
    \epsilon_{k,\gamma} = \mathrm{sign}(\gamma) \frac{\hbar^2 k^2}{2 m}
\end{equation}
with $\gamma = \pm 1$ for conduction and valence band respectively,
$v_{s}= 13.6 \mathrm{km}$ $\mathrm{s}^{-1}$ is the speed of sound for transverse acoustic phonons~\cite{sohier_phonon-limited_2014} and the deformation potential matrix element equals~\cite{borysenko_electron-phonon_2011}
\begin{equation}
|g_{\mathbf{k}+\mathbf{q},\mathbf{k}}|^{2} = \frac{(D_{ac})^{2} \hbar q}{2 \rho v_{s}} \left( \frac{1+ \cos(2\theta_{\mathbf{k},\mathbf{k}+\mathbf{q}})}{2}\right)
\end{equation}
with $D_{ac} = 10$-$30$ eV according to estimates in the literature~\cite{viljas_electron_2010,ochoa_temperature_2011} (we use $D_{ac} = 15$~eV for all our calculations in this section) and $\rho =  1.52 \times 10^{-24}$ $\mathrm{kg} / \mathrm{m}^{-2}$ is the areal mass density of bilayer graphene. 
$ N_{q}$ is the Bose-Einstein distribution 
\begin{equation}
N_{q} =\frac{1}{\exp(\frac{\hbar v_{s} q}{k_{B} T}) -1 }
\end{equation}
The scattering time $\tau \left( \epsilon_{k,\gamma} \right)$ above is easily computed numerically. 
To obtain a representative scattering rate for all the  electrons relevant to charge transport (i.e.~electrons within $~k_B T$ of the Fermi surface), we perform a thermal average~\cite{combescot_conductivity_1987} by inserting $\tau \left( \epsilon_{k,\gamma} \right)$ (with $\gamma =1 $ for electrons) into 
\begin{equation}
    \langle \tau \rangle =  \frac{ \int^{\infty}_{0} d\epsilon \tau(\pm \epsilon) |\epsilon|\left(-\frac{\partial f_{e/h}(\epsilon)}{\partial \epsilon}\right) }{\int^{\infty}_{0} d\epsilon |\epsilon| \left(-\frac{\partial f_{e/h} (\epsilon)}{\partial \epsilon}\right) },\label{thermalavrg}
\end{equation}
where the $\pm \epsilon$ and e/h correspond to $\gamma=\pm 1$ for electrons and holes respectively, with $f_{e/h} = (\exp((\epsilon \pm \mu )/ k_B T ) +1)^{-1}$.

\begin{figure}[t!]
\begin{center}$
\begin{array}{ccc}
\includegraphics[height=!,width=5cm]{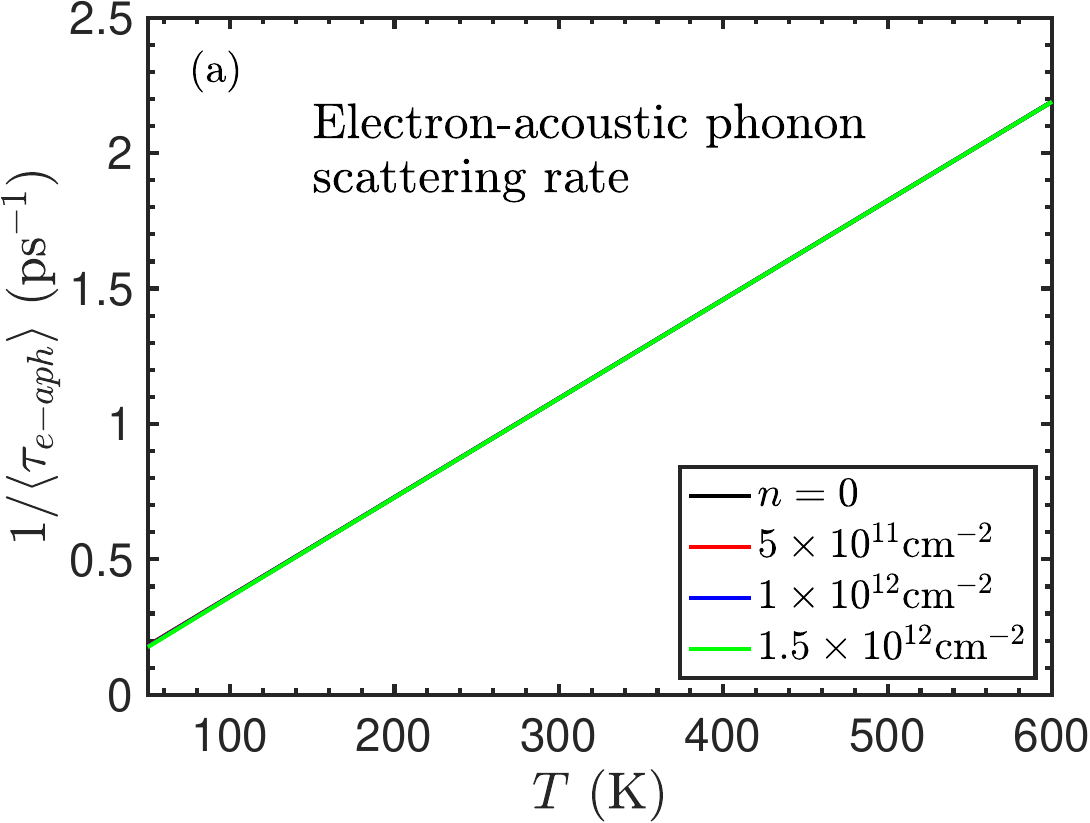} &
\includegraphics[height=!,width=5cm]{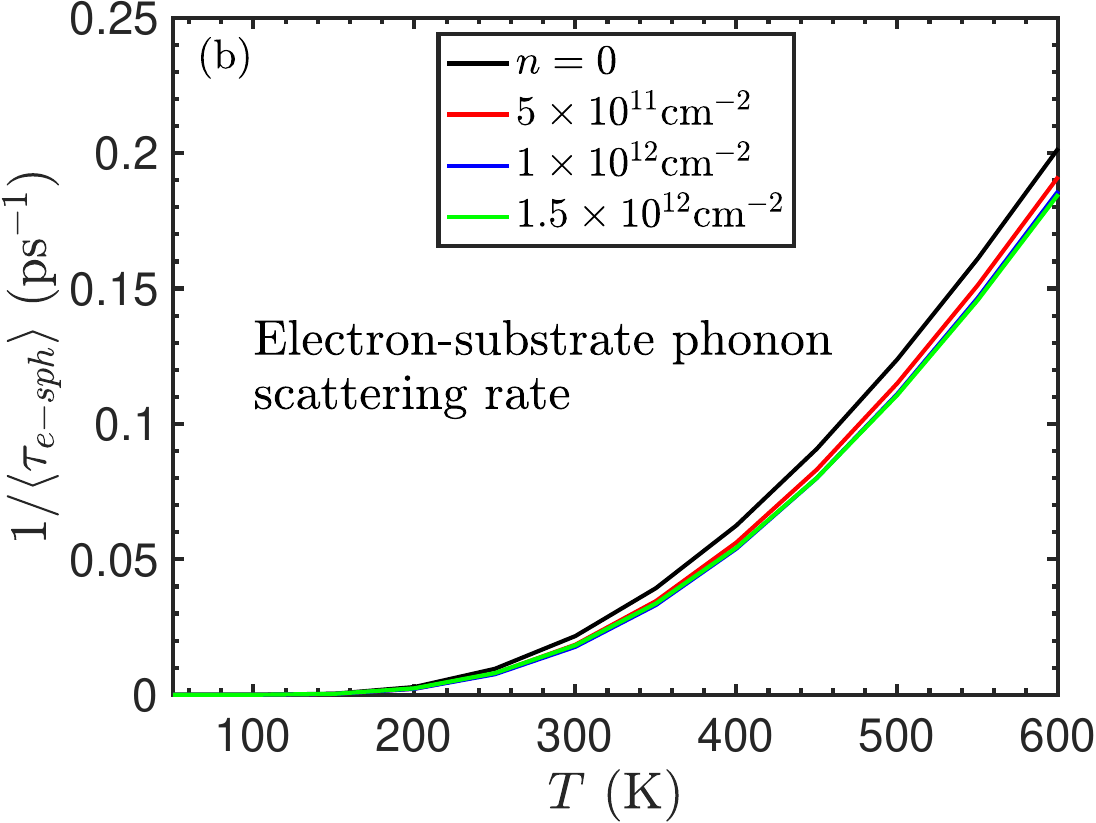} &
\includegraphics[height=!,width=5cm]{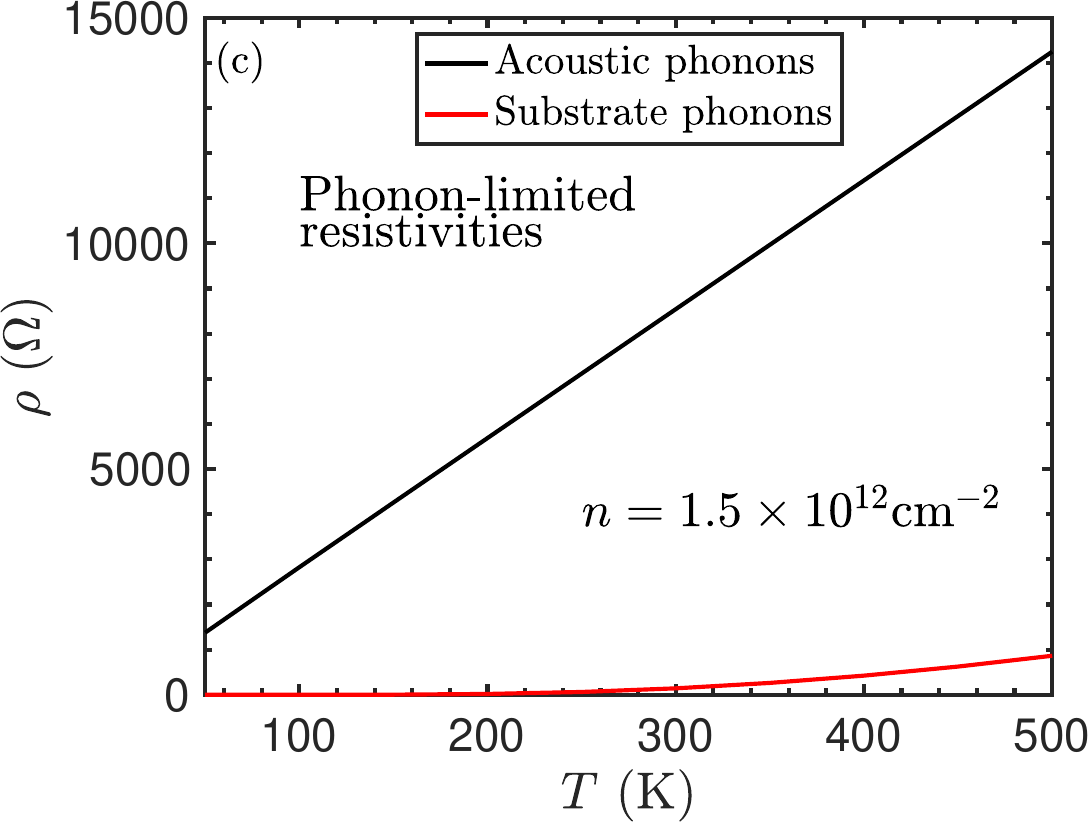}
\end{array}$
\end{center}
\caption{\textbf{Acoustic and optical phonon scattering}.  Comparison of electron scattering rates with (a) acoustic phonons and (b) optical polar substrate phonons. The contributions to resistivity are estimated in (c) using the standard expression $\rho = m /(ne^2 \langle \tau \rangle)$. Here we have used the same parameters for acoustic phonons as detailed in Sec. 3.1. For the substrate phonon calculations, we follow the parameters of Ref.~\cite{schiefele_temperature_2012}. 
Namely, the zero, intermediate and high frequency dielectric constants are taken to be 5.09, 4.575 and 4.10, respectively, and the two polar optical phonon energies of hBN are taken to be $101.6$~meV and $195.7$~meV respectively. 
}\label{acph-vs-subsph}
\end{figure}

We plot the inverse of the resulting thermally-averaged lifetime in Fig.~\ref{acph-vs-subsph}(a) below as a function of temperature for different densities.
Consistent with previous results in the literature~\cite{ochoa_temperature_2011,viljas_electron_2010}, we find that in the high temperature regime ($k_B T \geq \mu $), the scattering time (i.e.~inverse lifetime) is $1/\tau \left( \epsilon_{k,\gamma} \right) = \alpha_{ac} k_B T / \hbar$, with the proportionality constant $\alpha_{ac}$ determined by the value taken for the deformation potential. The estimated range for the deformation potential based on the literature is $D_{ac}=15$-$30$~eV, corresponding to $\alpha_{ac} = 0.028$-$0.11$.  In Fig.~\ref{invtausvsn} we used  $D_{ac} = 15$~eV (or $\alpha_{ac} = 0.03$). In the main text, $\alpha_{ac}$ is left as a fitting parameter and we found $0.030 \pm 0.008$ for electrons and $0.041 \pm 0.008$ for holes (see Table~\ref{paramtable} below) which are both well within the expected range.

\subsection{Optical phonon scattering}

The scattering rate of electrons with bilayer graphene optical phonons is three orders of magnitude lower than with acoustic phonons for all electron energies below 125~meV~\cite{li_electron_2011}, corresponding to a temperature of 1400~K and a density of $3.5\times 10^{12}\mathrm{cm}^{-2}$. Therefore intrinsic optical phonon scattering can be ignored for the purposes of this study.

We also consider scattering from polar optical phonons within the boron nitride substrate, using methods reported previously\cite{li_electron_2011, fratini_substrate_2008,schiefele_temperature_2012}. The scattering rate of an electron of energy $\epsilon_{\mathbf{k}} = \hbar^2 k^2 / 2m^{\star}$ with each of the two polar phonon modes of hBN (denoted $\nu=1,2$) is given by
\begin{eqnarray}
    \frac{1}{\tau_{e-sph}^{(\nu) } (k)} &=& \frac{2\pi}{\hbar} \sum_{\mathbf{q}} \left|\frac{M^{ss'}_{\nu}(q,\theta_{\mathbf{q}})}{\varepsilon(q,\omega=0,\mu,T)} \right|^2 [(N_{\nu} +n_{F} (\epsilon_{\mathbf{k}+\mathbf{q}}) ) \delta \left(\epsilon_{\mathbf{k}+\mathbf{q}} -\epsilon_{\mathbf{k}} - \hbar \omega_{sph}^{(\nu)} \right) \nonumber \\
    && +(N_{\nu} +1-n_{F} (\epsilon_{\mathbf{k}+\mathbf{q}})) \delta \left(\epsilon_{\mathbf{k}+\mathbf{q}} -\epsilon_{\mathbf{k}} + \hbar \omega_{sph}^{(\nu)} \right) ],
\end{eqnarray}
where 
\begin{eqnarray}
        |M^{ss'}_{\nu}(q,\theta_{\mathbf{q}})|^2 &=& e^2 \frac{\hbar \omega_{sph}^{(\nu)}}{2A\varepsilon_0} \left( \frac{1}{\kappa_{\nu}+1} - \frac{1}{\kappa_{\nu-1}+1}\right) \frac{e^{-2qd}}{q} \nonumber \\
          && \times \frac{1}{2} \left( 1 + ss' \cos (2 \theta_{\mathbf{k},\mathbf{k}+\mathbf{q}} )\right).
\end{eqnarray}
Above, $A$ is the area of the bilayer graphene and
$d= 3.5 \textup{~\AA}$~\cite{perebeinos_inelastic_2010} is the van der Waals distance between bilayer graphene and hBN and the substrate phonon energies are given by $\hbar \omega_{sph}^{(\nu=1,2)} = 101.6, 195.7$~meV respectively. 
The zero, intermediate and  high frequency dielectric constants $\kappa_{0,1,2} =5.09, 4.575, 4.10$ respectively.
$\varepsilon$ refers to the RPA dielectric function defined in Eq.~(\ref{dielectricfun}) below and $\varepsilon_0$ is the vacuum permittivity. $N_{\nu}$ and $n_F$ are the Bose-Einstein and Fermi-Dirac distributions respectively. 
The substrate phonon scattering rate shown in Fig.~\ref{acph-vs-subsph} above is obtained by taking the sum of scattering from the two modes $1/\tau^{(1)}_{e-sph} + 1/\tau^{(2)}_{e-sph}$ and performing the thermal average shown in Eq.~(\ref{thermalavrg}).

Figs.~\ref{acph-vs-subsph}(a) and (b) show the the scattering rates of electrons with acoustic and the substrate polar phonon modes, respectively, as a function of temperature. At all temperatures relevant to the experiments, the acoustic phonon scattering rate is more than an order of magnitude greater than the hBN substrate phonon scattering rate. Likewise, the resistivity due to acoustic phonon scattering far exceeds greatly exceeds the substrate phonon limited resistivity (Fig. ~\ref{acph-vs-subsph}c).  The insignificance of both intrinsic and substrate optical phonon scattering is easily understood qualitatively because the optical phonon energies of $\sim 100$~meV (~$\sim1200~K$) lead to a small population of these modes within the temperature range considered in this work~\cite{schiefele_temperature_2012,li_electron_2011}.

\subsection{Impurity scattering}

\begin{figure}[h!]
\begin{center}$
\begin{array}{cc}
\includegraphics[height=!,width=8cm]{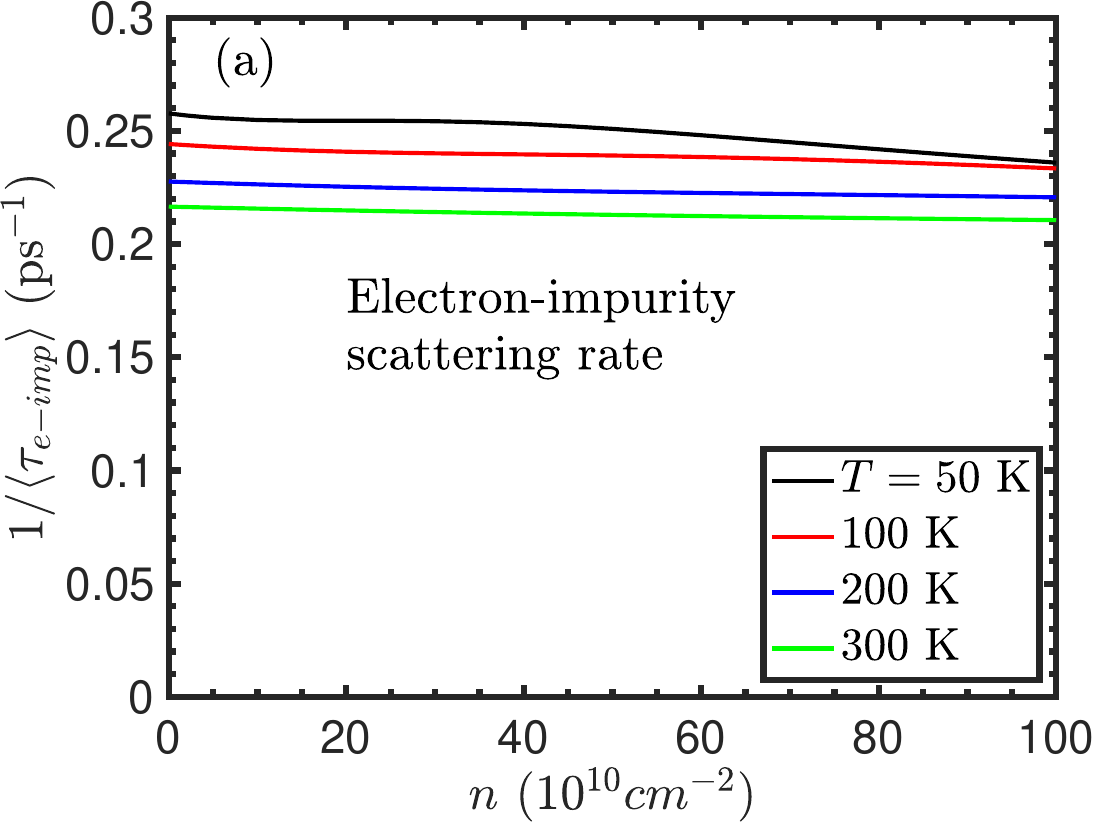} &
\includegraphics[height=!,width=8cm]{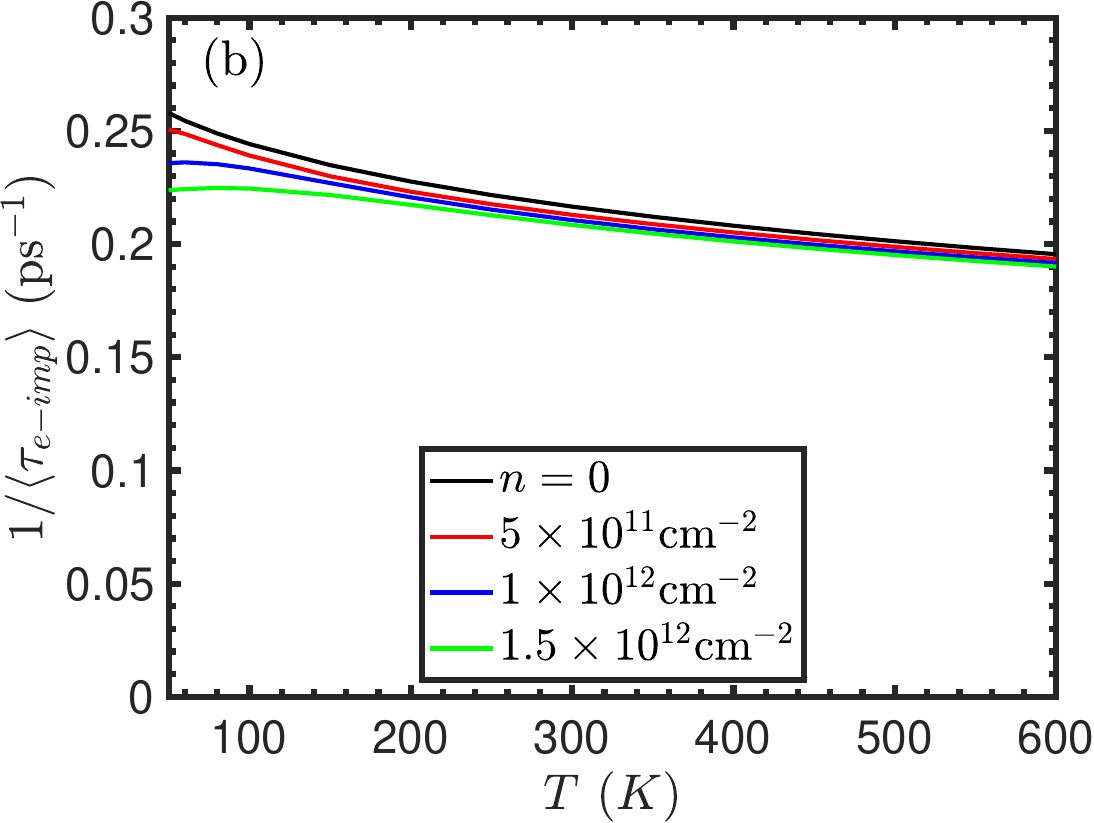} 
\end{array}$
\end{center}
\caption{\textbf{Electron-impurity scattering.}  Electron-impurity scattering rates as a function of (a) density and (b) temperature. The scattering rates change gradually with respect to both quantities.
}\label{imp-rates}
\end{figure}

We calculate the impurity scattering time using the standard expression for inverse quasiparticle lifetime due to charged impurities ~\cite{lv_screening-induced_2010, das_sarma_electronic_2011}
\begin{equation}
\frac{1}{\tau \left( \epsilon_{k,\gamma} \right)} = \pi n_{imp} \int \frac{d^{2} k^{'}}{(2\pi)^2} |\frac{v_{q}}{\varepsilon (q,\omega =0 ,\mu ,  T)}|^{2} \delta(\epsilon_{k,\gamma} - \epsilon_{k^{'},\gamma}) F_{\gamma, \gamma}(\theta_{\mathbf{k},\mathbf{k}-\mathbf{q}}) \label{imptranstime},
\end{equation}
where $n_{imp}$ is the charged impurity density, $\mathbf{k}^{'} = \mathbf{k} - \mathbf{q}$,
\begin{equation}
F_{\gamma, \gamma^{'}} (\theta_{\mathbf{k},\mathbf{k}-\mathbf{q}}) = \frac{1}{2} (1+ \mathrm{sign}(\gamma) \mathrm{sign}(\gamma^{'}) \cos (2 \theta_{\mathbf{k},\mathbf{k}-\mathbf{q}}) ),
\end{equation}
$\theta_{\mathbf{k},\mathbf{k}-\mathbf{q}}$ is the angle between $\mathbf{k}$ and $\mathbf{k}-\mathbf{q}$, and 
$\varepsilon$ refers to the RPA dielectric function defined in Eq.~(\ref{dielectricfun}) below.
We set $n_{\mathrm{imp}}=10^{10} \: \mathrm{cm}^{-2} $, $\gamma =1$ above for electrons ($-1$ is for holes) and perform the thermal average shown in Eq.~(\ref{thermalavrg}). 
We find that the scattering rate is largely independent of carrier density and temperature over the ranges relevant to this experiment (see Fig.~\ref{imp-rates}), and thus may be approximated as a constant value $\tau_{imp}$ for the purposes of fitting to experimental data (as is normally done in the literature).

Our Hall measurements in Fig.~\ref{fig_schematic} (upper inset) indicate that the charged impurity density is $n_{\mathrm{imp}} \sim 10^{10} \mathrm{cm}^{-2}$. This value corresponds to an impurity scattering time of $\tau_{imp} \sim 4$~ps (Fig.~\ref{imp-rates}). Therefore, in
Fig.~2A of the main text, we use a value of $\tau_{imp}= 8$~ps to calculate an upper bound for the impurity-limited conductivity at charge neutrality. We use a value of $\tau_{imp} = 0.8$~ps (corresponding to $n_{\mathrm{imp}}\sim 5\times 10^{10} \mathrm{cm}^{-2}$) to calculate the lower bound, since larger values of $n_{\mathrm{imp}}$ are ruled out by the Hall measurements as well as  existing literature~\cite{das_sarma_electronic_2011}.

In fitting the density- and temperature dependent resistivity data, $\tau_{imp}$ is kept as a free parameter. We find that the best match to the data is provided by $\tau_{imp} = 5$~ps, matching the value calculated above using the charged impurity density derived from the Hall measurements.



\subsection{Electron-hole scattering}\label{eh-tau-sec}

The finite temperature quasiparticle lifetime $\tau \left( \epsilon_{k,\gamma} \right)$ of a quasiparticle at energy $\epsilon_{k,\gamma} $ due to scattering with quasiparticles in the opposite band $\gamma^{'}$ is given within the Random Phase Approximation (RPA) by the expression~\cite{polini_quasiparticle_2014}
\begin{eqnarray}
&& \frac{1}{\tau \left( \epsilon_{k,\gamma} \right)}  \nonumber \\ 
&=& -\frac{2}{(2\pi)^{2}} \sum_{\gamma^{'}, \gamma^{''}} \int_{-\infty}^{\infty} d \omega \frac{1-f(\epsilon_{k,\gamma}-\omega)}{1-\exp(-\beta \hbar \omega)} \int_{0}^{\infty} dq \hspace{0.5mm} q |\frac{v_{q}}{\varepsilon (q,\omega ,\mu ,  T)}|^{2} \mathrm{Im} \chi^{(0)}_{-\gamma, \gamma^{''}} (q,\omega, \mu, T) A_{\gamma, \gamma^{'}}(k,q,\omega) \nonumber \\
& &  + \frac{2}{(2\pi)^{2}} \sum_{\gamma^{'},\gamma^{''}} \int_{-\infty}^{\infty} d \omega \frac{f(\epsilon_{k,\gamma}-\omega)}{1-\exp(\beta \hbar \omega)} \int_{0}^{\infty} dq \hspace{0.5mm} q |\frac{v_{q}}{\varepsilon (q,\omega ,\mu , T)}|^{2} \mathrm{Im} \chi^{(0)}_{-\gamma, \gamma^{''}} (q,\omega, \mu, T) A_{\gamma, \gamma^{'}}(k,q,\omega),\nonumber \\
\label{eh}
\end{eqnarray}
where 
\begin{equation}
v_{q} =\frac{2 \pi e^{2}}{\kappa q}
\end{equation}
is the bare Coulomb interaction and $\kappa=3.5$ is the dielectric constant for bilayer graphene mounted on hBN. 
$\mu$ denotes the chemical potential and $T$ denotes temperature.
$\varepsilon (q,\omega ,\mu , T)$ is the RPA dielectric function (not to be confused with the energy $\epsilon_{k,\lambda}$) given by
\begin{equation}
\varepsilon (q,\omega ,\mu , T) = 1 - \chi^{(0)}(q,\omega, \mu, T) v_{q}, \label{dielectricfun}
\end{equation}
and 
\begin{equation}
\chi^{(0)}(q,\omega, \mu, T) = \sum_{\nu, \nu^{'}} \chi^{(0)}_{\nu, \nu^{'}} \label{Lindhard}
\end{equation}
is the Lindhard polarizability function and the summation indices run over the carrier species $e$ and $h$, with components
\begin{equation}
\chi^{(0)}_{\nu, \nu^{'}}  = g \lim_{\eta \rightarrow 0} \int \frac{d^{2} k'}{(2\pi)^2} \frac{f(\epsilon_{\mathbf{k}',\nu}) - f(\epsilon_{\mathbf{k}'+\mathbf{q},\nu^{'}}) }{ \hbar \omega + \epsilon_{\mathbf{k}',\nu} - \epsilon_{\mathbf{k}'+\mathbf{q},\nu^{'}} + i \eta} F_{\nu, \nu'} (\theta_{\mathbf{k},\mathbf{k}-\mathbf{q}}), \label{Lindhardcompo}
\end{equation}
with $g=4$ being the degeneracy factor for bilayer graphene.
The first subscript of $\mathrm{Im} \chi^{(0)}_{\nu , \nu^{'}}$ is set to $-\gamma$ in Eq.~(\ref{eh}) for collisions between carriers from different bands. (To get the scattering rate for carriers in the same band, simply set $\nu$ to $\gamma$ instead.)

The $A_{\gamma, \gamma^{'}}(k,q,\omega)$ term is given by
\begin{equation}
A_{\gamma, \gamma^{'}}(k,q,\omega) \equiv \int_{0} ^{2\pi} d \theta_{\mathbf{q}} \delta (\epsilon_{k,\gamma} - \epsilon_{\mathbf{k}-\mathbf{q},\gamma^{'}} - \hbar \omega ) \times F_{\gamma, \gamma^{'}} (\theta_{\mathbf{k},\mathbf{k}-\mathbf{q}}),
\end{equation}
where 
\begin{equation}
F_{\gamma, \gamma^{'}} (\theta_{\mathbf{k},\mathbf{k}-\mathbf{q}}) = \frac{1}{2} (1+ \mathrm{sign}(\gamma) \mathrm{sign}(\gamma^{'}) \cos (2 \theta_{\mathbf{k},\mathbf{k}-\mathbf{q}}) )
\end{equation}
and $\theta_{\mathbf{k},\mathbf{k}-\mathbf{q}}$ is the angle between $\mathbf{k}$ and $\mathbf{k}-\mathbf{q}$.  
We substitute the above $\tau \left( \epsilon_{k,\gamma} \right)$ into the thermal average Eq.~(\ref{thermalavrg}) for $\gamma=1$ and display our results in black in Fig.~\ref{invtausvsn}.
At charge neutrality $\mu=0$, one can show by substitution of Eq.~(\ref{eh}) for both $\gamma=\pm 1$ into Eq.~(\ref{thermalavrg}) that $1/\langle \tau \rangle$ is linearly proportional to temperature in both cases, leading to $1/ \tau_0 \equiv  1/\langle \tau_{e}(\mu = 0) \rangle +1/\langle \tau_{h}(\mu = 0) \rangle= \alpha_{0} k_{B} T /\hbar $, with $\alpha_0 \sim 0.2$. 
This justifies the form $1/ \tau_0= \alpha_{0} k_{B} T /\hbar$ used in the main text with $\alpha_0$ as a constant fit parameter.
 Fig.~\ref{invtausvsn} also shows an exponential drop of $1/\tau_{eh}$ away from neutrality, consistent with the equation $\tau_{eh} = (n_e + n_h) / (n_{h}) \times \tau_0$ used in the main text.  For completeness, we show also results for the thermally averaged electron-electron scattering rate in red in Fig.~\ref{invtausvsn}.

\newpage 

\section{Extraction of relaxation times from experiment}

\subsection{Details of the fitting procedure}

\begin{figure}[t!]
\begin{center}
\includegraphics[height=!,width=16cm]{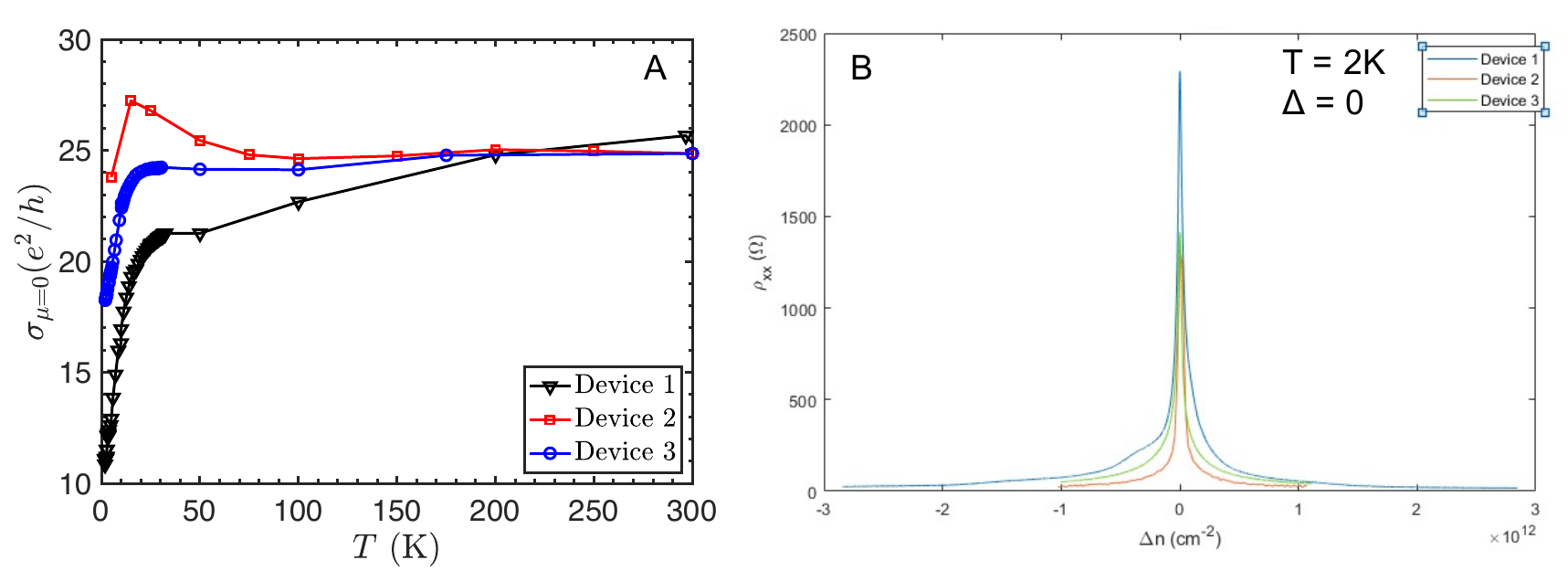}
\end{center}
\caption{\textbf{Sample variability.} (A) Measured conductivity (solid points) at $\Delta,\mu=0$ for devices 1,2, and 3, from 0~K to 300~K. The lines are a guide to the eye only. (B) Measured resistivity vs. net carrier density ($\Delta$n) at T=2K for the same three samples.  The cleanest samples with the narrowest peaks in $\rho_{xx}(\Delta n)$ (device 2 and 3) show the least variability.}
\label{dev123}
\end{figure}

Here we detail the procedure used to extract the relaxation rates from experiment. 
The hydrodynamic conductivity given by the sum of $\sigma_{\mathrm{c}} $ and $ \sigma_{\mathrm{dis}} $ in Eqs.~(\ref{sigmacasymm}) and (\ref{sigmadis}) contains three undetermined parameters $\alpha_0$ and $\tau_{\mathrm{e/h,dis}}$.  We first determine the electron-hole scattering parameter $\alpha_0$ by examining the temperature-independent conductivity at neutrality shown in Fig~2A of the main text. This data is shown in greater detail in figure in Fig.~\ref{dev123}A. As can be seen, devices 2 and 3 show near-identical conductivity between 100~K and 300~K, while device 1 shows slightly lower conductivity below 200 K. Fig. Fig.~\ref{dev123}B shows the measured resistivity across the charge neutrality point for the same samples. The width (FWHM) of the resistivity peak provides a good estimate of the device disorder~\cite{adam_self-consistent_2007, rhodes_disorder_2019}. All three devices show FWHM in the range of $10^{11}~\mathrm{cm}^{-2}$, in good agreement with the detailed Hall effect data shown for device 3 in Fig. S2. However, it is clear that device 1 has a slightly larger FWHM, indicating somewhat larger disorder. In addition, sample 1 shows a small extra peak in resistance at small negative density. This may be due to a superlattice moire pattern that modifies the low-energy bandstructure. For these reasons, the low-T data for device 1 may not precisely reflect the intrinsic behavior of bilayer graphene (at higher temperatures these effects are less important). 

We also note that the apparent anomalous behavior (i.e. slight rise in conductivity) of device 2 below 100~K is likely an experimental artifact. Its behavior was measured using a fixed grid of gate voltage points, which were not fine enough to precisely capture the exact point of charge neutrality when the resistivity peak becomes extremely narrow at low T.  Device 3 was measured by simultaneously sweeping top and bottom gates to vary the net density more finely, and was able to precisely follow the charge neutrality point even at low temperature.  Based on the above reasoning, we determine the value of $\alpha_0$ from the observed conductivity of devices 2 and 3 between 100~K and 300~K. This gives a value of $\alpha_0 = 0.225 \pm 0.002 $.   

\begin{figure}[t!]
\begin{center}$
\begin{array}{cc}
\includegraphics[height=!,width=8cm]{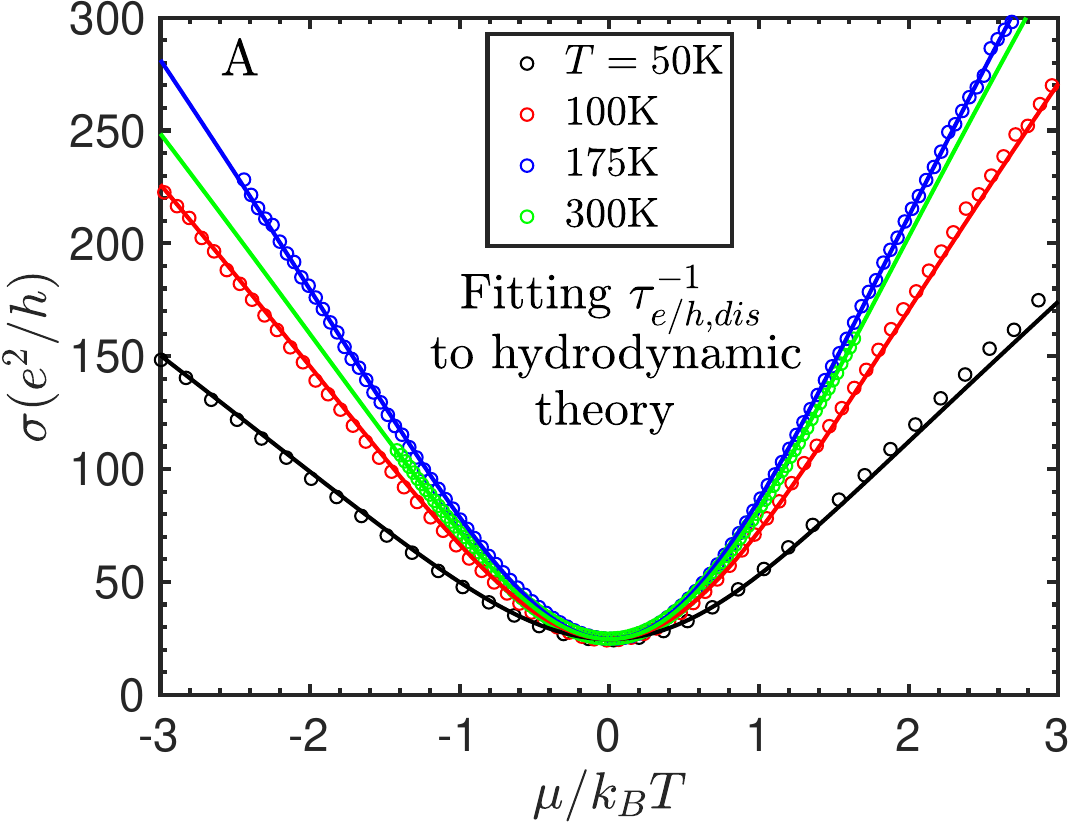} &
\includegraphics[height=!,width=8cm]{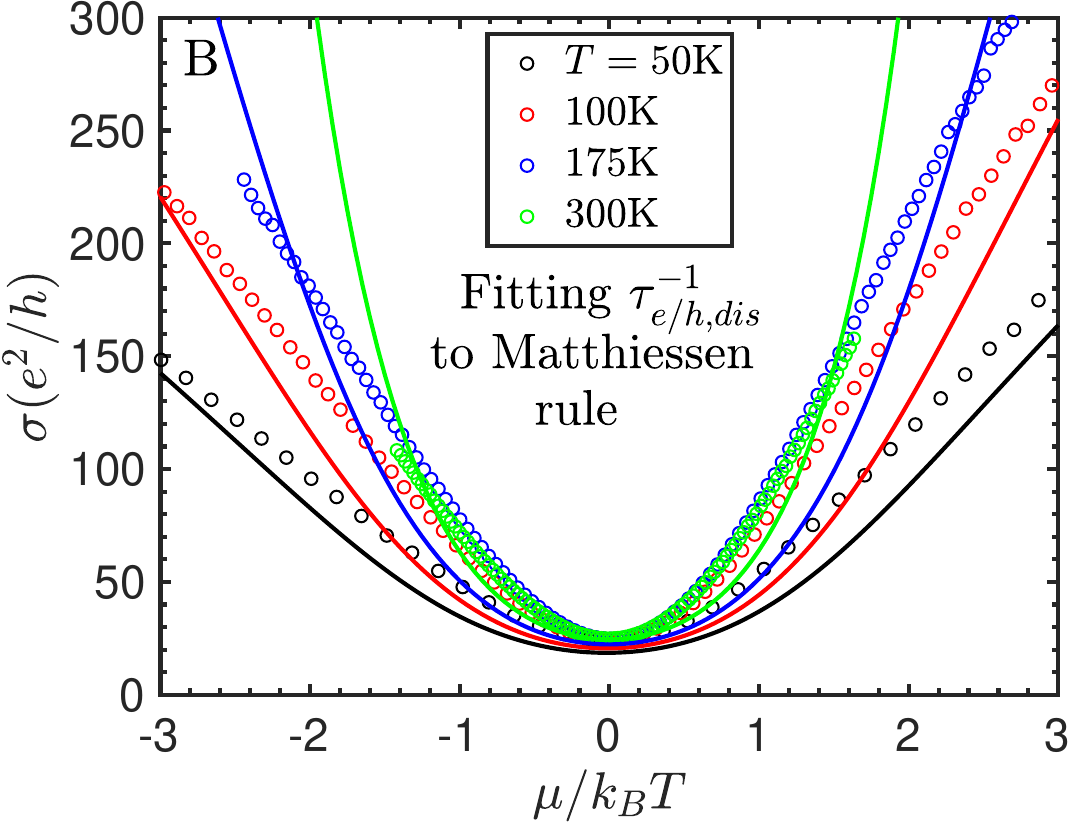} \\
\end{array}$
\end{center}
\caption{\textbf{Additional data for the gapless conductivity.} (A) Comparison of theoretical conductivity (solid lines) using hydrodynamic theory with experimental data (symbols). (B) The same using Matthiessen's rule. 
In both cases, we use the value $\alpha_0 = 0.225$ obtained from fitting to $\sigma_0$ in Fig.~2A of main text, and for $\tau_{\mathrm{e/h,dis}}^{-1}$ we use the values obtained by least square fitting to the data at each temperature. The values obtained from fitting using hydrodynamic theory are displayed in Table \ref{paramtable} below.
}\label{extracting-taus}
\end{figure}

We next measured $\sigma(\mu)$ for sample 3 at fixed temperatures of 50, 100, 175, and 300~K. This data is shown as discrete points in Figure~2B of the main text and Fig.~\ref{extracting-taus}. A least squares fit of hydrodynamic conductivity to the experimental conductivity is then performed at each temperature, with the two $\tau_{\mathrm{e/h,dis}}^{-1}$ as fit parameters and $\alpha_0 $ fixed at $0.225$. 
The resulting values obtained for $\tau_{\mathrm{e/h,dis}}^{-1}$ are shown in the left half of Table \ref{paramtable} below, and the resulting theoretical conductivities using these extracted parameters are displayed in Fig.~\ref{extracting-taus}A above.
For comparison, we repeat in Fig.~\ref{extracting-taus}B  the same fitting procedure for $\tau_{\mathrm{e/h,dis}}^{-1}$ using a phenomenological Matthiessen's rule
\begin{eqnarray}
\sigma = \frac{n_e e^2 }{m^{*}} \left(\frac{1}{\tau_{eh}}+\frac{1}{\tau_{e,dis}} \right)^{-1}  +\frac{n_h e^2 }{m^{*}} \left(\frac{1}{\tau_{he}}+\frac{1}{\tau_{h,dis}} \right)^{-1},   
\end{eqnarray}
where $\tau_{eh} = \tau_0 (n_e + n_h)/ (2 n_h) $, $\tau_{he} = \tau_0 (n_e + n_h)/ (2 n_e) $ and $\tau_0 = \hbar/ (\alpha_0 k_B T)$.
It is clear that Matthiesen's rule is unable to produce good agreement with experiment. Only hydrodynamic conductivity is able to reproduce the experimental data.
Note that because the hydrodynamic Eqs.~(\ref{sigmacasymm}) and (\ref{sigmadis}) agree with Matthiessen's rule at charge neutrality (up to terms of $O(\tau_0/ \tau_{e/h,dis} )$) where $\tau_{eh/he}^{-1} \gg \tau_{e/h,dis}^{-1}$, and also in the opposite high density regime where $\tau_{eh/he}^{-1} \ll \tau_{e/h,dis}^{-1}$, it is possible to obtain agreement with experiment at charge neutrality \emph{or} high density, but not across the the entire range.

In Fig.~3B of the main text, we perform a least squares fit of $\alpha_{ac}^{(e/h)} k_B T / \hbar + \tau_{imp}^{-1} $ to the $\tau_{\mathrm{e/h,dis}}^{-1}$ values obtained above, using $\alpha_{ac}^{(e/h)}$  and $\tau_{imp}^{-1}$ as fit parameters. 
The resulting values are displayed in the right half of Table~\ref{paramtable} below.
Note that $\tau_{imp}$ is the same for electrons and holes unlike $\alpha_{ac}^{(e/h)}$ due to the difference in effective mass between electrons and holes~\cite{zou_effective_2011}.

\FloatBarrier

\begin{table}[h!]
\caption{\textbf{List of all extracted fitting parameters.} (Left) Table of dissipative momentum relaxation rates extracted from experiment at each temperature from the fitted curves shown in Fig.~\ref{extracting-taus}A above. (Right)  Table of the phonon, impurity and electron-hole scattering parameters obtained from Fig.~3B of the main text by fitting the extracted dissipative momentum relaxation rates via $\tau^{-1}_{\mathrm{e/h,dis}} = \alpha_{ac}^{(e/h)} k_B T / \hbar + \tau_{imp}^{-1} $.  }\label{paramtable}
\vspace{0.1in}
\begin{center}$
\begin{array}{cc} 
   \begin{tabular}{|c|c|c|}
    \hline 
         T (K) & $\tau_{e,dis}^{-1}$ (ps$^{-1}$) & $\tau_{h,dis}^{-1}$ (ps$^{-1}$) \\
         \hline
         50  & 0.442 $\pm$ 0.002 & 0.519 $\pm$ 0.004  \\
         100 & 0.567 $\pm$ 0.002 & 0.685 $\pm$ 0.003 \\
         175 & 0.790 $\pm$ 0.002 & 0.961 $\pm$ 0.003 \\ 
         300 & 1.408 $\pm$ 0.002 & 1.869 $\pm$ 0.004 \\
         \hline 
    \end{tabular} &
    
 \begin{tabular}{|c|c|}
         \hline
         $\alpha_{ac}^{(e)}$ & 0.030 $\pm$ 0.008 \\
         \hline
         $\alpha_{ac}^{(h)}$ & 0.041 $\pm$ 0.008 \\
         \hline
         $\tau_{imp}^{-1}$ & 0.2 $\pm$ 0.2 ps$^{-1}$ \\
         \hline
         $ \alpha_0 $ & 0.225 $\pm$ 0.002 \\
         \hline 
    \end{tabular} 
    
    \end{array}$ 
\end{center}
\end{table}

\subsection{Ruling out ballistic transport}

Given that our samples are ultraclean, it is a valid concern as to whether electrons are simply traveling across the sample ballistically without undergoing any current-relaxing scattering events.
Indeed, Nam et al~\cite{nam_electronhole_2017} measured a negative bend resistance indicating ballistic transport for Fermi energies $E_F$ above $k_B T$ in their suspended samples that were $\sim 2 \mu \mathrm{m}$ in both width and length.  At $T=50$~K (the lowest temperature used for analysis of hydrodynamic transport) this corresponds to a density of about $3 \times 10^{11}\mathrm{cm}^{-2}$ and at at $T=300$~K this corresponds to a density of about $2 \times 10^{12}\mathrm{cm}^{-2}$.

The electron (hole) mean free path $l$ at carrier density $|n|$ can be extracted from the measured conductivity according to $\sigma = 2e^2/h \sqrt{\pi |n|} l$. Since the low-temperature conductivity increases linearly with $|n|$, the mean free path increases roughly as $\sqrt{|n|}$ and is greatest at high carrier density.  At $50$~K and at density of $|n|=3 \times 10^{11}\mathrm{cm}^{2}$, the measured conductivity of $150 \frac{2e^2}{h}$ gives a mean free path of roughly $0.75 \mu\mathrm{m}$. This is smaller than the smallest dimension of any of the samples studied here (see Table 1 of Section 1 above). Likewise, at $300$~K and density of $|n|=1 \times 10^{12}\mathrm{cm}^{2}$ (the highest density studied here), the conductivity is also roughly $150 \frac{2e^2}{h}$, corresponding to a mean free path of $0.4 \mu\mathrm{m}$. We note that at charge neutrality, $|n|$ is not given by the net charge but the total density of electrons and holes due to thermal excitation and static charge disorder. At $50$~K this value is roughly  $|n|=10^{11}\mathrm{cm}^{2}$, leading to a mean free path of $0.2 \mu\mathrm{m}$. Therefore we can conclude that the samples in this study are diffusive in the regime in which hydrodynamic transport was analyzed.

\subsection{Determination of the hydrodynamic transport regime }

While much experimental effort has been dedicated to the study of the electronic viscosity in the hydrodynamic regime, comparatively little has been dedicated to the influence of hydrodynamics on electrical transport, presumably because carrier-carrier scattering conserves total carrier momentum and thus cannot by itself yield a finite electrical resistance in general (see below for the exception). 
To put it more clearly, carrier-carrier scattering amounts to an internal force within the gas of charge carriers that cannot prevent its center-of-mass from accelerating to infinite velocity under a non-zero net force caused by an external electric field. 
The current induced by an external field in the presence of only carrier-carrier scattering is thus expected to be infinite.

An exception to this rule occurs in the case where electrons and holes co-exist in equal number, because the net force exerted on the gas of carriers (i.e.~the combined gas of electrons and holes) by the external electric field is zero.
In this special case, the internal forces within the combined gas due to electron-hole friction (i.e.~electron-hole scattering) are sufficient to prevent the charges from accelerating to infinite speed, thus leading to finite electrical resistance. 
In detail, upon switching on an external electric field, the individual electron and hole gases initially accelerate in opposite directions due to the field, but since the electron-hole frictional force increases with the difference in their relative velocity, this acceleration continues only up to the point where each gas experiences a frictional force that cancels the external field.
The velocity of each gas then remains at this constant value, yielding a finite current and conductivity. The hydrodynamic regime is thus expected to have a direct impact on electrical transport whenever electrons and holes are present in equal density, a situation that is naturally realised in  bilayer graphene at the charge neutrality point.

Nam \textit{et al.}~\cite{nam_electronhole_2017} were the first to experimentally realise this situation using ultraclean suspended bilayer graphene.
This work found a novel collapse of conductivity as a function of the ratio of Fermi energy to temperature over a range of finite but low charge densities near charge neutrality and at various temperatures, which the authors attributed solely to electron-hole scattering.
This conclusion was however problematic because as explained above, electron-hole collisions alone can only yield a well-defined resistance at precise charge neutrality.
Subsequent theoretical work~\cite{GlennWagnerDungX.Nguyen} showed that the conductivity collapse was in fact due to an interplay of electron-hole scattering and the next fastest scattering process- that of scattering between electrons and in-plane acoustic phonons.

Fig.~\ref{invtausvsn} above gives a sense of where the electron-hole dominant hydrodynamic regime lies.
It shows that a low-density window in which the electron-hole scattering rate $\tau_{eh}^{-1}$ dominates at all temperatures above $50$K, with the size of this widow increasing with temperature.

The electron-impurity scattering rate in monolayer graphene may be estimated by  $\tau_{\mathrm{imp}}^{-1} \approx 68 \times n_{\mathrm{imp}} [10^{9} \mathrm{cm}^{-2}] / T [K]$ THz~\cite{muller_collective_2008} assuming encapsulation in hexagonal boron nitride with a dielectric constant of $4$. 
According to a recent careful study of scattering times in ultraclean monolayer graphene~\cite{Gallagher2019}, the electron-hole scattering rate is given by $\tau_{eh}^{-1} \approx 0.20 k_B T / \hbar$ and $n_{\mathrm{imp}} = 2.1 \times 10^{9} \mathrm{cm}^{-2}$ in the cleanest samples.
Given the above estimates, electron-impurity and electron-hole scattering rates are equal at $\sim 70$~K. 
Achieving equality at $10$~K would then require $n_{\mathrm{imp}} \approx 4 \times 10^{7} \mathrm{cm}^{-2}$.
On the other hand, our calculation for bilayer graphene using the expression given above in Eq.~(S29) shows that a realistic impurity concentration of $n_{\mathrm{imp}} = 10^{10} \mathrm{cm}^{-2}$ corresponds to $\tau_{\mathrm{imp}}^{-1} = 0.25$~THz. Given $\tau_{eh}^{-1} \approx 0.15 k_B T / \hbar$~\cite{GlennWagnerDungX.Nguyen} in bilayer graphene, the electron-impurity and electron-hole scattering rates are equal at $\sim 10K$.

Using the model developed in this paper, we illuminate the relationship between electron-hole scattering and disorder. 
As in the main text, we use Eqs.~(\ref{sigmacasymm}) and (\ref{sigmadis}) to calculate the reduction in conductivity due to electron-hole scattering $\Delta \sigma = \sigma_{ac+i} - \sigma_{\mathrm{total}}$, where $\sigma_{ac+i}$ is the conductivity incorporating only acoustic phonons and impurities, and $\sigma_{\mathrm{total}}$ is the total conductivity.

Here we have kept all electron-hole and phonon scattering parameters consistent with those used in Fig. $5$ of the main text and varied only the impurity momentum relaxation time $\tau_\mathrm{imp}$, which takes the values $40$, $4$, $0.4$ and $0.04$ ps respectively for the aforementioned values of $n_\mathrm{imp}$.
The window for electron-hole scattering limited transport is greatly diminished across all temperatures with higher $n_\mathrm{imp}$ as impurity-limited transport dominates.
This result highlights the importance of ultra-clean bilayer graphene for hydrodynamic transport, and the degree to which hexagonal boron nitride encapsulation is needed to provide such a platform.
The electron-hole limited regime vanishes for impurity concentrations $\gtrsim 10^{11}$cm\textsuperscript{-2}, explaining why this regime was never observed in early samples possessing higher levels of disorder.

\section{Bilayer graphene with a bandgap}

Figure~\ref{fig_schematic}A shows the schematic bandstructure of bilayer graphene, which can be approximated as two hyperbolic bands, with dispersion $ \epsilon_{\pm}(k) = \pm \sqrt{\left( \hbar^2 k^2/ (2 m^{*}) \right)^2 + \left(\Delta/2 \right)^2 }$,
where $\pm$ denote the conduction and valence bands, $\hbar$ is Planck's constant, $k$ the wave vector, and $m^{*}$ the effective mass. Three relevant energy scales are shown: the field-tunable bandgap $\Delta$, and the chemical potential $\mu$ and thermal energy k$_B$T, which determine the density of thermally excited electrons and holes ($n_e$ and $n_h$). 
This dispersion is known to be valid for electrons of energy less than $0.4$~eV \cite{wang_coulomb_2010, mccann_landau-level_2006}, corresponding to density $n=10^{13}$~cm$^{-2}$ and temperature $4600$~K, making it applicable to our experiment that is restricted to $ n \lesssim 10^{12}$~cm$^{-2}$. 

\subsection{Device characterization and experimental control of carrier density and bandgap}

The devices were measured by biasing a small current (10 - 100 nA) between two outer leads, then measuring the voltages between the center longitudinal and transverse leads $V_{xx}$ and $V_{xy}$, respectively, as shown in Fig.~\ref{figdevice}. The currents $\sim 10$ nA ensure we are well within the range in which electrons may be considered to be in thermal equilibrium with the lattice even in the presence of strong electron-hole scattering~\cite{crossno_observation_2016}. We calculate the resistances by dividing the voltages by the current bias, using the results to derive the conductivity via the tensor relation:
\begin{equation}
\sigma = \frac{LR_{xx}}{W({R_{xx}^2+R_{xy}^2})} 
\label{sigmatensor}
\end{equation}
Where $L$ and $W$ are device dimensions.
The measured charge neutrality resistances of $\approx 10^5$ $ \mathrm{\Omega}$ in a current bias scheme show the system to be free from conducting defects such as strained soliton networks~\cite{Jiang2016}, consistent across devices in this work.
The dual gate devices allow us to independently control density and band gap in bilayer graphene using the top and bottom gates $V_\mathrm{TG}$ and $V_\mathrm{BG}$, respectively. 
The measured low-T resistance of one device as a function of applied top and bottom gate voltages ($V_{\mathrm{TG}}$ and $V_{\mathrm{BG}}$) is shown in Fig.~\ref{fig_schematic}C. The resistance peak along the diagonal tracks the charge neutrality point ($\mu=0$), with the emergence of a bandgap with displacement field appearing as increasing resistance toward top left and bottom right. These data, together with low-temperature Hall effect measurements (Fig.~\ref{fig_schematic}D insets), allow determination of the individual top and bottom gate capacitances (see supplementary material). 
We introduce two parameters $V_{\mathrm{eff}}$ and $\Delta_\mathrm{ext}$, which tune $\mu$ and $\Delta$ with one-to-one correspondence, respectively.
We calculate the electrostatic potential $V_\mathrm{eff}$ at a given constant displacement field $D$ (and therefore band gap $\Delta$) as:

\begin{equation}
V_{\mathrm{eff}} = \frac{(V_{\mathrm{TG}}-V_\mathrm{TG (CNP)}) - sV_{\mathrm{BG}}}{\sqrt{1+s^2}}.
\label{Veff}
\end{equation}

Here $s = -C_{\mathrm{TG}}/C_{\mathrm{BG}}$ is the negative ratio of the top and bottom gate capacitances and can be extracted from the slope of the charge neutrality point (CNP) (where $\Delta n = 0$) when plotting $R_\mathrm{xx}$ against $V_{\mathrm{TG}}$ and $V_{\mathrm{BG}}$. $V_{\mathrm{TG (CNP)}}$ is the CNP offset at $V_{\mathrm{BG}} = 0$. We also independently tune $\Delta$ with $\Delta_\mathrm{ext}$, which is calculated as:

\begin{equation}
\Delta_{\mathrm{ext}} = eDc_0 =\frac{ec_0}{2}\bigg[\frac{\epsilon_{\mathrm{TG}}(V_{\mathrm{TG}} - V_{\mathrm{TG0}})}{t_\mathrm{TG}} - \frac{\epsilon_{\mathrm{BG}}(V_{\mathrm{BG}}-V_{\mathrm{BG0}})}{t_\mathrm{BG}}\bigg],
\label{Deltaext}
\end{equation}
where $\Delta_{\mathrm{ext}} = 0$ at $V_\mathrm{TG0}$, $V_\mathrm{BG0}$, $e$ is the electron charge, $\epsilon_\mathrm{TG (BG)}$ is the top (bottom) boron nitride dielectric constant, $t_\mathrm{TG (BG)}$ is the top (bottom) boron nitride thickness, and $c_0$ is the BLG interlayer spacing.  The relation between $\Delta_{\mathrm{ext}}$ and $\Delta$ is monotonic but not straightforward, and theoretical studies of this relation vary depending on the model and details considered.
In the ranges considered in this paper, we can approximate the relation between the two as linear, $\Delta_\mathrm{ext} \approx 2.6\Delta$ as determined experimentally from Arrhenius fittings, in good agreement with tight-binding models~\cite{mccann_electronic_2013}.
We note that in the calculation of $V_{\mathrm{eff}}$ with various $\Delta_{\mathrm{ext}}$ the charge neutrality points do not align unless $s = -1$, and we've renormalized $V_{\mathrm{eff}}$ to $V_{\mathrm{eff}} - V_{\mathrm{eff (CNP)}}$ for ease of comparison. 

\begin{figure}[ht!]
\begin{center}$
\begin{array}{cc}
\includegraphics[height=!,width=8cm]{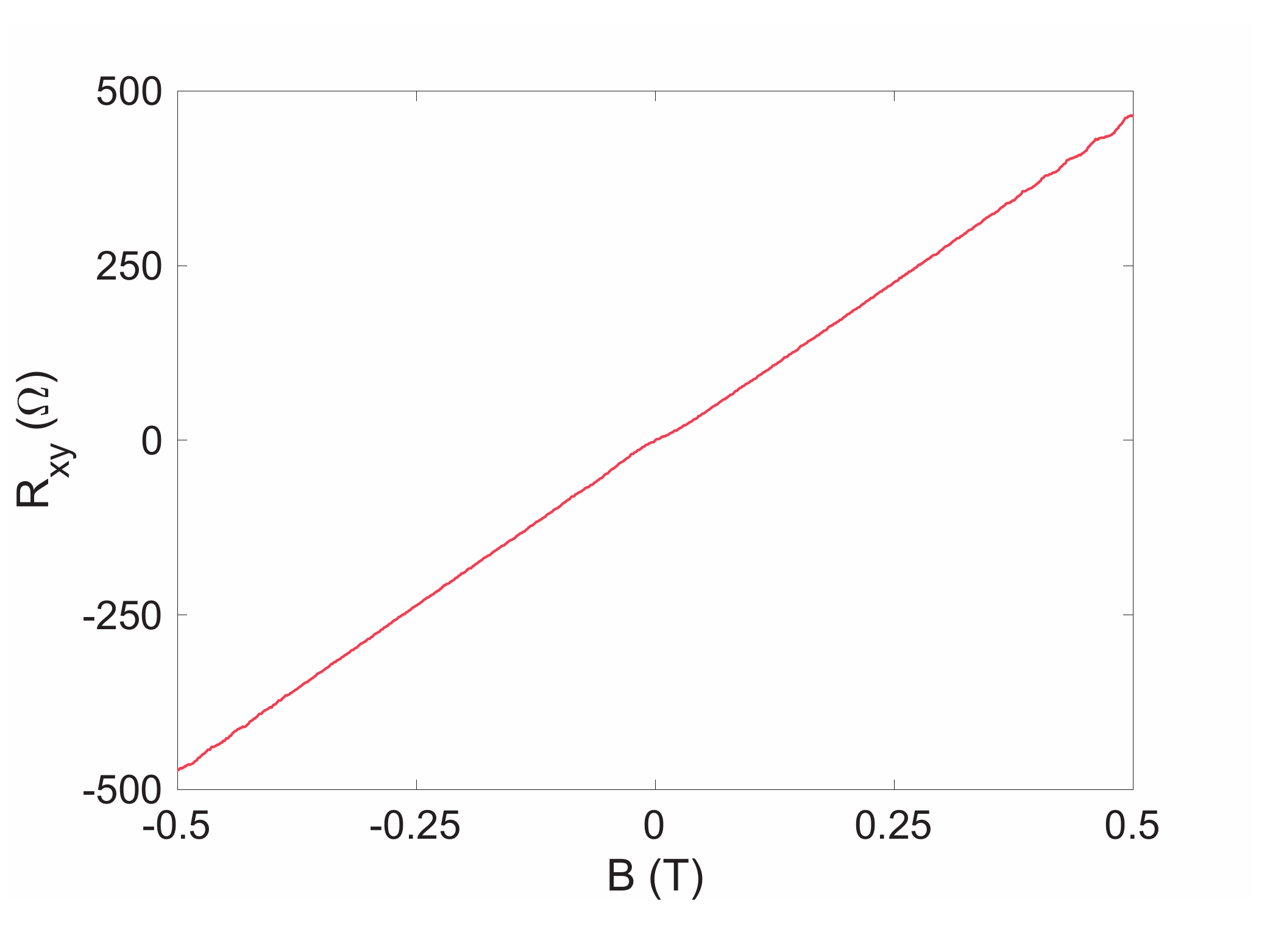} & \includegraphics[height=!,width=8cm]{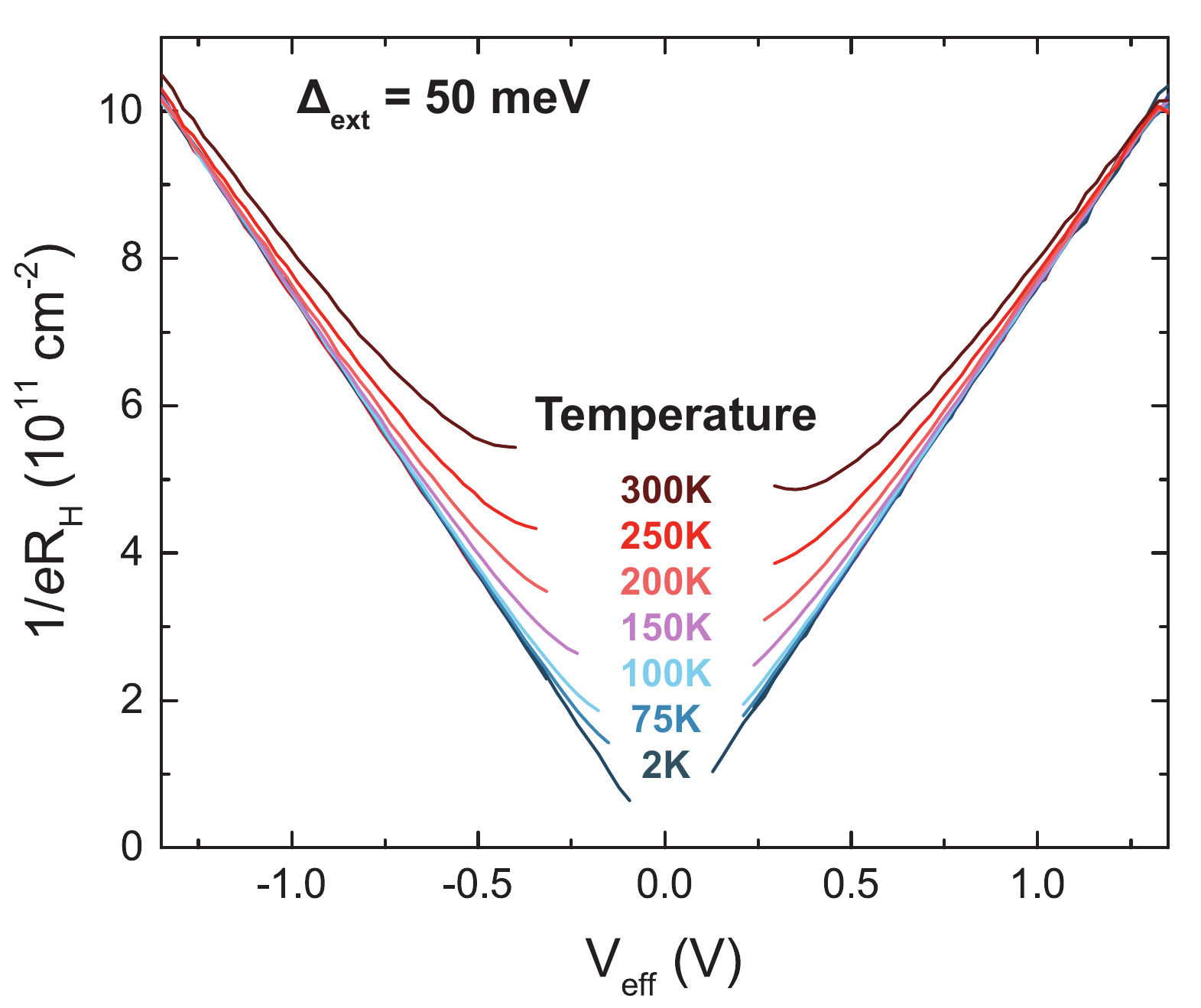} 
\end{array}$
\end{center}
\caption{\textbf{Measuring the charge density.} (a) A typical plot of $R_\mathrm{xy}$ vs $B$, from which the Hall coefficient $R_\mathrm{H}$ and charge density $\Delta n$ can be extracted.
(b) Plotted $1/eR_\mathrm{H}$ as a function of $V_\mathrm{eff}$. At low temperatures, the transport is single carrier dominated and the lines show the gate dependence of $\Delta n$. At higher temperatures, however, both electrons and holes exist near the charge neutrality point, and the single carrier transport model breaks down, resulting in non-linearity.} \label{figse1}
\end{figure}

 Fig.~\ref{fig_schematic}D show the measured conductivity as a function of $V_{\mathrm{eff}}$ for $\Delta_\mathrm{ext} = 0$ and $150$ meV, at temperatures from 5K to 300K.
Away from charge neutrality, both plots show metallic behavior. At charge neutrality, conductivity decreases upon cooling for  $\Delta_{\mathrm{ext}}=150$ meV, as expected from the opening of a band gap. Strikingly, for $\Delta_{\mathrm{ext}}=0$, the charge-neutral conductivity is large ($\sim20 \frac{e^2}{h}$) and temperature-independent.  The insets of Fig.~\ref{fig_schematic}D show $\Delta n$ as a function of $V_{\mathrm{eff}}$ at $T = 2$K for $\Delta_{\mathrm{ext}}=0$ and $\Delta_{\mathrm{ext}}=150$ meV, as  determined from Hall effect measurements.  The linear variation with $V_{\mathrm{eff}}$ to below $\approx 10^{11} \mathrm{cm^{-2}}$ sets an upper bound on charge disorder of $\sim3\times10^{10}$cm$^{-2}$, confirming that the devices are in the low-disorder limit.  The corresponding plot for $\Delta_{\mathrm{ext}}=150$ meV shows $\sim50$ meV separation between the electron and hole branches, reflecting the induced band gap.  At low temperatures such as $T = 2$K, carrier freeze out allow us to neglect minority carriers and use the single carrier model.
We calculate the single carrier charge density $\Delta n$ from the Hall coefficient $R_\mathrm{H}$ per the relation $R_\mathrm{H} = -1/\Delta ne$, where $e$ is the electron charge and $R_\mathrm{H}$ the slope of the linear relation between $R_\mathrm{xy}$ vs. $B$. A typical $R_\mathrm{xy}$ vs. $B$ plot is shown in Fig. ~\ref{figse1}.

At higher temperatures, carrier thermalization breaks down the single carrier picture near the charge neutrality point. As shown in Fig. \ref{figse1}, $1/eR_\mathrm{H}$ (the calculated single carrier charge density) deviates from a linear dependence upon $V_\mathrm{eff}$ near the charge neutrality point at higher temperatures.  This is due to the thermal population of minority carriers, even at considerably large band gaps. 
The single carrier model is therefore insufficient at higher temperatures, the region of most interest for hydrodynamic transport. 
Instead, we turn to the ambipolar transport model to extract the Hall density.
The conductivity for such a system is given by:
\begin{equation}
\sigma = e(n_eu_e+n_hu_h)
\label{sigmaeh}
\end{equation}
where the $n_e(h)$ and $u_e(h)$ are the electron (hole) density and mobility, respectively, and the mobility can be understood as $u = \frac{e\langle\tau\rangle}{m}$, where $\langle\tau\rangle$ is the carrier relaxation time.
Here we use $u$ instead of the standard $\mu$ for mobility as to avoid confusion with the chemical potential.
The Hall coefficient for the ambipolar system also becomes:
\begin{equation}
R_\mathrm{H} = \frac{n_h{u_h}^2-n_e{u_e}^2}{ e(n_eu_e+n_hu_h)^2}
\end{equation}
By assuming the carrier mobilities to be dominated by electron-hole scattering, we can then calculate the mobilities per:
\begin{equation}
    u_e = u_0\frac{n_e+n_h}{n_h},\;u_h = u_0\frac{n_e+n_h}{n_e}
\end{equation}

\begin{figure}[h!]
\begin{center}$
\begin{array}{c}
\includegraphics[height=!,width=16cm]{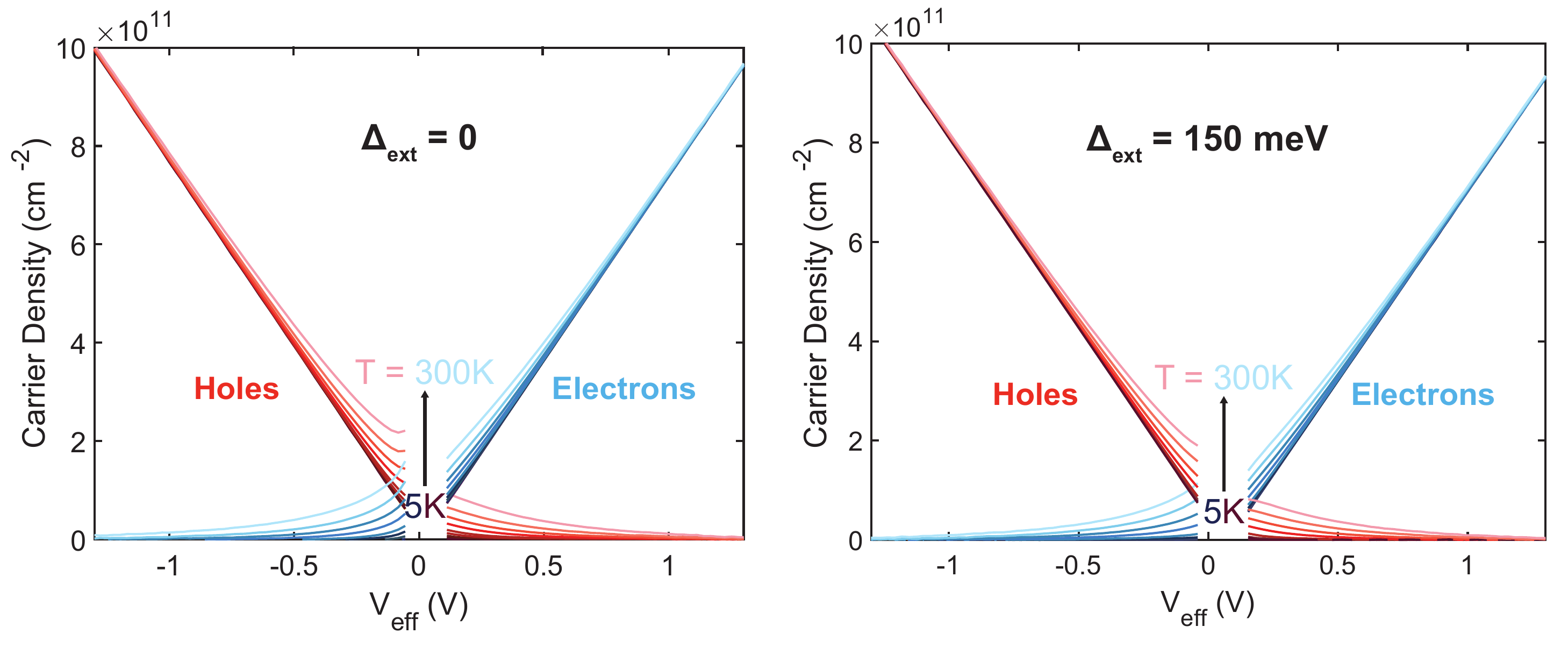} 
\end{array}$
\end{center}
\caption{\textbf{Measuring the electron and hole densities.} The extracted electron and hole densities for $\Delta_\mathrm{ext} =  0$ meV (left) and $\Delta_\mathrm{ext} =  150$ meV (right), for $T = 5\rightarrow300$K. The thermalization of carriers is shown to increase both minority and majority carriers near the charge neutrality point.
}
\label{figse4}
\end{figure}

Where $u_0=\frac{e\langle\tau\rangle_0}{m}$ is the carrier mobility at charge neutrality, $\langle\tau\rangle_0$ is calculated below.
We can then use the above equations to calculate the electron and hole densities from the measured conductivity and Hall coefficient, provided that $\Delta n = n_e-n_h$ remains consistent with the single carrier Hall density measured at $T = 2$K. We note that while this holds for the gapless case due to the parabolic approximation, it is an extremely rough approximation at large $\Delta_{ext}$, due to quantum capacitance. Nonertheless, this allows us to calculate the minority carrier concentration without the assumption of a band structure. The resulting equation is then:
\begin{equation}
    R_\mathrm{H} = \frac{(n_e-\Delta n)^3 - {n_e}^3}{e(n_e^2+(n_e-\Delta n)^2)^2}
\end{equation}
From which we numerically calculate the carrier concentrations using the experimentally extracted $R_\mathrm{H}$ at all temperatures and $\Delta n$ at $T = 2$K
The extracted carrier concentrations for varying temperatures are shown in Fig. \ref{figse4} for differing $\Delta_\mathrm{ext}$. 
Even with a sizable band gap, there is sufficient thermal population of minority carriers near the charge neutrality point at higher temperatures, bringing about hydrodynamic transport. 

\begin{figure}[h!]
\begin{center}$
\begin{array}{cc}
\includegraphics[height=!,width=8cm]{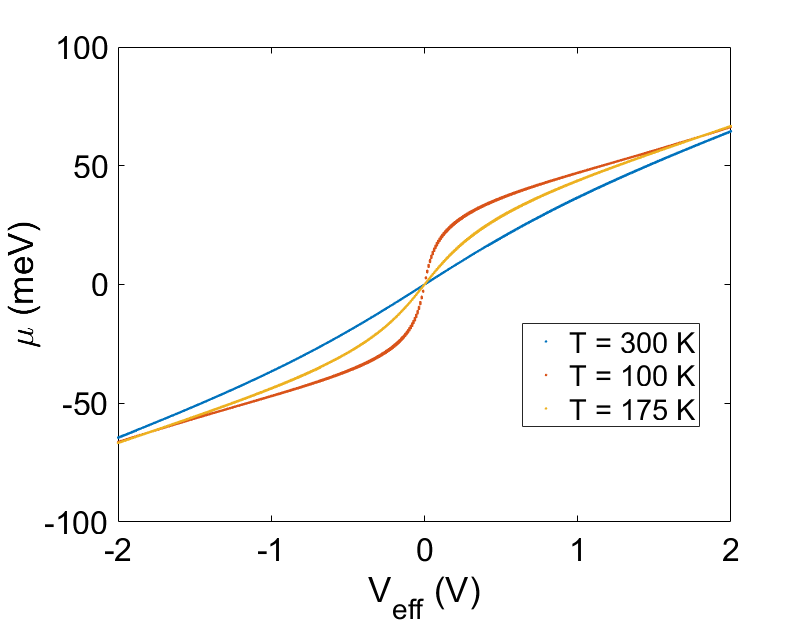} & 
\includegraphics[height=!,width=8cm]{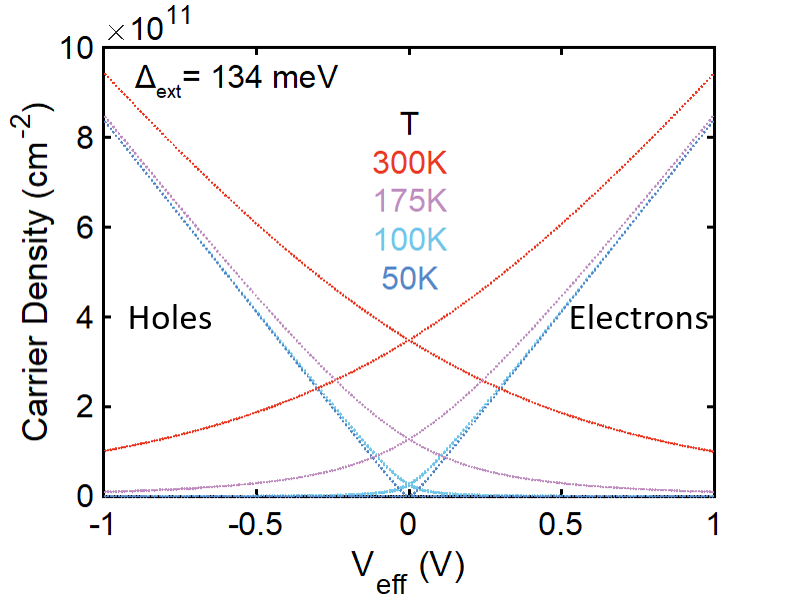}
\end{array}$
\end{center}
\caption{\textbf{Calculated chemical potential and carrier density}. (a) Calculated chemical potential plotted against $V_\mathrm{eff}$ for $\Delta_{ext} = 134$ meV for three distinct temperatures $T = 100, 175, 300$ K. (b) Calculated carrier density as a function of $V_\mathrm{eff}$ for $\Delta_\mathrm{ext} = 134$ meV at temperatures $T = 50, 100, 175, 300$ K. }
\label{muvsVeff}
\end{figure}

A key assumption of the ambipolar Hall model requires $\Delta n = n_e-n_h$ at a given $V_\mathrm{eff}$ to remain constant with temperature. 
While this assumption is valid for $\Delta = 0$, it does not hold in the case of a finite band gap.
A more involved calculation can be done, assuming a band structure.
In this section we present how we numerically calculated the chemical potential $u$ from $V_\mathrm{eff}$ with a hyperbolic band gap. 
The total capacitance of the system can be written as
\begin{equation}
    C = e\frac{\partial n}{\partial V_\mathrm{eff}} = \frac{C_\mathrm{eff}C_q}{C_\mathrm{eff}+C_q}
\end{equation}
where
\begin{equation}
    \frac{\partial n}{\partial V_\mathrm{eff}} = \frac{\partial n}{\partial \mu}\frac{\partial \mu}{\partial V_\mathrm{eff}} =
    \frac{C_q}{e^2}\frac{\partial \mu}{\partial V_\mathrm{eff}}
\end{equation}
giving the relation between $V_\mathrm{eff}$ and $\mu$ as 
\begin{equation}
    eV_\mathrm{eff} = \mu+ \frac{1}{C_\mathrm{eff}}\int_{0}^{\mu} C_q(\mu) \partial \mu 
    \label{muVeff}
\end{equation}
$C_\mathrm{eff}$ is the effective geometric capacitance as a function of $V_\mathrm{eff}$, extracted from Hall measurements. 
$C_q$ is the quantum capacitance, calculated from the density of states and derivative of the Fermi-Dirac distribution:
\begin{equation}
    C_q = \frac{e^2}{kT}\Bigg[\int_{-\infty}^{-\Delta/2}g(E, \Delta)\frac{\exp{(\frac{\mu-E}{kT})}}{\big(1+\exp{(\frac{\mu-E}{kT})}\big)^2}dE+\int_{\Delta/2}^{\infty}g(E, \Delta)\frac{\exp{(\frac{E-\mu}{kT})}}{\big(1+\exp{(\frac{E-\mu}{kT})}\big)^2}dE\Bigg]
    \label{Cq}
\end{equation}
The density of states $g(E,\Delta)$ is dependent on the bandgap $\Delta$ and given as:
\begin{equation}
    g(E,\Delta) = \frac{2m^*}{\pi\hbar^2}\frac{E}{\sqrt{E^2-(\frac{\Delta}{2})^2}}
    \label{dos}
\end{equation}
Where we approximate the effective mass $m^*$ as 0.033$m_e$.  Combining Equations \ref{muVeff}, \ref{Cq}, and \ref{dos}, we can then solve numerically for $\mu$ given $V_\mathrm{eff}$.
We plot in Fig. \ref{muvsVeff} the calculated $\mu$ as a function of $V_\mathrm{eff}$ for $\Delta_\mathrm{ext} = 134$ meV, considering temperatures $T = 100, 175, 300$ K. At high temperatures, $\mu$ is approximately linear with $V_\mathrm{eff}$ due to thermalization of carriers.

\begin{figure}[h!]
\begin{center}$
\begin{array}{cc}
\includegraphics[height=!,width=8cm]{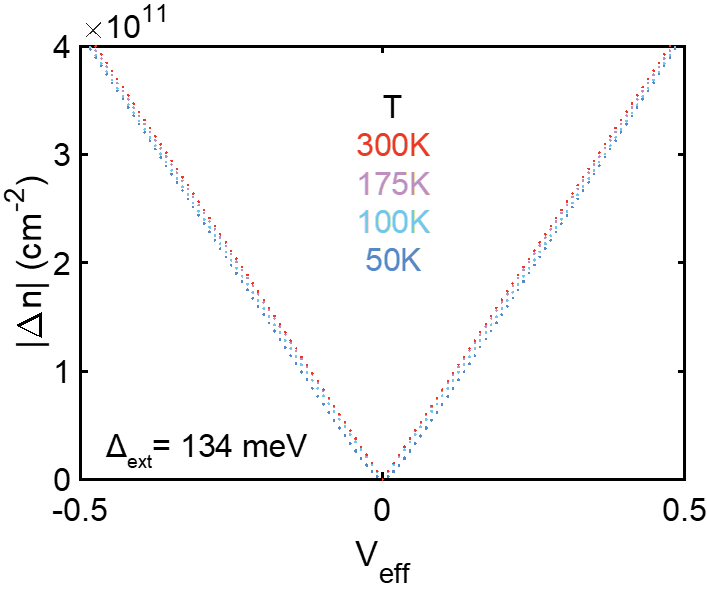} &
\includegraphics[height=!,width=8cm]{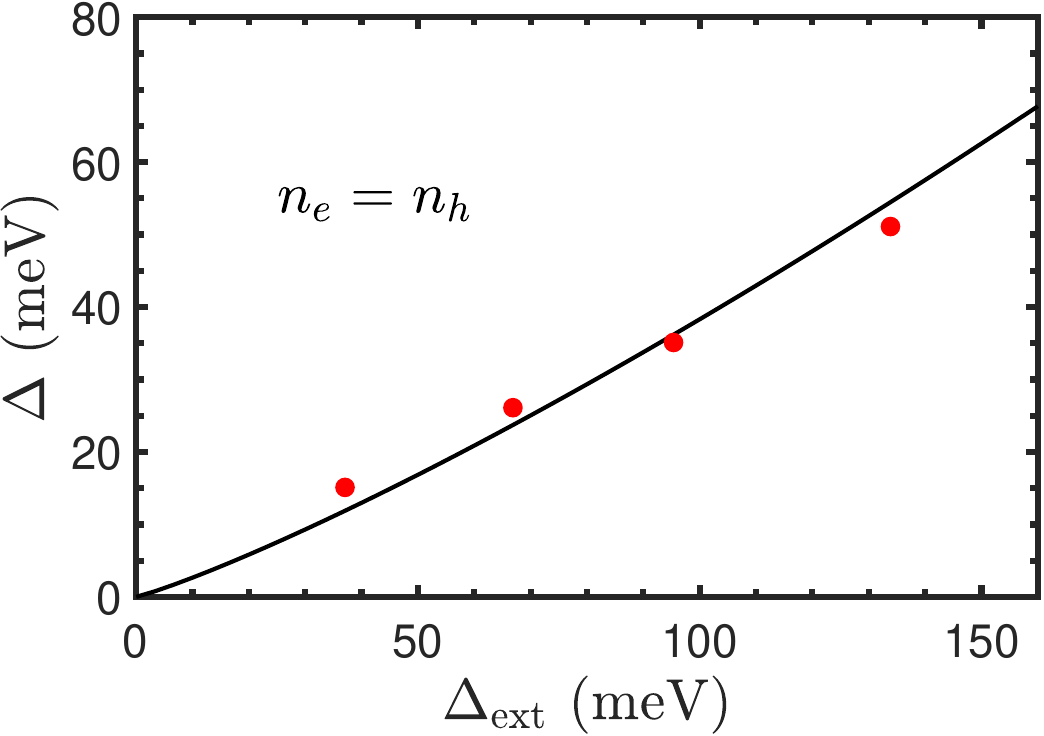} 
\end{array}$
\end{center}
\caption{\textbf{Calculated charge density and band gaps.} (a) Calculated charge density as a function of $V_\mathrm{eff}$ for $\Delta_\mathrm{ext} = 134$ meV at temperatures $T = 50, 100, 175, 300$ K.
(b) Self-consistent calculation of the gap size $\Delta$ as a function of the external bias strength $\Delta_{\mathrm{ext}}$ at charge neutrality where $n_e = n_h$. We plot the experimental band gaps extracted per the Arrhenius relationship alongside in red.}
\label{DeltanvsVeff}
\end{figure}

From $\mu$ we can then calculate the carrier densities. Fig. \ref{muvsVeff} plots the calculated carrier densities as a function of $V_\mathrm{eff}$ for $\Delta_\mathrm{ext} = 134$ meV, temperatures $T =  50, 100, 175, 300$ K. Compared with the experimentally determined carrier densities in Fig. \ref{figse4}, the calculated carrier densities are higher at each $V_\mathrm{eff}$ when compared to the gapless case.  We can then calculate the charge density and check the breakdown of the initial assumption of $\Delta n = n_e-n_h$ and its temperature dependence. 
Fig. \ref{DeltanvsVeff} shows that taking quantum capacitance into account, the charge density indeed varies with increasing temperature, as expected due to thermal excitation.

\subsection{Self-consistent determination of gap from displacement field}
When a transverse displacement field $D$ is applied across bilayer graphene, a band gap of size $\Delta$ opens up and the resulting dispersion is approximately hyperbolic, given by~\cite{mccann_electronic_2013}
\begin{equation}
\epsilon^{(hyp)}_{\pm} (k) = \pm \sqrt{ \left( \frac{\hbar^{2} k^{2}}{2 m^{*}} \right)^{2} + \left( \frac{\Delta}{2} \right)^{2} }.
\end{equation}

In this work, we determine the size of the gap for a given displacement field $D$ and density $n=n_{e}-n_{h}$ self-consistently using the method of Ref.~\cite{mccann_electronic_2013}. 
This involves evaluating the potential drop across the two graphene layers $\Delta_{\mathrm{ext}} = e c_{0} D$, where $c_{0} =3.35 $ Angstrom is the interlayer separation, and determining $\Delta$ numerically in the equation
\begin{equation}
\Delta(n) = \Delta_{\mathrm{ext}} \left[ 1 - \frac{\Lambda}{2} \log \left( \frac{|n|}{2 n_{\perp} } + \frac{1}{2} \sqrt{\left(\frac{n}{n_{\perp}} \right)^{2} + \left(\frac{\Delta}{2 \gamma_{1}} \right)^{2} } \right) \right]^{-1},
\end{equation}
where $n_{\perp} = 1.1 \times 10^{13} \mathrm{cm}^{-2}$ is a characteristic density scale. 
Solving this equation self-consistently at charge neutrality for $\Delta$ reveals the almost perfectly linear relationship between $\Delta$ and $\Delta_{\mathrm{ext}}$ in Fig.~\ref{DeltanvsVeff} below.
Experimentally, the relation between $\Delta_\mathrm{ext}$ and $\Delta$ is extracted via a linear fit, yielding the $\Delta_\mathrm{ext}\approx 2.6\Delta$ relation used in the main text.

\subsection{Modification of scattering times in the presence of a bandgap}

Within the relaxation time approximation, the non-equilibrium electron distribution $g(\vec{k})$ in the presence of a driving field $\vec{E}$ is 
 \begin{equation}
     g(\vec{k}) = f(\epsilon_{k}) - e \vec{E} \cdot \vec{v}(\epsilon_{\vec{k}}) \tau(\epsilon_{k}) \left( - \frac{\partial f (\epsilon_k ) }{\partial \epsilon_k} \right), \label{gfun}
\end{equation}
 where $\tau(\epsilon_{k})$ is the transport scattering or momentum relaxation time for a charge carrier at energy $\epsilon_k$ whose form depends on the scattering mechanism, $\vec{v}(\epsilon_k)$ the group velocity at $\vec{k}$ and $f(\epsilon_{k}) = 1/ \left( \exp[(\epsilon_{k}-\mu)/k_B T] +1 \right)$.
Following standard steps~\cite{Ashcroft}, we integrate over the Brillouin zone to find the total current density $\vec{j}$ and read off the conductivity $\sigma$ from $j = \sigma E$ to obtain
the semiclassical conductivity as before (Eq.~\ref{conductivity}).   We separate the integral into two parts ranging from $-\infty$ to $0$ and $0$ to $\infty$ and refer to these as $\sigma_{h}$ and $\sigma_{e}$ respectively. 
 The total conductivity $\sigma$ is then given by the sum $\sigma_h + \sigma_e$.
 The momentum relaxation time for each carrier species is obtained by manipulating $\sigma_{e/h}$ into the standard Drude form (Eq.~\ref{drude}) and reading off $\tau_{e/h}$. Here, 
 \begin{equation}
     n_{e/h} = \frac{2m^{*}}{\pi \hbar^2} \int^{\infty}_{\Delta/2} d \epsilon \sqrt{\epsilon^2 - \left(\frac{\Delta}{2} \right)^2} \left( - \frac{\partial f_{e/h}(\epsilon) }{\partial \epsilon} \right), \label{neh}
 \end{equation}
with $f_{e/h}(\epsilon) = 1/ \left( \exp[(\epsilon \mp \mu)/k_B T] +1 \right)$.  Following this procedure, we obtain
 \begin{equation}
      \tau_{e/h}(\Delta) = \frac{ \int^{\infty}_{\Delta /2} d\epsilon \frac{\epsilon^2 -(\Delta/2)^2}{|\epsilon|} \tau(\pm \epsilon) \left(-\frac{\partial f_{e/h}(\epsilon)}{\partial \epsilon}\right) }{\int^{\infty}_{\Delta /2} d\epsilon \sqrt{\epsilon^2 -(\Delta/2)^2} \left(-\frac{\partial f_{e/h} (\epsilon)}{\partial \epsilon}\right) }, \label{tauavrg}
 \end{equation}
 where the `+' and `$-$' are for electrons and holes respecively. 
 The collision times $\tau_{e/h} (\Delta)$ for each carrier species is obtained by substituting $\tau(\pm \epsilon)$ (i.e.~the transport scattering time for a quasiparticle at a particular energy $\epsilon$) into Eq.~(\ref{tauavrg}). 
 Note that for electron-hole scattering, Eq.~(\ref{tauavrg}) is only valid at the charge neutrality point (CNP) $\mu=0$ as that is the only density at which the relaxation time approximation is valid for total-momentum-conserving electron-hole scattering. 
 The average electron-hole collision time away from the CNP is obtained using the conservation of momentum condition $\tau_{e/h} = (n_e + n_h) / (n_{h/e}) \times \tau_0$. 

The calculation of $\tau_{e/h}$ according to Eq.~(\ref{tauavrg}) may be carried out exactly for all scattering mechanisms without difficulty in the zero-gap $\Delta= 0 $ situation for all scattering mechanisms we consider. 
 These calculations lead to the simple expressions for $\tau_{e/h}$ of the forms detailed in the main text. Evaluating Eq.~(\ref{tauavrg}) at finite gap $\Delta \neq 0 $ is difficult since evaluating $\tau(\pm \epsilon) $ with a gap followed by the energy integral in Eq.~(\ref{tauavrg}) is very demanding computationally.
 To circumvent this, we assume that for all scattering mechanisms, the expression obtained for the collision time at zero gap, denoted henceforth by $\tau_{e/h}$, may be used as a transport scattering time in place of $\tau(\epsilon_k)$ in Eq.~(\ref{gfun}) in the presence of a gap. 
 Physically, this corresponds to assuming that $\tau(\epsilon)$ for electron-phonon, electron-impurity, and electron-hole scattering are determined primarily by the phonon population, impurity concentration and overall phase space availability, all of which are unaffected by the introduction of a gap.
 The above procedure then yields for the gapped collision time 
 \begin{equation}
     \tau_{e/h}(\Delta) = \tau_{e/h} \times \frac{ \int^{\infty}_{\Delta /2} d\epsilon \frac{\epsilon^2 -(\Delta/2)^2}{|\epsilon|} \left(-\frac{\partial f_{e/h}(\epsilon)}{\partial \epsilon}\right) }{\int^{\infty}_{\Delta /2} d\epsilon \sqrt{\epsilon^2 -(\Delta/2)^2} \left(-\frac{\partial f_{e/h} (\epsilon)}{\partial \epsilon}\right) }.\label{gaptau}
 \end{equation}
 In this manner, we are able to evaluate collision times $\tau_{e/h}(\Delta)$ numerically at arbitrary gap $\Delta$ given their values at $\Delta=0$.
 The validity of the above procedure is validated by the good agreement with experiment demonstrated in the main text.
 We remind that for electron-hole scattering, Eq.~(\ref{gaptau}) may only be applied to the electron-hole scattering time $\tau_0$ at CNP. 
 The electron-hole scattering time away from CNP is then obtained by $\tau_{e/h} = (n_e + n_h) / (n_{h/e}) \times \tau_0$, where $n_{e/h}$ are given by Eq.~(\ref{neh}).
 
\begin{figure}[!ht]
\includegraphics[width=15cm]{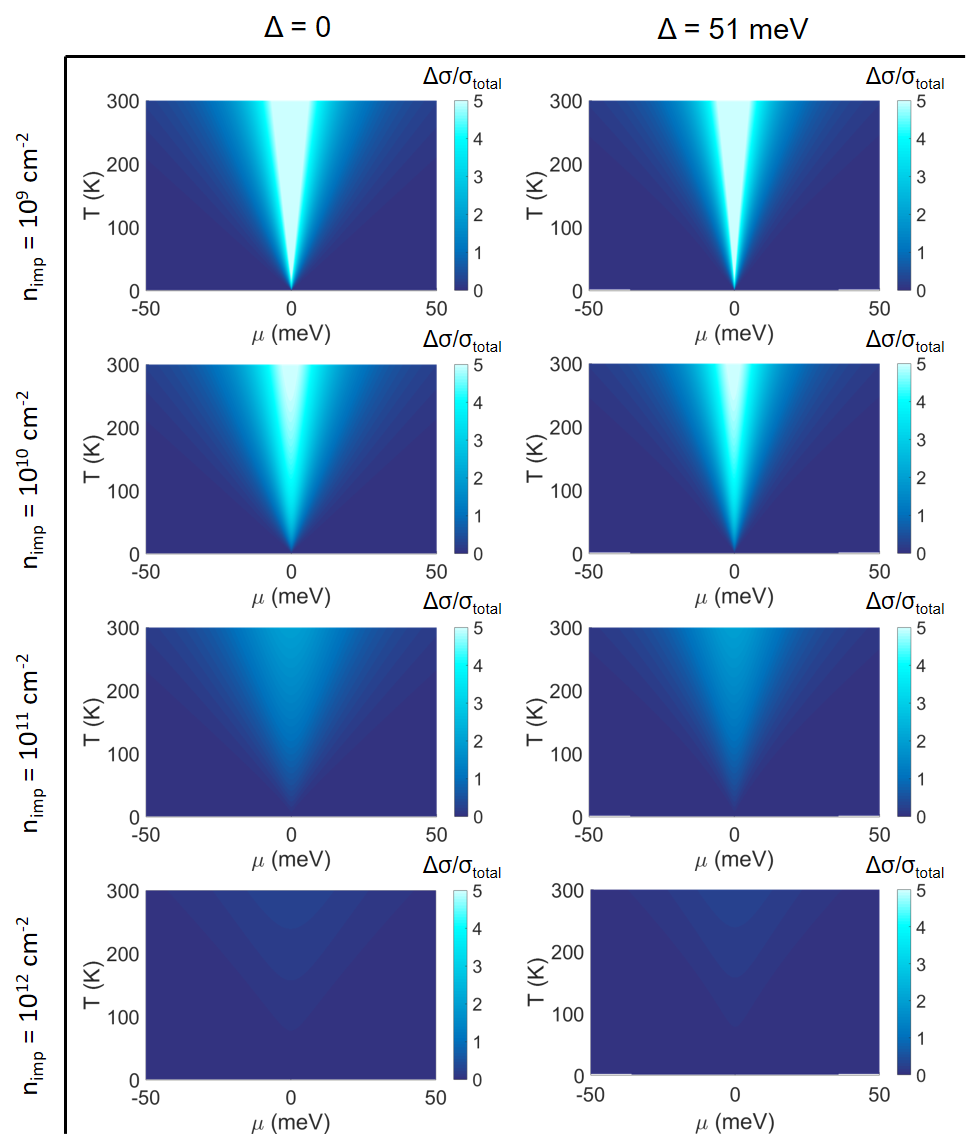}
\caption{\label{hydro_imp} \textbf{Hydrodynamic window in a semiconductor.}  Normalized reduction in conductivity due to electron-hole scattering $\Delta\sigma/\sigma_\mathrm{total}$ as a function of chemical potential $\mu$ and temperature $T$ for $\Delta$ = $0$ and $51$ meV at various impurity densities $n_\mathrm{imp}$. Increasing impurity density shrinks and eventually closes the window for hydrodynamic conductivity.}
\end{figure}

 The calculation for the gapped case $\Delta= 51$~meV is performed as follows.
For electron-impurity and electron-acoustic-phonon scattering, we insert the gapless momentum relaxation time $\tau$ as given by the fit parameters in Table 2 into Eq.~\ref{gaptau} to find $\tau_{e/h}(\Delta)$. This is the generalisation of Eq.~(\ref{thermalavrg}) to the case of finite gap.
Physically, this corresponds to assuming that $\tau(\epsilon)$ for electron-phonon, and electron-impurity are determined primarily by the phonon population and impurity concentration respectively.
For electron-hole scattering, we use Eq.~(\ref{gaptau}) at charge neutrality to obtain $\tau_0(\Delta)$. From there, we obtain the electron-hole momentum relaxation time at finite densities using the usual $\tau_{e/h} = (n_e + n_h) / (n_{h/e}) \times \tau_0$, where $n_{e/h}$ are given by Eq.~\ref{neh}.  We plot in Fig.~\ref{hydro_imp} the value of $\Delta\sigma/\sigma_\mathrm{total}$ against chemical potential $\mu$ and temperature $T$ for band gaps $\Delta = $ $0$, $51$ meV (see below for details of the calculation for $\Delta$) and impurity densities $n_\mathrm{imp} = 10^9$, $10^{10}$, $10^{11}$, $10^{12}$ cm\textsuperscript{-2}.

\FloatBarrier

\section{Evidence of universality}

The term ``universal'' is used in different contexts to mean different things.  The purpose of this section is to define how our hydrodynamic transistor is universal.  To situate this issue in a broader context, we note that in some communities the term universal is reserved for phenomena that are metrologically precise -- for example, the quantum Hall effect is now used for the SI definition of the Ohm \cite{klitzing_the_1986}.  The driving force for this switch was after the observation of the quantum Hall effect in monolayer graphene, after which it was understood that the phenomena remained universal regardless of whether the material had linear or quadratic bands.  However, in the mesoscopic community, phenomena is considered universal when it becomes independent of impurity concentration.  Since the concentration of defects varies from sample-to-sample, in this context, universal implies that the phenomena would not exhibit sample-to-sample fluctuations.  For example, ``universal conductance fluctuations'' (UCF) \cite{lee_universal_1985} are considered universal because the variance of the conductance is independent of impurity concentration (provided the temperature is sufficiently low that phase-coherence length is larger than the sample-size).  In practice, the magnitude of UCF does depend on factors like the degree of spin-orbit coupling and the applied magnetic field, and since many experiments are done at temperatures where the phase-coherence length is smaller than the sample size, the observed UCF depends on material parameters like the effective mass and impurity concentrations through the phase-coherence length \cite{altshuler_effects_1982}.  Another example is the minimum conductivity of graphene.  The first experimental transport measurements showed remarkable insensitivity to disorder\cite{geim_the_2007}, and for this reason many at the time believed that a universal mechanism was responsible.  However, we showed that this apparent universality arose from a delicate cancellation between carriers induced by impurities and the scattering of these carriers off these impurities~\cite{adam_self-consistent_2007}.  We predicted that the mechanism was not universal, and that there would be a logarithmic increase of the minimum conductivity with decreasing disorder, and effect largely confirmed experimentally~\cite{rhodes_disorder_2019}.

It is in this context that the hydrodynamic conductivity should be understood.  As the impurity concentration is further reduced such that electron-hole scattering becomes the dominant scattering mechanism, there emerges two 
unconventional contributions to the hydrodynamic conductivity.  The first contribution depends on extrinsic electron scattering mechanisms such as charged impurities or acoustic phonons.  However, it is not the usual scattering of electrons (or holes) off impurities, but the collective scattering of the electron-hole plasma.  Far from neutrality it reduces to the usual Drude diffusive transport.  This dissipative contribution, while unusual, is not universal.  It dominates away from charge neutrality and depends sensitively on both the choice of platform (bilayer graphene in our case) as well as impurity concentration.  For the second contribution, however, all the extrinsic factors, such as impurity concentration and electron-phonon coupling constant drop out, and we also demonstrate that the $\sigma = (e^2/h) (8 \log 2/\alpha_0)$ at charge neutrality.  As in the UCF example, that independence of the phenomena on impurity concentration (or sample-to-sample variation), is an example of a universal phenomena.  However, the universality in our work is even stronger since in addition to no sample-to-sample variation, material parameters like effective mass also drop out. In Figure~\ref{alpha0}, we show the inverse lifetime by temperature, i.e.~$\hbar / (\tau k_B T)$ from G$_0$W-RPA calculation for gapless parabolic bands as a function of $m^\star e^4/(\hbar^2 \kappa^2 k_B T)$.  In the limit of sufficiently strong Coulomb interaction and large effective mass, i.e~$m^\star e^4/(\hbar^2 \kappa^2 k_B*T) \gg 1$, $\hbar/\langle \tau \rangle = 0.35 k_{\rm B} T$.  This is a stronger example of universal in which the phenomena is independent of material parameters, and it is common in the literature to call such phenomena universal (e.g.~Planckian resistivity \cite{bruin_similarity_2013}) once it becomes independent of effective mass. 

\subsection{Theoretical evidence of universality: generalization to all strongly interacting hydrodynamic materials with a hyperbolic dispersion}\label{qp-vs-trans}

For elastic processes such as electron-impurity and electron-acoustic phonon scattering, the collision operator in the Boltzmann equation may be manipulated to a ``relaxation time form" i.e. $- (f(\epsilon)- f_0(\epsilon)) / \tau(\epsilon)$, where $f(\epsilon)$ and $f_0(\epsilon)$ are the non-equilibrium and equilibrium distribution functions respectively and $\tau(\epsilon)$ is the transport scattering time at energy $\epsilon$.  Then, one can use standard methods~\cite{das_sarma_electronic_2011} to obtain the conductivity in the form of 
 \begin{equation}
\sigma = e^{2} \int_{-\infty}^{\infty} d \epsilon D(\epsilon) \frac{ (v_{F}(\epsilon))^{2} \tau(\epsilon) }{2} \left(- \frac{\partial f (\epsilon ) }{\partial \epsilon} \right), \label{conductivity}
\end{equation}
which can then be expressed in the Drude form
\begin{equation}
\sigma_{e/h} = \frac{n_{e/h} e^{2}  \tau_\mathrm{e/h}  }{m^{*}},  \label{drude}
\end{equation}
giving the energy-averaged transport scattering time as shown in Eq.~(\ref{thermalavrg}).  However, for inelastic processes such as electron-hole scattering, in general, this procedure is not possible.    Instead, a reasonable way to define the momentum relaxation time would be to first calculate the conductivity $\sigma$ by rigorously solving the Boltzmann equation containing the electron-hole collision operator, then substitute this into the Drude expression $\sigma = n e^2 \tau / m^{*}$ and read off $\tau$. For monolayer graphene, this has been done numerically using a convergent set of basis functions (see e.g. Ref.~\cite{xie_transport_2016} and references therein), but to our knowledge, the analogous calculation for bilayer graphene has not yet been done.  For monolayer graphene, the conductivity is log-divergent introducing a non-universal scale $t/T$ (where $t$ is the intralayer hopping) and so the numerical value of the conductivity at a given temperature is not so interesting.  
For bilayer graphene, as we show in the main text, there is no such log-divergence, the conductivity is universal and temperature independent.  Moreover, in what follows we argue that energy-averaged electron-hole momentum relaxation time $\tau_0 = \hbar /(\alpha_0 k_B T) $ with $\alpha_0 = 0.225 \pm 0.002$ extracted from experiment should be universally observable in all strongly interacting ambipolar hydrodynamic materials with the same dispersion.

\begin{figure}[ht!]
\begin{center}
\includegraphics[height=!,width=8cm]{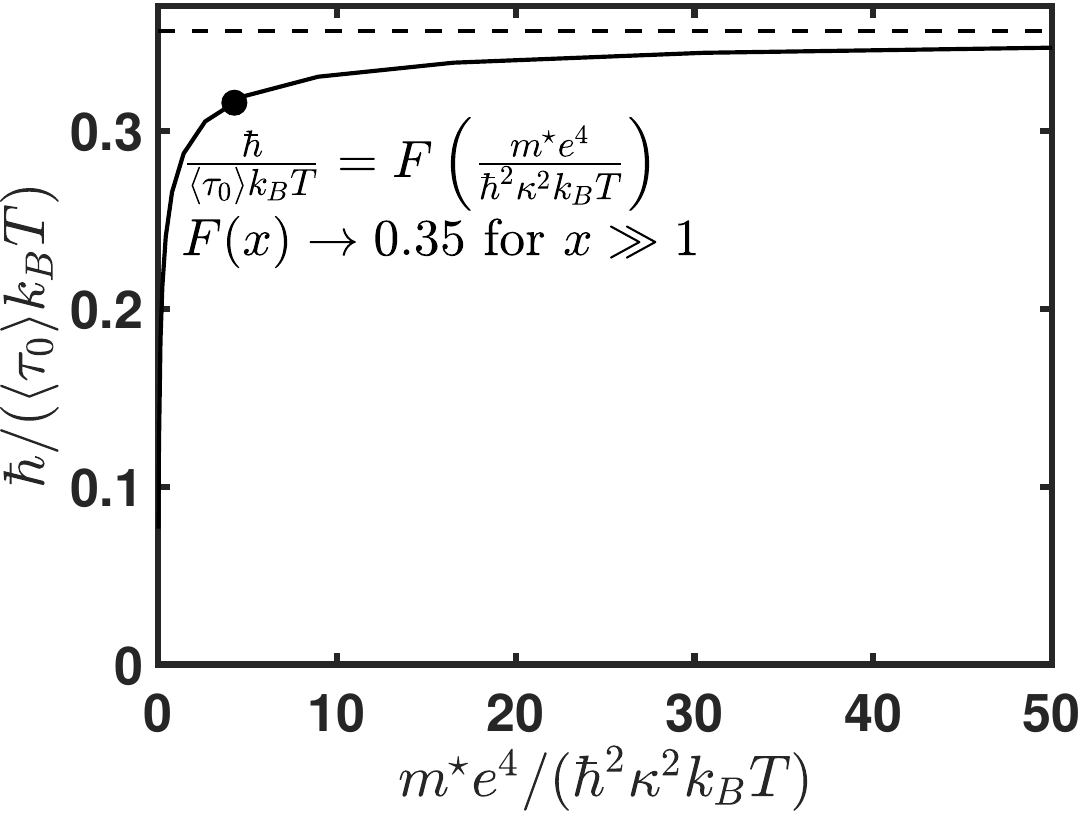} 
\end{center}
\caption{\textbf{Universal limit for the electron-hole lifetime.}  The inverse energy-averaged electron-hole quasiparticle lifetime for gapless parabolic bands $\epsilon_{\pm} = \pm \hbar^2 k^2 / 2 m^{*}$ tends to a universal form $1 / \langle \tau_0 \rangle = 0.35 k_B T / \hbar$ in the limit of strong Coulomb interaction and large effective mass $m^{*} e^4 /( \hbar^2 \kappa^2) \gg k_B T$. The black circle denotes the bilayer graphene with $\kappa = 3.5$, $T = 200$~K and $m^{*} = 0.033 m_{e}$.
}\label{alpha0}
\end{figure}

In the absence of a fully convergent numerical solution of the quantum Boltzmann equation, a common approach is to expand the non-equilbrium distribution function using a finite number of modes. This was done for bilayer graphene by Ref.~\cite{zarenia_breakdown_2019-1} using the Thomas-Fermi approximation and in Ref.~\cite{GlennWagnerDungX.Nguyen} using the leading order temperature expansion for the polarizability.  These two contemporaneous studies found $\alpha_0 = 0.15$ and $0.29$, respectively.  In our approach, we use another common method which is to approximate the energy-averaged electron-hole transport scattering time using the energy-averaged electron-hole quasiparticle lifetime which can then be obtained using the full Random Phase Approximation (RPA).  This is equivalent to assuming a relaxation time of the form $- (f(\epsilon)- f_0(\epsilon)) / \tau(\epsilon)$ for the electron-hole collision operator in the Boltzmann equation, and substituting the electron-hole quasiparticle lifetime in place of $\tau(\epsilon)$ as an effective transport scattering time.  In Fig.~\ref{alpha0} we show the inverse quasiparticle lifetime calculated using the finite-temperature RPA polarizability.  As expected, for sufficiently large Coulomb interactions $m (e^2/\kappa)^2/T \gg 1$, our numerics show that it quickly saturates to a constant universal value $0.356$, and that bilayer graphene (black dot) is already approaching this limit.  All of these different (and contemporaneous) estimates for $\alpha_0$ agree to within a factor of $\sim 2$.  An intermediate approach is to map the electron-hole collision operator onto a ``relaxation time form" using a generalization of the Bhatnagar-Gross-Krook formalism, where the quasiparticle lifetime is always smaller than the transport scattering time and found (numerically) that they differ by at most a factor of $\sim 3$.  Given this history, we find it reasonable to use the energy-averaged electron-hole quasiparticle lifetime (see Sec.~\ref{eh-tau-sec}) within a full RPA approximation to estimate the energy-averaged transport scattering time.

We note that at charge neutrality and zero gap, there exist only two energy scales in the system -- temperature $k_B T$ and the Coulomb energy $m^{*} e^{4} / \hbar^2 \kappa^2$.  Therefore,  $1/\langle \tau_0 \rangle$ must be a function of only these two energy scales. (Here, $\langle \tau_0 \rangle$ refers to Eq.~(\ref{eh}) evaluated at charge neutrality and energy-averaged using Eq.~(\ref{thermalavrg}).  Within the RPA, we find that $\hbar / (k_B T \langle \tau_0 \rangle) $ is a function only of the ratio of these two energy scales $m^{*} e^{4} / (\hbar^2 \kappa^2 k_B T)$, and becomes universal when the ratio is large.  Using the same argument, we believe that the energy-averaged transport scattering time also shows the same universal behavior as the quasiparticle lifetime.  Moreover, we can make the same argument using dimensional analysis.  Since the charge-neutral conductivity $\sigma_0$ is temperature-independent, it is therefore an experimental fact that the energy-averaged transport scattering time $\tau_0$ goes as $1/T$ since density increases linearly in $T$.  The only functional dependence of $\tau_0$ consistent with $1/T$-linearity is $1/\tau_0 = F \left( m^{*} e^{4} / \hbar^2 \kappa^2 k_B T \right) k_B T / \hbar$, where $F(x)$ is a dimensionless function that can depend on $x$ only through $x^{0}$. In other words, $F$ cannot depend on system-specific parameters such as $m^{*}$ and $\kappa$.  This convinces us that our experimentally observed value of $\alpha_0 = 0.225 \pm 0.002$ is universal.

Having established that the quasiparticle lifetime is a good estimate of the transport scattering time, we substitute $\tau(\epsilon_{k,\gamma})$ in Eq.~(\ref{eh}) into Eq.~(\ref{thermalavrg}) at charge neutrality and zero gap and evaluate the resulting expression over a range of temperatures to read off the prefactor $\alpha_{0,qp}$ in $\tau_{0,qp}^{-1} = \alpha_{0,qp} k_B T / \hbar$.  This is to be compared against the actual $\alpha_0$ in the energy-averaged transport scattering time $\tau_0$, for which $\tau_{0}^{-1} = \alpha_{0} k_B T / \hbar$.  We find that $\alpha_{0,qp}= 0.32$ for bilayer graphene at $200~$K.
  We note that the $\alpha_{0,qp}$ obtained from quasiparticle lifetime is $1.5$ times \emph{larger} than that obtained from our experiment ($\alpha_0 = 0.225$), as expected since transport scattering time is always larger than lifetime as mentioned above.  In the limit of strong Coulomb interaction $m^{*} e^{4} /(\kappa^2 \hbar^2) \gg k_B T$ and large effective mass $m^\star$, the value of $\tau_{0,qp}^{-1} $ is independent of material-specific parameters $m^{*}$ and $\kappa$, making the value of $\alpha_{0,qp} = 0.356$ in $\tau_{0,qp}^{-1} $ universally applicable in all strongly interacting materials with the same dispersion as bilayer graphene.  Consistent with our expectations, the value for $\alpha_0$ resulting from transport scattering time is slightly smaller than that from the quasiparticle lifetime as explained earlier.  We therefore expect that the universal conductivity demonstrated in this work will be reproducible in all strongly interacting ambipolar hydrodynamic materials with the same dispersion.
  
\subsubsection{Analytic calculation of $\alpha_0$}

Substituting $\tau(\epsilon)$ in Eq.~(\ref{eh}) into the energy average formula in Eq.~(\ref{thermalavrg}), we numerically evaluate the average  scattering rate $\langle \tau \rangle_{0}$ at charge neutrality and find that it is given by $0.356 k_{B} T / \hbar$ as shown in Fig.~\ref{alpha0}.
We can make some approximations to obtain analytical results that shed light on how this limit emerges.  In particular, Eq.~(\ref{eh}) is the sum of a quasi-electron term, 
\begin{eqnarray}
\frac{1}{\tau^{(eh)}_{\mathrm{qe}}(\epsilon_{\lambda}(\vec{k}) )} & =& \frac{2\pi g}{\hbar} \int\frac{d^2 q}{(2\pi)^2} \int\frac{d^2 k'}{(2\pi)^2} \sum_{\lambda^{'},\lambda^{"}} |W_{\lambda,\lambda^{"}} |^2 \nonumber \\
& & \times \left\lbrace n_{\vec{k}',-\lambda} \left( 1 - n_{\vec{k}'+\vec{q},\lambda^{'}} \right)  \left( 1 - n_{\vec{k}-\vec{q},\lambda^{"}} \right) \right\rbrace F^{-\lambda, \lambda^{'} }_{\vec{k}^{'}, \vec{k}^{'} +\vec{q}} F^{\lambda, \lambda^{"} }_{\vec{k}, \vec{k}-\vec{q}} \nonumber \\
&& \delta\left( \epsilon_{\vec{k}-\vec{q},\lambda^{"}} + \epsilon_{\vec{k}^{'}+\vec{q},\lambda^{'}} - \epsilon_{\vec{k},\lambda} - \epsilon_{\vec{k}^{'},-\lambda} \right) \label{quasiel} 
\end{eqnarray}
and a quasi-hole term that is given by the above equation with all the Fermi distribution functions replaced by one minus themselves (i.e. the probability of an electron being present is replaced by the probability of a hole being present). The quantity $|W_{\lambda,\lambda^{"}} |^2$ is the screened Coulomb interaction, 
\begin{equation}
W_{\lambda,\lambda^{"}} = \frac{\frac{2\pi e^2}{\kappa q}}{1 - \chi^{(0)}(q,\omega)\frac{2\pi e^2}{\kappa q} }.\nonumber
\end{equation}
We simplify Eq.~(\ref{quasiel}) in the low energy regime $\epsilon_{\lambda}(\vec{k}) \rightarrow 0$ and show that linear-in-$T$ behavior results.
Because the most important contributions to Eq.~(\ref{quasiel}) are expected to come from the $q \rightarrow 0$ regime, we approximate the screened Coulomb interaction $W_{\lambda,\lambda^{"}} = - \left[ \chi^{(0)}(0,0) \right]^{-1}$, and approximate $\chi^{(0)}(0,0)$ as the density of states $g m^{*} / 2 \pi \hbar^2$.
The two integrals are cut off by temperature since scattering can only take place within a window of size $k_B T$ due to Pauli-blocking. The upper limits of the radial integrals are thus defined by $\hbar^2 q^2 / 2m^{*} = k_{B} T$ and $\hbar^2 k'^2 / 2m^{*} = k_{B} T$ and the Fermi functions are approximated as equal to $1/2$ within this range of integration. 
The chirality factors are set to unity as they are not expected to cause any difference to the order of magnitude of the final result.
Finally, we consider $\lambda =1$, and neglect interband transitions due to their small contribution compared to intraband. This means that $\lambda^{''} = 1$ and $\lambda^{'} = -1$. 
Under these assumptions, Eq.~(\ref{quasiel}) becomes
\begin{equation}
\frac{1}{\tau^{(eh)}_{\mathrm{qe}}(0)} =\frac{k_{B} T}{\hbar} \frac{1}{ 4 \pi g} \int_{0}^{1} d\tilde{q} \int_{0}^{2\pi} d \theta_{\tilde{q}}  \int_{0}^{1} d\tilde{k}^{'} \int_{0}^{\pi} d \theta_{\tilde{k}^{'}} \delta\left( \cos(\theta_{\tilde{k}'} )\right), \nonumber 
\end{equation}
where we have defined $\tilde{k'}^2 =\hbar^2 k'^2 / 2 m^{*} k_B T$, $\tilde{q}^2 =\hbar^2 q^2 / 2 m^{*} k_B T$. 
Working out the integrals yields $1 / \tau^{(eh)}_{\mathrm{qe}}(0) = (1/8) k_{B} T / \hbar$.
It can be verified that the quasi-hole term contribution is equal to the quasi-electron term, as it must by electron-hole symmetry. 
Adding together the quasi-electron and quasi-hole terms in Eq.~(\ref{eh}), we obtain the result that 
$1 / \tau^{(eh)}(0) = (1/4)  k_{B} T / \hbar$.
This analytical result of dissipation at $1/4$ is very close to our full numerical solution of 0.356, which is reasonable because the energy average in Eq.~(\ref{thermalavrg}) only involves energies close to zero. 
 
\subsection{Experimental evidence of universality: Conductivity collapse curve at $\mu=0$} 
Here we derive the collapse curve in Fig.~4(C) of the main text. 
Upon substituting the appropriate expressions for density and electron-hole momentum relaxation time $\tau_{e/h}(\Delta)$ into Eq.~(1) of the main text, the electron-hole limited conductivity at charge neutrality $\sigma_{eh}(\Delta)$ normalized by its gapless value $\sigma_{eh}(\Delta=0)$ collapses as a function of the ratio $k_B T / \Delta $ according to the fit parameter-free function 

\begin{figure}[h!]
\begin{center}$
\begin{array}{c}
\includegraphics[height=!,width=8cm]{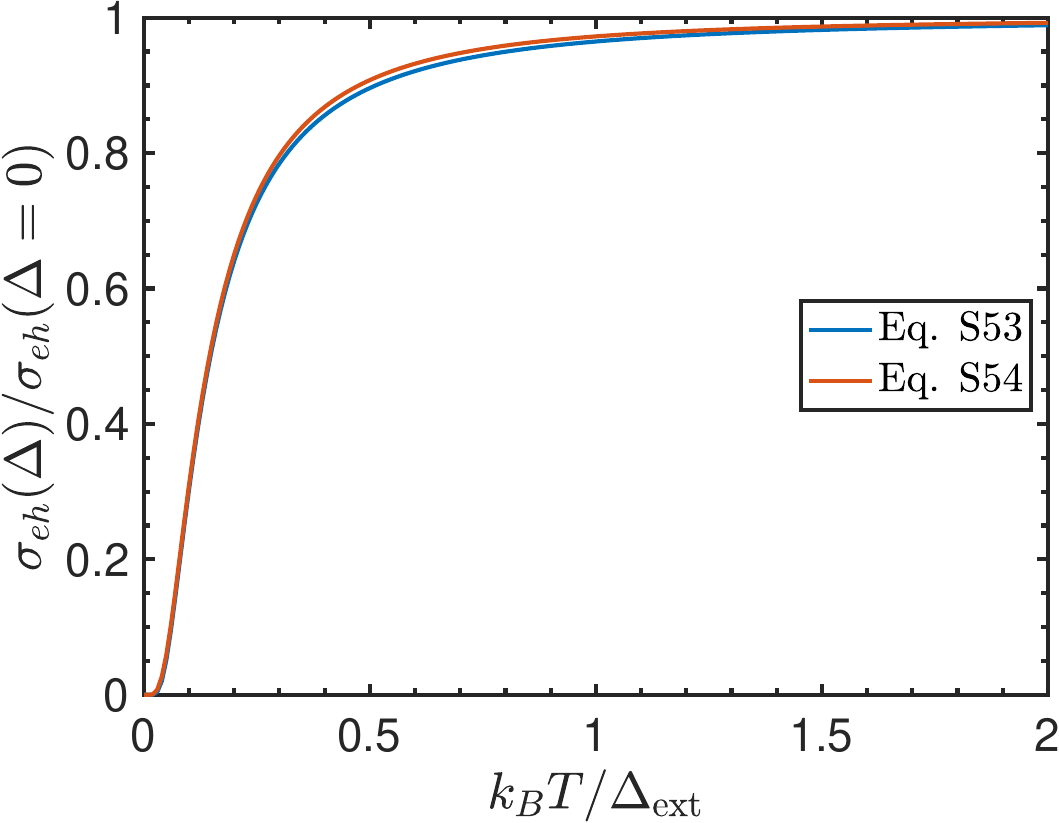} 
\end{array}$
\end{center}
\caption{\textbf{Validation of Equation~\ref{collapse2}.}
Comparison of the collapse curves of conductivity at charge neutrality against $k_BT/\Delta_\mathrm{ext}$ shown in Eqs.~\ref{collapse} (numerical) and \ref{collapse2} (analytic).  The two curves are in excellent agreement.}
\label{comparecollapse}
\end{figure}

\begin{eqnarray}
\frac{\sigma_{eh}(\Delta) }{\sigma_{eh}(\Delta=0) } &=& 1 + \frac{1}{\log(2)} \left[ \log\left( \cosh\left(\frac{ \Delta}{4 k_B T}\right) \right) - \frac{ \Delta }{4 k_B T} \tanh \left(\frac{ \Delta }{4 k_B T} \right) - \right. \nonumber \\
&& \left. \left(\frac{\Delta }{2 k_B T} \right)^2 \int^{\infty}_{\Delta / 2k_B T } dx \frac{1}{4 |x|} \frac{1}{\cosh^{2}(x/2)} \right]. \label{collapse}
\end{eqnarray}
While the integral in the final term has no exact analytical solution, it is trivial to work out numerically and may be excellently approximated using a function of the form $ F(x) = A x^2 \exp(- B x) $, where $x \equiv \Delta / k_B T $ and $A, B$ are numerical fit parameters.
We find that setting $A=1/8$ and $B=5/8$ reproduces the exact expression to an extent almost indistinguishable to the eye.
Making use of this approximation for the final term in Eq.~(\ref{collapse}), we obtain
\begin{eqnarray}
\frac{\sigma_{eh}(\Delta) }{\sigma_{eh}(\Delta=0) } = & 1 + \frac{1}{\log(2)} \left[ \log\left( \cosh\left(\frac{ \Delta}{4 k_B T}\right) \right) - \frac{ \Delta }{4 k_B T} \tanh \left(\frac{ \Delta }{4 k_B T} \right) \right. \nonumber\\
& \left. - \frac{1}{8}\left(\frac{\Delta }{k_B T} \right)^2 \exp\left( - \frac{5 \Delta}{8 k_B T} \right)\right], \label{collapse2}
\end{eqnarray}
which is Eq.~(2) of the main text. We stress that there are no fits to experimental data here and the parameters $A$ and $B$ are introduced only to remove the inconvenience of the numerical integral in Eq.~(\ref{collapse}). 
We show in Fig.~\ref{comparecollapse} that the curves produced by equations (\ref{collapse}) and (\ref{collapse2}) are almost identical.
Here we have used the same linear relationship $\Delta_\mathrm{ext} \approx 2.6\Delta$ as Fig.~4(A) of the main text.

We note that a similar collapse occurs for the commonly encountered case of gapped parabolic bands $\epsilon_{\pm} (k) = \pm \left( \Delta/2 + \hbar^2 k^2 / (2m^{*}) \right) $. 
In this case, following the same steps outlined above, the same collapse obtains but with the last term in square brackets removed. That is, 
\begin{equation}
\frac{\sigma^{(para)}_{eh}(\Delta) }{\sigma^{(para)}_{eh}(\Delta=0) } = 1 + \frac{1}{\log(2)} \left[ \log\left( \cosh\left(\frac{ \Delta}{4 k_B T}\right) \right) - \frac{ \Delta }{4 k_B T} \tanh \left(\frac{ \Delta }{4 k_B T} \right) \right]. \label{collapse3}
\end{equation}

\begin{figure}[h!]
\begin{center}$
\begin{array}{cc}
\includegraphics[height=!,width=8cm]{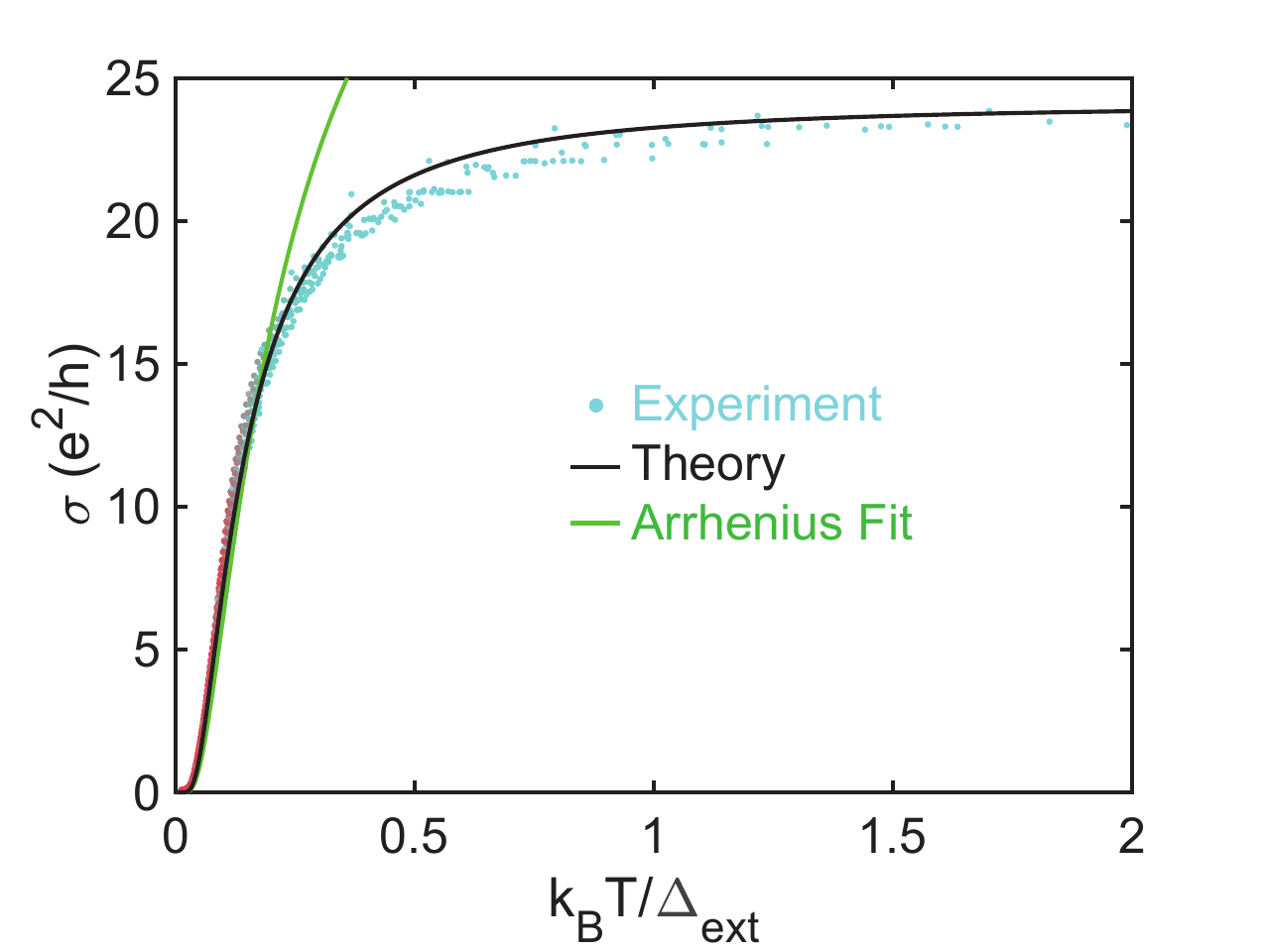} &
\includegraphics[height=!,width=8cm]{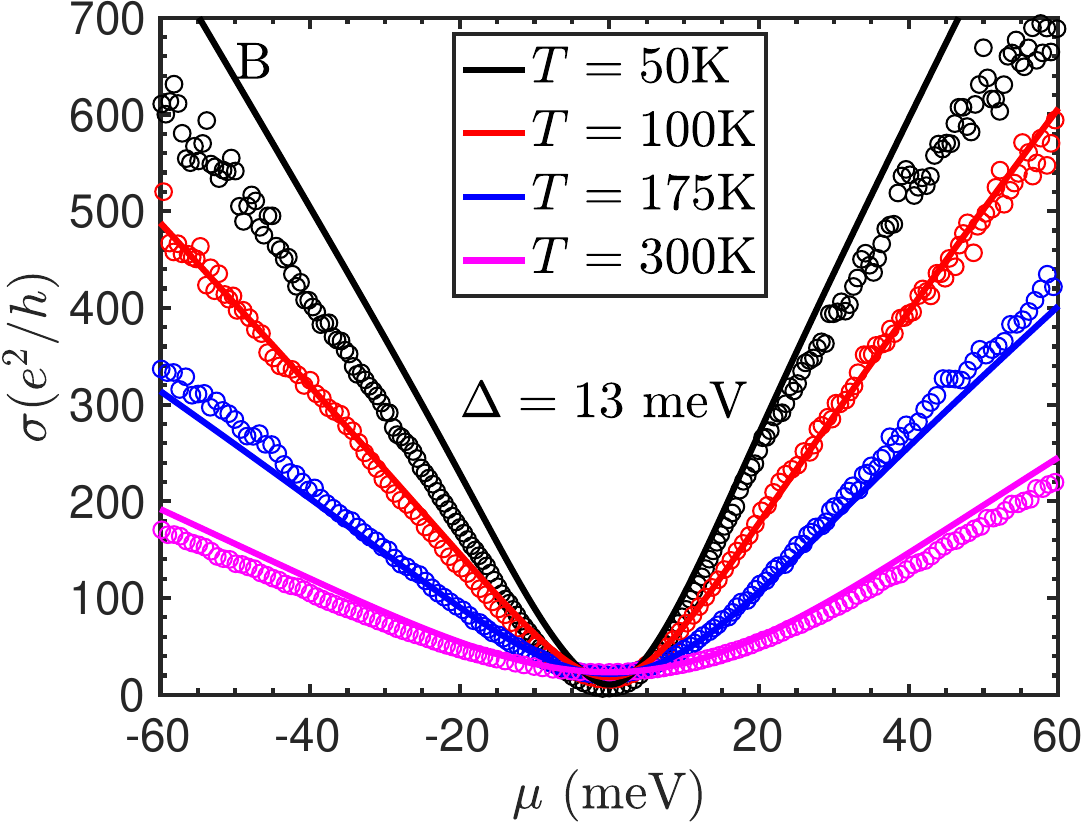} \\
\includegraphics[height=!,width=8cm]{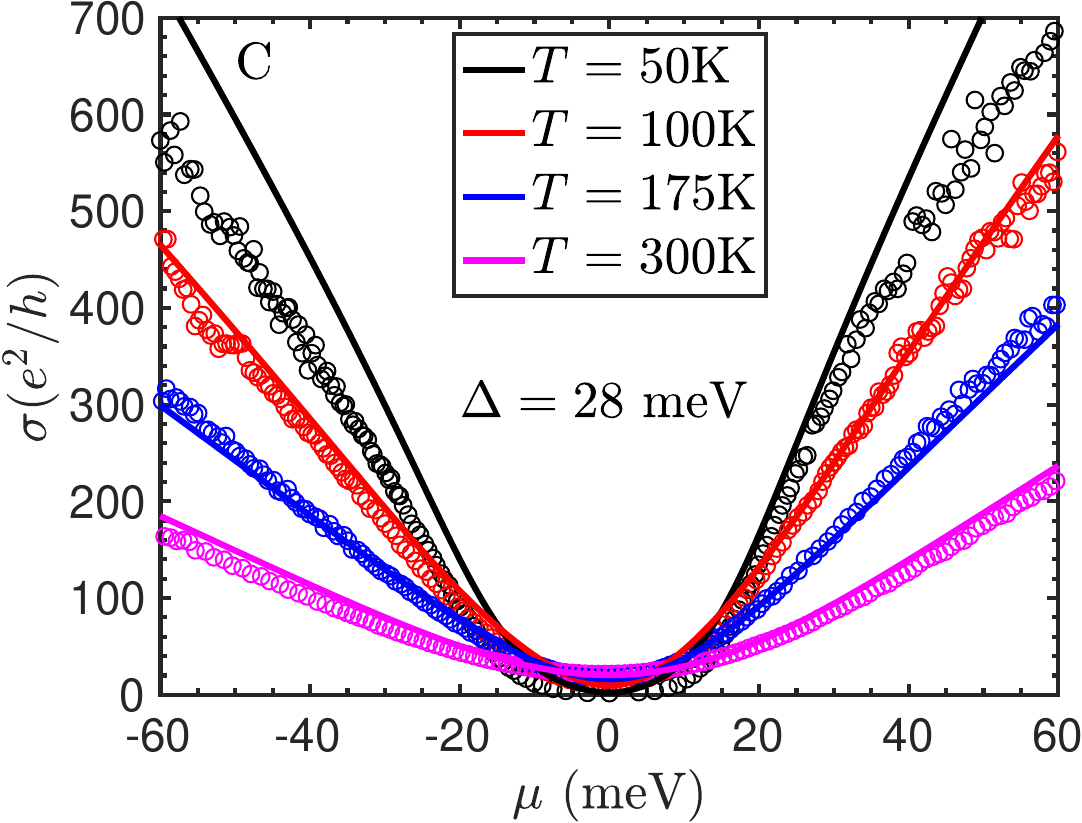} &
\includegraphics[height=!,width=8cm]{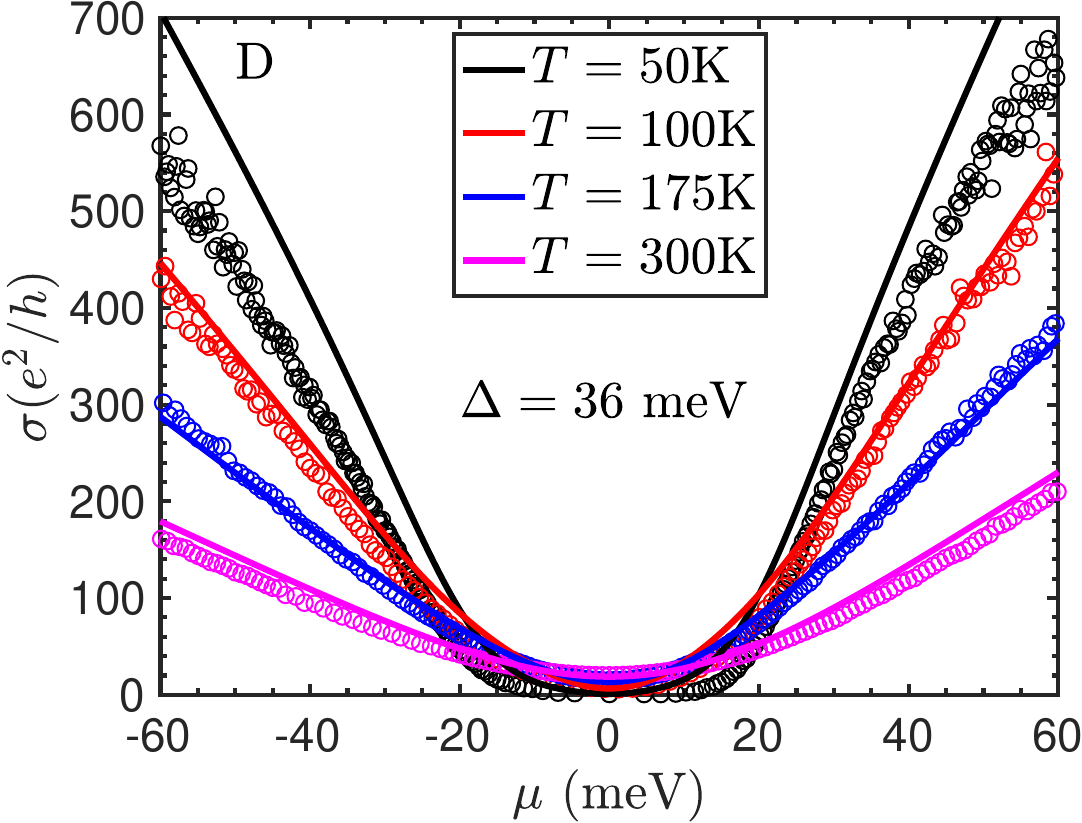} 
\end{array}$
\end{center}
\caption{\textbf{Additional data for the gapped conductivity.}  (A) Renormalized charge neutral conductivity plotted against $k_BT/\Delta_\mathrm{ext}$. The data is observed to collapse in agreement with ambipolar hydrodynamic conductivity in the strong electron-hole scattering regime, and in disagreement with the Arrhenius fit.  (B-D) Comparison of theory (solid lines) in the strong electron-hole scattering limit against experimental data (circles) at four different temperatures and three different band gaps. The same single set of fit parameters is used throughout.
}\label{gappedsigmavsmu}
\end{figure}
\FloatBarrier

\subsection{Further comparisons of theory and experiment at different gaps} 

We present in Fig. \ref{gappedsigmavsmu} the collapse of the charge neutral conductivity as a function of $k_BT/\Delta_\mathrm{ext}$ up to room temperature without normalization. 
 Here we have $\alpha_0 =0.225$ as obtained by fitting to $\sigma_0$ in the main text. 
 Note that all the other scattering parameters  $\alpha^{e/h}_{ac}$, and $\tau_{i}$ play no role at charge neutrality regardless of the gap. 
 As seen in the figure, theory collapses in excellent agreement with experiment.
 We note that this collapse deviates from Arrhenius behavior, as shown from fitting the experimental results to the Arrhenius equation $\sigma = A \exp(- \Delta / 2 k_B T)$, where $A$ is a numerical fit parameter and $\Delta = \Delta_{\mathrm{ext}} / 2.6$.  As stated in the text, we compare theory against experiment at different gaps in Fig.~\ref{gappedsigmavsmu} below, using the same single set of four fit parameters featured in Table \ref{paramtable} in Eqs.~(\ref{sigmacasymm}) and (\ref{sigmadis}).
At all gaps considered, experiment agrees with theory using only the fit parameters extracted at zero gap.

\newpage 


\bibliographystyle{Science}

\end{document}